\documentclass[nofootinbib,aps,prd,groupedaddress,preprintnumbers,%
  showpacs,showkeys,floatfix,amssymb,amsfonts]{revtex4-1}  
 
\usepackage[usenames]{color}

\usepackage{ulem}
\usepackage{graphicx}
\usepackage{bm}
\usepackage{amsmath}
\usepackage{amssymb}
\usepackage{amsfonts}
\usepackage{float}
\usepackage{dsfont}  
\usepackage{slashed}  
\usepackage{booktabs}
\usepackage{multirow}
\usepackage{subfigure}
\usepackage{slashed}
\usepackage[sort&compress]{natbib}

\usepackage{xr-hyper}

\usepackage{hyperref}

\makeatletter
\newcommand*{\addFileDependency}[1]{
  \typeout{(#1)}
  \@addtofilelist{#1}
  \IfFileExists{#1}{}{\typeout{No file #1.}}
}
\makeatother

\newcommand{\be}{\begin{equation}}  
\newcommand{\ee}{\end{equation}}  
\newcommand{\beq}{\begin{eqnarray}}  
\newcommand{\eeq}{\end{eqnarray}}

\newcommand{\bea}{\begin{eqnarray}}
\newcommand{\eea}{\end{eqnarray}}

\newcommand{\MSb}{{\overline{\rm MS}}}

\begin{document}

\title{Transversity GPDs of the proton from lattice QCD}

\author{Constantia Alexandrou$^{1,2}$,
Krzysztof Cichy$^3$,
Martha Constantinou$^4$,\\
Kyriakos Hadjiyiannakou$^{1,2}$,
Karl Jansen$^5$,
Aurora Scapellato$^4$,
Fernanda Steffens$^6$
}
\affiliation{
  \vskip 0.25cm
$^1$Department of Physics, University of Cyprus,  P.O. Box 20537,  1678 Nicosia, Cyprus\\
  \vskip 0.05cm
$^2$Computation-based Science and Technology Research Center,
  The Cyprus Institute, 20 Kavafi Str., Nicosia 2121, Cyprus \\
  \vskip 0.05cm
$^3$Faculty of Physics, Adam Mickiewicz University, Uniwersytetu Pozna\'nskiego 2, 61-614 Pozna\'{n}, Poland  \\
  \vskip 0.05cm
$^4$Department of Physics,  Temple University,  Philadelphia,  PA 19122 - 1801, USA   \\
  \vskip 0.05cm
$^5$NIC, DESY,
  Platanenallee 6,
  D-15738 Zeuthen,
  Germany   \\
  \vskip 0.05cm
$^6$Institut f\"ur Strahlen- und Kernphysik, Rheinische
  Friedrich-Wilhelms-Universit\"at Bonn, Nussallee 14-16, 53115 Bonn 
  \vspace*{-0.25cm}
  \centerline{\includegraphics[scale=0.19]{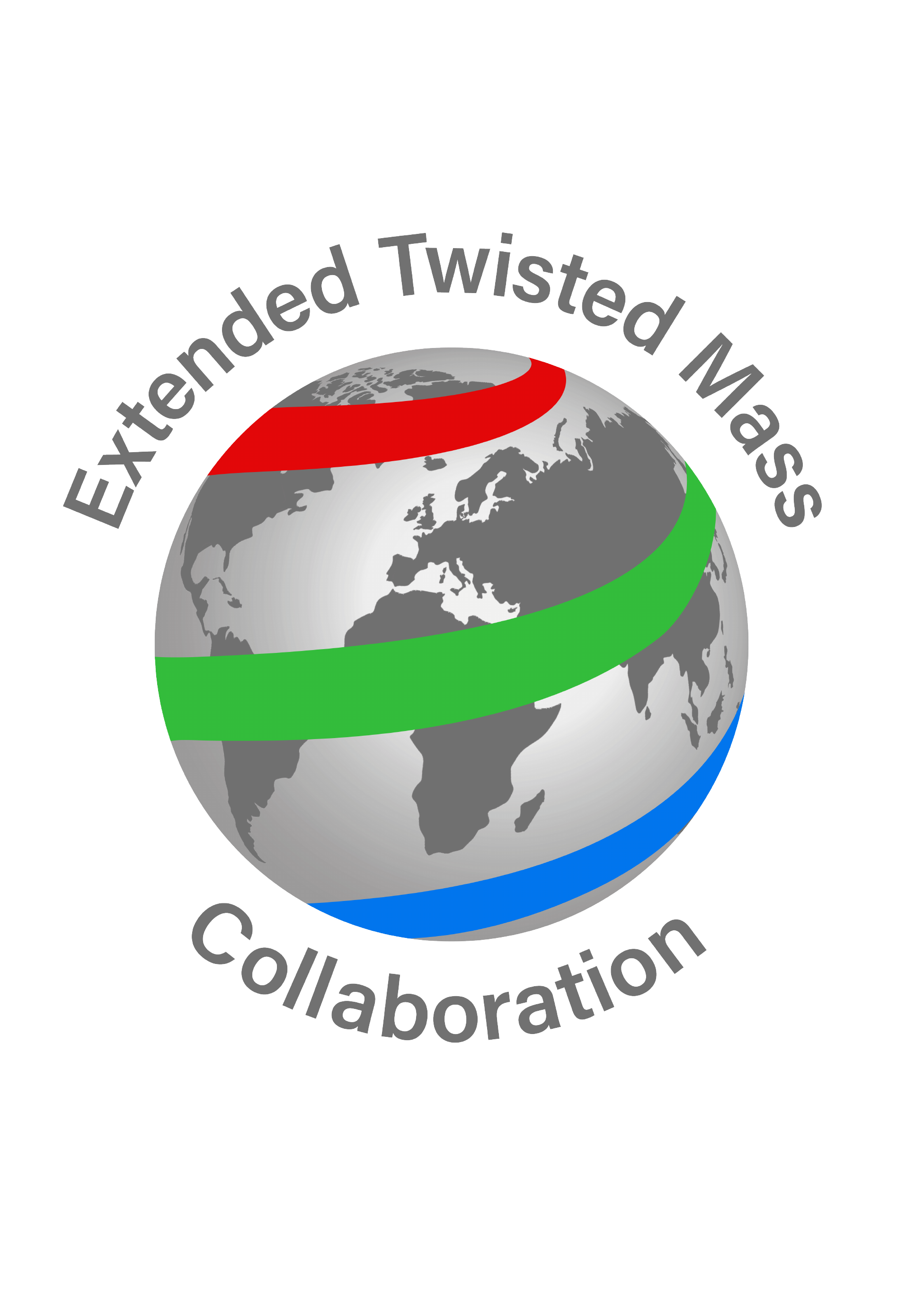}}
   }
  
\begin{abstract}
\vspace*{-.7cm}
\noindent We present the first calculation of the $x$-dependence of the isovector transversity generalized parton distributions (GPDs) for the proton within lattice QCD. We compute the matrix elements with non-local operators containing a Wilson line. The calculation implements the Breit symmetric frame. The proton momenta are chosen as $0.83,\,1.25,\,1.67$ GeV, and the values of the momentum transfer squared are $0.69,\,1.02$ GeV$^2$. These combinations include cases with zero and nonzero skewness. The calculation is performed using one ensemble of two degenerate-mass light, a strange and a charm quark of maximally twisted mass fermions with a clover term. The lattice results are renormalized non-perturbatively and finally matched to the light-cone GPDs using one-loop perturbation theory within the framework of large momentum effective theory. The final GPDs are given in the $\overline{\rm MS}$ scheme at a scale of 2 GeV. In addition to the individual GPDs, we form the combination of the transversity GPDs that is related to the transverse spin structure of the proton. Finally, we extract the lowest two moments of GPDs and draw a number of important qualitative conclusions.

\vspace*{0.75cm}

\end{abstract}

\maketitle

\section{Introduction}
 
 The current picture on the nucleon structure stems from decades of increasingly 
 precise measurements of form factors (FFs) and parton distribution 
 functions (PDFs), which, in turn, are special cases of more general 
 functions, the generalized parton distributions (GPDs). At a given hard scale $Q^2$, 
 GPDs depend on three variables: the longitudinal momentum fraction of 
 the parent nucleon carried by a given parton, $x$, the square of the 
 four-momentum transferred to the target in a given reaction, $t$, 
 and on the skewness $\xi$, which represents the change in the longitudinal 
 momentum fraction induced by the momentum transfer. Physically, GPDs 
 can be seen as correlations between the longitudinal momentum of partons, 
 with a given spin, and their position in the transverse spatial plane of 
 the parent hadron. Together with the transverse-momentum-dependent PDFs, 
 these functions give an overall, three-dimensional, picture of the nucleon,
 whose comprehension is one of the main goals of the high-energy nuclear physics community. 
 
 GPDs have been proposed in the 1990s~\cite{Muller:1994ses,Ji:1996ek,Radyushkin:1996nd,Ji:1996nm}, 
 but they are still relatively unknown when compared to their FFs and 
 PDFs counterparts. Experimentally, the access to GPDs is through exclusive 
 reactions, such as deeply virtual Compton scattering (DVCS) and deeply 
 virtual meson production (DVMP). As in the case of collinear PDFs, 
 GPDs can be separated into chiral-even and chiral-odd distributions. 
 In the chiral-even sector, there are two unpolarized, $H(x,\xi,t)$ 
 and $E(x,\xi,t)$, and two helicity, $\widetilde{H}(x,\xi,t)$ and 
 $\widetilde{E}(x,\xi,t)$, GPDs. While $H$ and $\widetilde{H}$ are helicity-preserving functions, $E$ and $\widetilde{E}$ carry information on 
 the helicity flip of the parent hadron, and contribute to the quark 
 angular momentum whilst preserving its helicity. In the forward limit, 
 $\xi, t \rightarrow 0$, $H(x,0,0) = f_1(x)$ and $\widetilde{H}(x,0,0)=g_1(x)$, 
 with $f_1(x)$ and $g_1(x)$ the unpolarized and the helicity PDFs, respectively. 
 Most of the experimental activity has been so far in the determination 
 of the helicity preserving, chiral-even distributions, 
 see Ref.~\cite{ Kumericki:2016ehc} for a comprehensive review.
 In the chiral-odd sector, there are four transversity GPDs~\cite{Diehl:2001pm}, $H_T(x,\xi,t), E_T(x,\xi,t), \widetilde{H}_T(x,\xi,t)$, and 
 $\widetilde{E}_T(x,\xi,t)$. All twist-2 GPDs are even under the 
 replacement $\xi \rightarrow -\xi$, except for $\widetilde{E}_T$, 
 which is odd. As a result, the integral over $x$ of $\widetilde{E}_T$ vanishes. 
 Of the four chiral-odd GPDs, only one survives in the forward limit~\cite{Diehl:2001pm}, $H_T(x,0,0) = h_1(x)$, where $h_1(x)$ is 
 the transversity PDF. Chiral-odd GPDs are, thus, objects 
 describing the correlation between the parton momentum and its position 
 in the transverse plane of a transversely polarized nucleon. In fact, 
 as shown in Ref.~\cite{Diehl:2005jf}, GPDs describe the density of polarized 
 partons in the impact parameter plane for both longitudinal and transverse 
 polarizations. M. Burkardt then proposed~\cite{Burkardt:2005hp} that 
 chiral-odd impact-parameter-dependent PDFs are related to the chiral-odd GPDs, 
 making possible a decomposition of the quark angular momentum with 
 respect to quarks with definite transversity. Such relations have been also explored in Refs.~\cite{Burkardt:2006ev,Bhoonah:2017olu}. As a result, a
 combination of chiral-odd GPDs can be used to calculate the correlation 
 between the quark spin and the quark angular momentum in an unpolarized nucleon. 
 In particular, the quark contribution to the nucleon transverse anomalous 
 magnetic moment can be computed from the combination $E_T(x,0,0)+ 2\widetilde{H}_T(x,0,0)$.
 Transversity GPDs are, thus, remarkably interesting objects. 
 Because they are chiral odd, they cannot be measured in DVCS, 
 making them largely unexplored. However, they can be measured using DVMP, 
 either through photon production of vector mesons~\cite{Boussarie:2016qop}, 
 or from the diffractive production of two vector mesons~\cite{Ivanov:2002jj,Cosyn:2020kfe}. 
 Notably, chiral-odd GPDs are the leading-twist contributions in $\gamma\rho$ 
 photoproduction~\cite{Beiyad:2010qg,Boussarie:2016qop} and simulations are presently being performed~\cite{Pire:2019hos} in the kinematic range of the future Electron Ion Collider (EIC) to be built at Brookhaven National Laboratory in the U.S. Also, transversity GPDs appear in the exclusive neutrino and antineutrino production of a $D$ pseudoscalar charmed meson on an unpolarized nucleon~\cite{Pire:2015iza,Pire:2017lfj}.
 
 Until recently, the study of GPDs using lattice QCD was restricted to 
 the computation of the first few Mellin moments (see, e.g., the reviews of Refs.~\cite{Syritsyn:2014saa,Constantinou:2014fka,Green:2018vxw,Lin:2017snn,Constantinou:2020hdm}). The reason is that the correlation functions defining GPDs are non-local operators sitting on the light front, and such a computation is not amenable to lattice QCD, 
 because the latter is formulated in Euclidean spacetime. Conversely, GPDs could 
 be defined in the infinite-momentum frame, in which case the hadron under 
 study receives an infinite boost, which is also not attainable in lattice QCD. 
 However, as proposed  by X. Ji~\cite{Ji:2013dva}, one can define a purely spatial 
 correlation and apply a large, but finite, momentum boost into a given direction. Then, one can use perturbation theory to connect the resulting distributions to 
 the light-front ones~\cite{Xiong:2013bka,Stewart:2017tvs,Izubuchi:2018srq}, 
 in the context of a large momentum effective theory 
 (LaMET)~\cite{Ji:2014gla,Ji:2020ect}. After LaMET was proposed, 
 other approaches, which are simultaneously concurring and complementary to 
 LaMET, have been put forward, namely pseudo-PDFs~\cite{Radyushkin:2017cyf}, 
 good lattice cross sections~\cite{Ma:2014jla,Ma:2014jga,Ma:2017pxb}, and 
 the ``OPE without OPE''~\cite{Chambers:2017dov} approaches.
 Moreover, some earlier proposed methods \cite{Liu:1993cv,Detmold:2005gg,Braun:2007wv} have been reinvestigated and further developed.
 The different approaches have been applied to the computation of a variety of quantities, markedly to quark isovector and 
 isoscalar distributions in the nucleon, to the gluon distribution in the 
 nucleon and pion, and to isovector distributions in the pion and kaon, see, e.g., Refs.~\cite{Lin:2014zya,Alexandrou:2015rja,Chen:2016utp,Alexandrou:2016jqi,Chambers:2017dov,Alexandrou:2017huk,Orginos:2017kos,Ishikawa:2017faj,Ji:2017oey,Radyushkin:2018cvn,Alexandrou:2018pbm,Chen:2018fwa,Alexandrou:2018eet,Liu:2018uuj,Karpie:2018zaz,Zhang:2018diq,Bhattacharya:2018zxi,Li:2018tpe,Ji:2018hvs,Chen:2019lcm,Sufian:2019bol,Karpie:2019eiq,Alexandrou:2019lfo,Izubuchi:2019lyk,Cichy:2019ebf,Joo:2019jct,Radyushkin:2019owq,Joo:2019bzr,Chai:2020nxw,Ji:2020baz,Braun:2020ymy,Bhat:2020ktg,Alexandrou:2020zbe,Alexandrou:2020uyt,Bringewatt:2020ixn,Liu:2020rqi,DelDebbio:2020rgv,Alexandrou:2020qtt,Liu:2020krc,Zhang:2020rsx,Huo:2021rpe,Detmold:2021uru,Karpie:2021pap,Alexandrou:2021oih,HadStruc:2021wmh}. A summary of these approaches together with lattice results can be found in the recent reviews of Refs.~\cite{Cichy:2018mum,Ji:2020ect,Constantinou:2020pek}. Very recently, LaMET has also been used to the realm of transverse-momentum-dependent PDFs~\cite{Ebert:2018gzl,Ebert:2019okf,Ji:2019sxk,Ji:2019ewn,Shanahan:2020zxr,Shanahan:2021tst,Schlemmer:2021aij}, with first results for the associated soft function being already reported~\cite{LatticeParton:2020uhz,Li:2021wvl}.  
 Even more recently, LaMET has been extended to the exploration of twist-3 PDFs in the
 nucleon~\cite{Bhattacharya:2020cen,Bhattacharya:2020xlt,Bhattacharya:2020jfj,Bhattacharya:2021moj}, as well as, twist-3 GPDs~\cite{Bhattacharya:2021rua}.
 
 Although still in their infancy due to constraints in increasing the momentum boost, as well as from systematic effects, such as discretization and volume effects, the lattice QCD calculations in the field of PDFs have advanced enormously. Returning to GPDs, the first perturbative 
 calculation of the matching equations, which relate the distributions with 
 finite momentum boost to the ones with infinite momentum, appeared soon 
 after the original proposal by X. Ji~\cite{Ji:2015qla,Xiong:2015nua}. 
 However, it took a few years until the first study of GPDs of the proton within lattice 
 QCD was performed~\cite{Alexandrou:2020zbe}. In that work, the focus was on the chiral-even GPDs for both the unpolarized and helicity case. Here, we extend this work for the chiral-odd GPDs of the proton, following the same methodology as Ref.~\cite{Alexandrou:2020zbe}.

The paper is organized as follows. In Sec.~\ref{sec:method}, we outline the general methodology of the GPDs and present the relations between the matrix elements and the GPDs that are needed to disentangle the latter. In Sec.~\ref{sec:latt_details}, we give the lattice details for the isolation of the ground state, the control of statistical uncertainties and the kinematic setup. In separate subsections, we summarize the renormalization procedure, the reconstruction of the $x$-dependence using the Backus-Gilbert method, as well as the matching formalism. The main results for the matrix elements are presented in Sec.~\ref{sec:results}, and the final GPDs are given in Sec.~\ref{sec:xGPDs}. We also present a comparison with the unpolarized and helicity GPDs. Sec.~\ref{sec:moments} shows our results for the Mellin moments of GPDs and quasi-GPDs, and Sec.~\ref{sec:summary} summarizes our findings.

\section{Methodology}
\label{sec:method}

The most computationally expensive aspect of this work is the calculation of the proton matrix elements of non-local operator containing a Wilson line. Without loss of generality, the Wilson line is in the $z$-direction, $W(0,z)$, which is the same as the direction of the momentum boost for the proton. The operator under study is the tensor with a Dirac structure of the form $\sigma^{3j}$, where $j$ is in the $x$- or $y$-direction.  Under these constraints, the matrix element reads
\begin{equation}
\label{eq:ME}
h^j_T(\Gamma_\nu,z,P_f,P_i,\mu_0)\equiv Z_T(z,\mu_0)\cdot \langle N(P_f)|\bar\psi\left(z\right) \sigma^{3j} W(0,z)\psi\left(0\right)|N(P_i)\rangle\,,\quad j=1,2\,,\,\,\,\,\,\nu=0,1,2,3.
\end{equation}
$|N(P_i)\rangle$ and $|N(P_f)\rangle$ represent the initial (source) and final (sink) state of the proton labeled by its momentum. We calculate the matrix elements $h^1_T$ and $h^2_T$ separately, because they do not contribute to the same kinematic setup, and can be used as independent equations for disentangling the GPDs (see, e.g., Eqs.~\eqref{eq:h12_zero} - \eqref{eq:h21_xi}). The matrix elements have dependence on the parity projection, $\Gamma_\nu$, which is implied in the right-hand-side of Eq.~\eqref{eq:ME} for simplicity. We will discuss this below and in Sec.~\ref{sec:latt_details}. Also, in this discussion, we consider $h^j_T$ as the renormalized matrix element in a given scheme and at a scale $\mu_0$, entering through the renormalization function, $Z_T(z,\mu_0)$. More details on the renormalization procedure are given in Sec.~\ref{sec:Zfactors}.

GPDs require off-forward kinematics, that is, $ \mathbf{P}_f - \mathbf{P}_i \equiv \mathbf{\Delta} \ne 0$. In fact, GPDs depend on the 4-vector momentum transfer squared, $t$, and not on the individual nucleon momenta. We note that the matrix element $h^j_T$ depends on the source and sink momenta. In the boosted frame, $t$ is defined as $t \equiv - \mathbf{\Delta}^2 + (E(P_i) - E(P_f))^2$. $E(p)$ is the energy of the proton at momentum $p$ given by the dispersion relation, $E(p)=\sqrt{m^2+\mathbf{p}^2}$, and $m$ is the mass of the proton.  

The standard definition of the light-cone GPDs is in the symmetric (Breit) frame, which requires that $\mathbf{P}_f=\mathbf{P} + \frac{\mathbf{\Delta}}{2}$ and $\mathbf{P}_i=\mathbf{P} - \frac{\mathbf{\Delta}}{2}$, where $\mathbf{P}$ represents the proton momentum boost, $\mathbf{P} = (0,0,P_3)$. Besides $t$, the GPDs have implicit dependence on the momentum transfer in the direction of the boost via the parameter skewness. On the lattice, the relevant quantity is the quasi-skewness defined as 
\begin{equation}
    \xi= -\frac{{P_f}_3 - {P_i}_3 }{{P_f}_3 + {P_i}_3 }= - \frac{\Delta_3}{2P_3}\,.
\end{equation}
The skewness is an important parameter of GPDs, as it separates the $x$ region into two parts, that is the Dokshitzer-Gribov-Lipatov-Altarelli-Parisi (DGLAP) region~\cite{Dokshitzer:1977sg,Gribov:1972ri,Lipatov:1974qm,Altarelli:1977zs}, and the Efremov-Radyushkin-Brodsky-Lepage (ERBL)~\cite{Efremov:1979qk,Lepage:1980fj} region, defined as
\begin{align}
{\rm DGLAP\,region:} \hspace*{1cm}  x>|\xi|\,, \nonumber \\[1ex]
{\rm ERBL\,region:}   \hspace*{1cm}  x<|\xi| \,. \nonumber
\end{align}
Each region has a physical interpretation~\cite{Ji:1998pc}. In the positive-$x$ (negative-$x$)  DGLAP region, the GPDs correspond to the amplitude of removing a quark (antiquark) of momentum $p$ from the hadron, and then inserting it back with momentum $p +\Delta$ ($\Delta$: Minkowski momentum transfer). In the ERBL region, the GPD is the amplitude for removing a quark-antiquark pair with momentum $-\Delta$. By definition, the ERBL region becomes trivial at $\xi=0$.

\medskip
As mentioned above, the matrix elements depend on the details of the kinematic setup, $P_f,\,P_i$, while the GPDs depend on $t$ and $\xi$; the remaining dependence on the setup is absorbed into the coefficients of the GPDs that appear in the decomposition. Since there are four transversity GPDs, $H_T$, $E_T$,  $\widetilde{H}_T$, and $\widetilde{E}_T$, one needs four independent matrix elements $h^j_T(\Gamma_\nu,z,P_f,P_i)$ to disentangle them; this can be controlled by the choice of the operator ($j$), parity projector ($\nu$), initial and final momenta. Note that the decomposition is independent of $z$, and is applied at each value of $z$ separately.
The decomposition of the matrix elements is based on continuum parametrizations, which for the transversity case take the following form in Euclidean space
\begin{eqnarray}
\label{eq:decomp}
h^j_T(\Gamma_\nu,z,P_f,P_i) &=& 
\langle\langle\sigma^{3j}\rangle\rangle F_{H_T}(z,\xi,t,P_3) +
 \frac{i}{2m} \langle\langle\gamma^3 \Delta_j - \gamma^j \Delta_3 \rangle\rangle F_{E_T}(z,\xi,t,P_3) \nonumber \\[0.5ex]
&& +
\frac{ P_3 \Delta_j - P_j \Delta_3 }{m^2} \langle\langle \hat{1} \rangle\rangle F_{\widetilde{H}_T}(z,\xi,t,P_3) +
 \frac{1}{m} \langle\langle \gamma^3 P_j - \gamma^j P_3 \rangle\rangle F_{\widetilde{E}_T}(z,\xi,t,P_3) \,,
\end{eqnarray}
where $\langle\langle {\cal O} \rangle\rangle \equiv \Gamma_\nu \bar{u}_N(P_f,s')\, {\cal O}  \,u_N(P_i,s)$ with $u_N$ the proton spinors. Also, $P=\frac{P_f+P_i}{2}$ and $\Delta = P_f-P_i$. $F_G$ plays the role of a form factor, which gives the quasi-GPD of $G$, $G_q$, once the Fourier transform is taken ($G:\,H_T,\,E_T,\,\widetilde{H}_T,\,\widetilde{E}_T$). The parametrization of Eq.~\eqref{eq:decomp}, in its general form, is very complicated. Here, we give the relevant expressions for the class of momentum transfer that we use. We apply four different parity projectors, that is, the unpolarized, $\Gamma_0$, and three polarized, $\Gamma_k$,
\begin{eqnarray}
\Gamma_0 &=& \frac{1}{4} (1 + \gamma^0)\,, \\
\Gamma_k &=& \frac{1}{4} (1 + \gamma^0) i \gamma^5 \gamma^k\,, \quad k=1,2,3\,.
\end{eqnarray}

The first class of momenta we employ is $\mathbf{\Delta}=(0,q,0)$, which correspond to zero skewness. In this case, the initial and final momenta are $\mathbf{P}_i=(0,-\frac{q}{2},P_3)$ and $\mathbf{P}_f=(0,\frac{q}{2},P_3)$, respectively. For these momenta, we have nonzero contributions from four matrix elements, that is:
\begin{eqnarray}
\label{eq:h12_zero}
h^1_T(\Gamma_2,z,P_f,P_i) &=& 
- i \,C_0\frac{ E ( E +m)}{2 m^2} \, F_{H_T}\,, \\[2ex]
\label{eq:h13_zero}
h^1_T(\Gamma_3,z,P_f,P_i) &=& 
- i \,C_0\frac{q\,P_3\,( E +m)}{4 m^3} \, F_{\widetilde{E}_T}\,, \\[2ex]
\label{eq:h20_zero}
h^2_T(\Gamma_0,z,P_f,P_i) &=& 
i \,C_0 \left(
\frac{q\, P_3 }{4 m^2}\, F_{H_T} 
+ \frac{q\, P_3 \,  ( E +m)}{4 m^3}\, F_{E_T} 
+\frac{q\, P_3 \,  \left( E ( E +m)- P_3 ^2\right)}{2 m^4} \, F_{\widetilde{H}_T}
\right)\,, \\[2ex]
\label{eq:h21_zero}
h^2_T(\Gamma_1,z,P_f,P_i) &=& 
i \,C_0\, \left(
\frac{  \left(m( E +m)+ P_3 ^2\right)}{2 m^2}\,F_{H_T}
-\frac{q^2\, ( E +m)}{8 m^3}\,F_{E_T}
+\frac{q^2 \,P_3^2 }{4 m^4}\, F_{\widetilde{H}_T} \right)\,,
\end{eqnarray} 
where $C_0=\frac{2 m^2}{E (E+m)}$ for zero skewness, and $E$ denotes the energy ($E_f=E_i\equiv E$). Note that $F_{H_T}$ and $F_{\widetilde{E}_T}$ are obtained directly from Eq.~\eqref{eq:h12_zero} and Eq.~\eqref{eq:h13_zero}, respectively. $F_{E_T}$ and $F_{\widetilde{H}_T}$ are disentangled using Eqs.~\eqref{eq:h20_zero} - \eqref{eq:h21_zero} together with Eq.~\eqref{eq:h12_zero}.

For nonzero skewness, we employ $\mathbf{\Delta}=(0,q_y,q_z)$ with $|q_z|=|q_y|=q$ ($q>0$), that is, $\mathbf{P}_f=(0,\frac{q_y}{2},P_3+\frac{q_z}{2})$ and  $\mathbf{P}_i=(0,-\frac{q_y}{2},P_3-\frac{q_z}{2})$. For these momenta, we have nonzero contributions from four matrix elements, that is:

\begin{eqnarray} 
\label{eq:h12_xi}
\hspace*{-.5cm}
h^1_T(\Gamma_2,z,P_f,P_i) &=& 
 i \,C\,  \Bigg( 
 {-} \frac{(E_f+m) (E_i+m) + P_3^2}{4 m^2}\,F_{H_T}
+ \frac{{P_{fz}}\,(E_i+m)-{P_{iz}}\, (E_f+m)}{8 m^3}\, 
\left(2 P_3\,  F_{\widetilde{E}_T} + q \, F_{E_T} \right)
 \Bigg)\,, \hspace*{0.5cm}
\\[5ex]
\label{eq:h13_xi}
h^1_T(\Gamma_3,z,P_f,P_i) &=& 
 i \,C\, {\rm sign}{(q_z)}\, \Bigg(
-\frac{q^2}{8 m^2}\, F_{H_T} 
- \frac{\,q\,P_3\, (E_i+E_f+2m)}{8 m^3}\, F_{\widetilde{E}_T} 
- \frac{ q^2 \,(E_i+E_f+2m)}{16 m^3}\, F_{E_T}  \Bigg)\,,
\\[5ex]
\label{eq:h20_xi}
\hspace*{-1.5cm}
h^2_T(\Gamma_0,z,P_f,P_i) &=& 
C\,  \Bigg(
 \frac{q\,P_3}{4 m^2}\,F_{H_T} 
 +\frac{q\,P_3\, ({E_f}+{E_i}+2 m) }{8 m^3}\, F_{E_T}
+ \frac{q\,P_3\,({E_f}-{E_i})}{8 m^3}\,F_{\widetilde{E}_T}  \nonumber \\
&&\hspace*{0.75cm}    
+ F_{\widetilde{H}_T}\frac{q\,P_3\,
    (2(E_f+m) (E_i+m) + q^2 - 2 P_3^2)}{8 m^4}    \Bigg)\,, \\[5ex]
\hspace*{-1.5cm}
\label{eq:h21_xi}
h^2_T(\Gamma_1,z,P_f,P_i) &=& 
i \,C\,  \Bigg(
 \frac{
    (2 (E_f+m) (E_i+m)+2 P_3^2 - q^2)}{8 m^2} \,F_{H_T} 
    -  \frac{q\, (E_f (q-P_3) + E_i (q+P_3) + 2 \,m \,q)}{8 m^3}\,F_{E_T} \nonumber \\
&&\hspace*{0.75cm}    
+ \frac{q^2\,P_3^2}{4 m^4} \, F_{\widetilde{H}_T} 
-  \frac{P_3 ( 2 \, P_3\,(E_i-E_f) + q\,(E_i+E_f+2m))}{8 m^3}\,F_{\widetilde{E}_T} 
\Bigg)\,,
    \label{eq:transv2}
\end{eqnarray} 
where $\displaystyle C=\frac{2 m^2}{\sqrt{E_f E_i (E_f+m) (E_i+m)}}$, $P_{fz}=P_3+\frac{q_z}{2}$, and  $P_{iz}=P_3-\frac{q_z}{2}$. Unlike the case $\mathbf{\Delta}=(0,q,0)$, here all matrix elements enter in the decomposition of all four GPDs. Since we are using positive and negative values for the momentum transfer, one has to be careful with the signs in the decomposition. In addition to the signs of the kinematic factors, one also has to consider that $H_{Tq}$, $E_{Tq}$, and $\widetilde{H}_{Tq}$ are even functions of $\xi$, while $\widetilde{E}_{Tq}$ is odd~\cite{Diehl:2003ny,Meissner:2007rx}, which also holds for the quasi-GPDs~\cite{,Bhattacharya:2019cme} and $F_G$. For example, $F_{\widetilde{E}_T}(z,-\xi,t,P_3)= - F_{\widetilde{E}_T}(z,\xi,t,P_3)$, which is taken into account in the decomposition for the negative value of $\xi$.

Once $F_{H_T}$, $F_{E_T}$, $F_{\widetilde{H}_T}$, and $F_{\widetilde{E}_T}$ are disentangled from the renormalized matrix elements, we transform them in momentum ($x$) space to obtain the $x$-dependence using the  Backus-Gilbert (BG) method~\cite{BackusGilbert}, as described in Sec.~\ref{sec:reconstruction}. This procedure gives the quasi-GPDs, $H_{Tq}$, $E_{Tq}$, $\widetilde{H}_{Tq}$, and $\widetilde{E}_{Tq}$. Note that we use the subscript $q$ to denote the quasi-GPDs. Finally, the light-cone GPDs $H_T$, $E_T$, $\widetilde{H}_T$, and $\widetilde{E}_T$ are obtained after application of the matching procedure, outlined in Sec.~\ref{sec:matching}.

\section{Lattice Calculation}
\label{sec:latt_details}

\subsection{Matrix elements} 

\begin{figure}[ht]
\centerline{\includegraphics[scale=0.9]{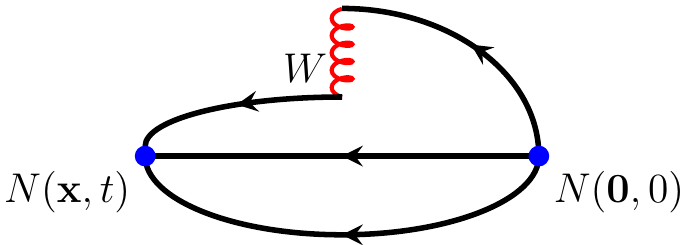}}
\vspace*{-0.15cm}
\begin{minipage}{15cm}
\caption{Connected diagram entering the calculation of the three-point functions. The initial and final states with the quantum numbers of the nucleon are indicated by $N(\mathbf{0},0)$ and $N(\mathbf{x},t)$, respectively. The red curly line indicates the Wilson line, $W$, of the non-local operator.}
\label{fig:diagram}
\end{minipage}
\end{figure}

In this work, we focus on the isovector flavor combination $u-d$ for the transversity GPDs, which requires calculation of only the connected diagram shown in Fig.~\ref{fig:diagram}. The matrix elements are constructed from the two-point and three-point correlation functions,
\begin{equation}
C^{\rm 2pt}(\mathbf{P},t) =2 ({\Gamma_0})_{\alpha\beta}\sum_\mathbf{x}e^{-i\mathbf{P}\cdot \mathbf{x}}\langle 0\vert N_\alpha(\mathbf{x},t) N_\beta(\mathbf{0},0)\vert 0 \rangle\,, 
\end{equation}

\begin{equation}
C_{T}^j(\Gamma_\nu,z,\mathbf{P}_f,\mathbf{P}_i,t_s,\tau) {=} ({\Gamma_\nu})_{\alpha\beta}\sum_{\mathbf{x},\mathbf{y}}\,e^{i(\mathbf{P}_f{-}\mathbf{P}_i)\cdot \mathbf{y}}\,e^{{-}i\mathbf{P}_f\cdot \mathbf{x}}\, \langle 0\vert N_{\alpha}(\mathbf{x},t_s) \, 
 \bar{\psi}(\mathbf{y}{+}z\hat{z},\tau) \sigma^{3j}\, W(\mathbf{y}{+}z\hat{z},\mathbf{y})\psi(\mathbf{y},\tau) 
 N_{\beta}(\mathbf{0},0)\vert 0\rangle,\,\,\,\,
\end{equation}
where $N_{\alpha}(x)=\epsilon ^{abc}u^a _\alpha(x)\left( d^{b^{T}}(x)\mathcal{C}\gamma_5u^c(x)\right)$ is the interpolating field for the proton, and $\tau$ is the current insertion time. Without loss of generality, we take the source to be at $(\mathbf{0},0)$. The three-point functions are calculated for the up- and down- quark/antiquark fields, $(\psi,\bar{\psi})$, which are combined to form the $u-d$ isovector contribution.
 
For nonzero momentum transfer, one must form an optimized ratio to cancel the time dependence in the exponentials and the overlaps between the interpolating field and the nucleon states, namely
\begin{equation}
R^j_T(\Gamma_\nu,z,\mathbf{P}_f,\mathbf{P}_i,t_s,\tau) = \frac{C_T^j(\Gamma_\nu,z,\mathbf{P}_f,\mathbf{P}_i,t_s,\tau\
)}{C^{\rm 2pt}(\mathbf{P}_f,t_s)} \times 
 \sqrt{\frac{C^{\rm 2pt}(\mathbf{P}_i,t_s-\tau)\, C^{\rm 2pt}(\mathbf{P}_f,\tau)\, C^{\rm 2pt}(\mathbf{P}_f,t_s)}{C^{\rm 2pt}(\mathbf{P}_f,t_s-\tau)\, C^{\rm 2pt}(\mathbf{P}_i,\tau) \,C^{\rm 2pt}(\mathbf{P}_i,t_s)}}\,.
\label{Eq:ratio}
\end{equation}
In the limit $(t_s-\tau) \gg a$ and $\tau \gg a$, the ratio of Eq.~(\ref{Eq:ratio}) becomes time-independent and the ground state matrix element is extracted from a constant fit in the plateau region, that is
\begin{equation}
\label{Eq:ratio2}
R^j_T(\Gamma_\nu,z,\mathbf{P}_f,\mathbf{P}_i,t_s,\tau) \,\, \xrightarrow[\tau\gg a]{t_s-\tau\gg a} \,\,
h^{j,B}_T(\Gamma_\nu,z,\mathbf{P}_f,\mathbf{P}_i)\,.
\end{equation}
In this work, we choose $t_s=12 a$, which was also used in our previous work for the unpolarized and helicity GPDs~\cite{Alexandrou:2020zbe}. An extensive study on the excited-states effect for the forward limit of non-local operators was done in Ref.~\cite{Alexandrou:2019lfo}. In that work, we demonstrated that a source-sink separation above 1 fm is sufficient to obtain the ground state contribution within the reported uncertainties. $h^{j,B}_T$ denotes the bare matrix element, while Eq.~\eqref{eq:ME} is the renormalized one. These are related multiplicatively, using the renormalization function, $Z_T(z,\mu_0)$, obtained non-perturbatively,
\begin{equation}
h^{j}_T(\Gamma_\nu,z,\mathbf{P}_f,\mathbf{P}_i,\mu_0) = Z_T(z,\mu_0) \cdot h^{j,B}_T(\Gamma_\nu,z,\mathbf{P}_f,\mathbf{P}_i)\,.   
\end{equation}
Note that the multiplication is complex. We refer the reader to Sec.~\ref{sec:Zfactors} for more details.

\medskip
To improve the overlap with the proton ground state, we construct the proton interpolating field using momentum-smeared quark fields~\cite{Bali:2016lva}, on APE-smeared gauge links~\cite{Albanese:1987ds}. The momentum smearing technique is essential to suppress gauge noise in matrix elements with boosted hadrons, and in particular for non-local operators~\cite{Alexandrou:2016jqi}. The momentum smearing approach allowed us to obtain GPDs for protons boosted up to $P_3=1.67$ GeV. Beyond that momentum, it is unfeasible to obtain the matrix element with controlled statistical uncertainties at a reasonable computational cost. This is in agreement with other calculations of non-local operators with boosted hadrons (see, e.g., Table 1 of Ref.~\cite{Constantinou:2020pek}). The momentum smearing function $\mathcal{S}$ on a quark field, $\psi$, reads
\begin{equation}
\mathcal{S}\psi(x)=\frac{1}{1+6\alpha_G}\left(\psi(x)+\alpha_G\sum_{j=1}^{3}U_j(x)e^{i\overline{\xi}\mathbf{P}\cdot\mathbf{j}}\psi(x+\hat{\mathbf{j}})\right),   
\label{eq:mom_smearing_function}
\end{equation}
where $\alpha_G$ is a parameter of the Gaussian smearing~\cite{Gusken:1989qx,Alexandrou:1992ti}, $U_j$ is a gauge link in a spatial $j$-direction. $\mathbf{P}$ is the momentum of the proton (either at the source, or at the sink) and $\overline{\xi}$ is a free parameter that can be tuned to achieve maximal overlap with the proton boosted state. For $\overline{\xi}=0$, Eq.~(\ref{eq:mom_smearing_function}) reduces to the Gaussian smearing function. The fact that the exponent in Eq.~\eqref{eq:mom_smearing_function} depends on the momentum of the proton state, means that separate quark propagators are needed for every $\mathbf{\Delta}$, because the gauge links are modified every time by a different complex phase. In our implementation, we keep $\overline{\xi} \mathbf{P}$ parallel to the proton momentum at the source and at the sink. This strategy avoids potential problems due to rotational symmetry breaking. It also has the benefit that every correlator entering the ratio of Eq.~(\ref{Eq:ratio}) is optimized separately.
The effectiveness of the momentum smearing has been demonstrated in our previous work for PDFs~\cite{Alexandrou:2016jqi,Alexandrou:2019lfo}, as well as for GPDs~\cite{Alexandrou:2020zbe}. In fact, for the unpolarized GPDs, we found that the statistical noise is suppressed by a factor of 4-5 in the real part, and 2-3 in the imaginary part, depending on the value of $z$.

\medskip
The analysis is performed using a gauge ensemble of twisted-mass fermions with a clover improvement, and Iwasaki-improved gluons. The ensemble has two dynamical degenerate light quarks plus a strange and a charm quark ($N_f=2+1+1$)~\cite{Alexandrou:2018egz} in the sea. The quark masses have been tuned so that the pion mass is about 260 MeV. The lattice volume is $32^3 \times 64$ and the lattice spacing $a=0.093$ fm. The three-point correlators are obtained for a source-sink separation of $t_s=1.12$ fm.
In Table~\ref{tab:stat}, we summarize the statistics for each value of the nucleon momentum boost $P_3$, momentum transfer $\mathbf{\Delta}$ and $t$, as well as skewness $\xi$. The GPDs have definite symmetry with respect to $\xi \to -\xi$, and therefore, we combine the data at $+1/3$ and $-1/3$.
Below, we also compare with the transversity PDF, $h_1$, obtained on the same ensemble with the statistics shown in Table~\ref{tab:stat_PDFs}.

\begin{table}[h!]
\begin{center}
\renewcommand{\arraystretch}{1.4}
\begin{tabular}{cccc|cc}
\hline
$P_3$ [GeV] & $\quad \mathbf{\Delta}$ $[\frac{2\pi}{L}]\quad$ & $-t$ [GeV$^2$] & $\xi$ & $N_{\rm confs}$ & $N_{\rm meas}$\\
\hline
0.83 &(0,2,0)  &0.69  &0      & 519  & 4152\\
1.25 &(0,2,0)  &0.69  &0      & 1315  & 42080\\
1.67 &(0,2,0)  &0.69  &0      & 1753  & 112192\\
1.25 &(0,2,2)  &1.02  & 1/3   & 417  & 40032\\
1.25 &(0,2,-2) &1.02  & -1/3  & 417  & 40032 \\
\hline
\end{tabular}
\begin{minipage}{15cm}
\caption{Statistics for the transversity GPDs at each momentum boost, momentum transfer and skewness.}
\label{tab:stat}
\end{minipage}
\end{center}
\end{table}

\begin{table}[h!]
\begin{center}
\renewcommand{\arraystretch}{1.4}
\begin{tabular}{c|cc}
\hline
$P_3$ [GeV] & $N_{\rm confs}$ & $N_{\rm meas}$\\
\hline
0.83 & 194 & 1560\\
\hline 
1.25  &  731  & 11696\\
\hline 
1.67  & 1644  & 105216\\
\hline
\end{tabular}
\begin{minipage}{15cm}
\caption{Statistics for the transversity PDF at the three values of the proton momenta.}
\label{tab:stat_PDFs}
\end{minipage}
\end{center}
\end{table}

\subsection{Renormalization} 
\label{sec:Zfactors}

The matrix elements $h_T^j$ are renormalized non-perturbatively with the renormalization function $Z_T$, which is defined in an RI-type scheme at some scale $\mu_0$. The vertex functions of the non-local tensor operator are calculated using the momentum source method~\cite{Gockeler:1998ye,Alexandrou:2015sea} that suppresses statistical noise. We work in the twisted basis to calculate the matrix elements $h_T^j$, and, therefore, the operator $\sigma^{3j}$ (physical basis) renormalizes with the operator $\gamma^5\sigma^{3j}$. We apply the following condition to the vertex functions of $\gamma^5\sigma^{3j}$ at each value of $z$ separately,
\begin{eqnarray}
\label{renorm}
{\cal Z}_q^{-1}\, {\cal Z}_T(z) &&  {\rm Tr} \left[{\cal V}_T(p,z) \, \slashed{p} \right] \Bigr|_{p^2{=} \mu_0^2} =
{\rm Tr} \left[{\cal P}_T \, {\cal V}_T^{{\rm Born}}(p,z) \right] \Bigr|_{p^2{=} \mu_0^2}
\, . 
\end{eqnarray}
We also calculate the quark propagator that is needed for the quark field renormalization, ${\cal Z}_q$,
 \begin{eqnarray}
&& \hspace*{-0.5cm} {\cal Z}_q =\frac{1}{12} {\rm Tr} \left[(S(p))^{-1}\, S^{\rm Born}(p)\right] \Bigr|_{p^2= \mu_0^2}\,.
\end{eqnarray}
${\cal V}(p,z)$ ($S(p)$) is the amputated vertex function of the operator (fermion propagator) and $S^{{\rm Born}}(p)$ is the tree-level of the propagator. 
In the prescription of Eq.~\eqref{renorm}, the vertex functions are projected with the so-called minimal projector, which defines ${\cal Z}_T(z)$. The use of this definition is necessary, as the matching formalism is only known for this scheme~\cite{Liu:2019urm}. Note that we use the symbol ${\cal Z}_T$ for the renormalization function prior to taking the chiral limit of Eq.~\eqref{eq:Zchiral_fit}. Similarly for the fermion field renormalization, ${\cal Z}_q$. In a nutshell, the vertex function of the tensor non-local operator $\sigma^{0l}$ ($l\ne j\ne3\ne l$)~\footnote{This is equivalent to the operator we are interested in the twisted basis, $\gamma^5\sigma^{3j}$.} with a Wilson line in the $z$-direction contains contributions from three structures, that is
\begin{equation}
 {\cal V}_T =  \sigma^{0l}\, S_1  +\frac{1}{p_3^2}(\gamma^0 \,p_l - \gamma^l \,p_0) \slashed{p}\,  S_2  + \frac{1}{p_3} (\sigma^{30} p_l - \sigma^{3l} p_0)  \,S_3
\end{equation}
(see, e.g., Eq.~(76) of Ref.~\cite{Constantinou:2017sej}). The projector is defined such that it isolates the tree-level contribution of the vertex function, $S_1$. For the operator under study, we use the projector
\begin{equation}
\label{eq:PT}
{\cal P}_T = \frac{1}{4} \left(-\sigma^{0l} + \frac{(p_0^2+p_l^2)}{p_0\,p_j}\,\sigma^{jl} \right) \,.
\end{equation}
In this work, we calculate the vertex functions with the Wilson line in all spatial directions projected with the equivalent ${\cal P}_T$. Since $S_1$ is independent of the direction of the Wilson line, we average over the three directions. Numerically, we find that the estimates of ${\cal Z}_T(z)$ using the minimal projector are similar to the estimates obtained by projecting with the tree-level value. This is an indication that the contamination from $S_3$ in the vertex function is small~\footnote{The structure $S_2$ is automatically eliminated with the minimal projector, and is therefore, irrelevant in this discussion.}.

The prescription of Eq.~\eqref{renorm} is mass-independent, and therefore ${\cal Z}_T$ should not depend on the quark mass. However, there might be residual cut-off effect of the form $a m_q$. To eliminate any systematics related to such an effect, we extract ${\cal Z}_T$ using five degenerate-quark-mass ensembles ($N_f=4$) with the same lattice spacing as the ensemble we use for $h_T^j$. These $N_f=4$ ensembles correspond to a pion mass in the range 350 - 520 MeV. The estimates of ${\cal Z}_T$ from each ensemble are used for a chiral extrapolation. More details on this procedure can be found in Ref.~\cite{Alexandrou:2019lfo}. 

${\cal Z}_T$ is scheme- and scale-dependent, and therefore, is defined at some RI scale $ \mu_0$. We use several values of $\mu_0$, chosen to be isotropic in the spatial directions, which suppresses discretization effects. Furthermore, the vertex momentum is such that the ratio $\frac{p^4}{(p^2)^2}$ is less than 0.35~\cite{Constantinou:2010gr}. In this work, we use different values of $ \mu_0$ ($(a\, \mu_0)^2 \in [1,5]$) to check the dependence of the matching formalism on $ \mu_0$. 
For each value of $ \mu_0$, we apply a chiral extrapolation using the fit
\begin{equation}
\label{eq:Zchiral_fit}
{\cal Z}^{\rm RI}_T(z, \mu_0,m_\pi) = {Z}^{\rm RI}_{T}(z, \mu_0) + m_\pi^2 \,{\bar{Z}}^{\rm RI}_{T}(z, \mu_0) \,,
\end{equation}
to extract the mass-independent ${Z}^{\rm RI}_{T}(z, \mu_0)$. For our final results, we use the renormalization functions defined on a single ${\rm RI}$ renormalization scale, $(a  \mu_0)^2\approx2.57$. This scale also enters the matching equations, which connect the quasi-GPDs in the RI at a scale of $\mu_0$ to the GPDs in the $\overline{\rm MS}$ at a scale of 2 GeV. We find negligible dependence in the final GPDs when varying the initial scale $\mu_0$ in the quasi-GPDs.

\subsection{Reconstruction of $x$-dependence}
\label{sec:reconstruction}
The quantities $F_G$~\footnote{In this discussion, we show explicitly the dependence of $F_G$ on the renormalization scale, $\mu_0$, as it refers to renormalized quantities.} , where $G=H_T,\,E_T,\,\widetilde{H}_T,\,\widetilde{E}_T$, are related to the quasi-distributions, $G_q$, via a Fourier transform, as the latter are expressed in momentum space,
\begin{equation}
\label{eq:X2F}
 G_q(x,\xi,t,\mu_0,P_3) = \int_{-\infty}^\infty dz \, e^{-i x P_3 z} \, F_G(z,\xi,t,P_3,\mu_0)\,.
\end{equation}
Therefore, extracting the quasi-GPDs requires integration over a continuum range of $z$, while the lattice provides only a discrete set of determinations of $F_G$, for integer values of $z/a$ up to roughly half of the lattice extent in the direction of the boost, $L/2a$.
Thus, obtaining the quasi-GPDs, or for that matter any $x$-dependent distributions,  poses a mathematically ill-defined problem, as discussed in detail in Ref.~\cite{Karpie:2018zaz}. The inverse problem originates from incomplete information, i.e.\ attempting to reconstruct a continuous distribution from a finite number of input data points.
As such, its solution necessarily requires making additional assumptions that provide the missing information.
These assumptions should be mild and preferably model-independent -- else, the reconstructed distribution may be biased.

In this work, we use the Backus-Gilbert (BG) method~\cite{BackusGilbert}, which was also proposed in Ref.~\cite{Karpie:2018zaz}. The method relies on a model-independent criterion to choose from among the infinitely many possible solutions to the inverse problem, namely that the variance of the solution with respect to the statistical variation of the input data should be minimal. While the BG method is superior to the naive Fourier transform, there are limitations due the small number of the lattice data, and the BG would be improved if a larger volume and finer lattice spacing ensemble is used. The reconstruction is done separately for each value of $x$. In practice, we separate the exponential of the Fourier transform into its cosine and sine parts, related to the real and imaginary parts of the matrix elements, respectively.
We define a vector $\textbf{a}_K(x)$, where $K$ is either the cosine or sine kernel, of dimension $d$ equal to the number of available input matrix elements, i.e.\ $d=z_{\rm max}/a+1$; the matrix elements for $z$ beyond $z_{\rm max}$ are neglected, assuming that they are approximately zero within uncertainties.
The BG procedure consists in finding the vectors $\textbf{a}_K(x)$ for both kernels according to the variance minimization criterion.
The vector $\textbf{a}_K(x)$ is an approximate inverse of the cosine/sine kernel function $K(x)$, that is
\begin{equation}
\Delta(x-x')=\sum_{z/a=0}^{d-1} a_K(x)_{z/a} K(x')_{z/a}\,,
\end{equation}
where $K(x')_{z/a}=\cos(x'P_3z)$ or $K(x')_{z/a}=\sin(x'P_3z)$ are elements of a $d$-dimensional vector of discrete kernel values corresponding to integer values of $z/a$ entering the reconstruction. Therefore, the function $\Delta(x-x')$ is an approximation to the Dirac delta function $\delta(x-x')$. The quality of this approximation depends on the achievable dimension $d$ at given simulation parameters.

The vectors $\textbf{a}_K(x)$ are identified from optimization conditions based on the BG criterion.
For more details, see Ref.~\cite{Karpie:2018zaz}. Below, we summarize the methodology.
We define a $d\times d$-dimensional matrix $\textbf{M}_K(x)$, with matrix elements
\begin{equation}
M_K(x)_{z/a,z'/a}=\int_0^{x_c} dx'\,(x-x')^2 K(x')_{z/a} \, K(x')_{z'/a}+\rho\,\delta_{z/a,z'/a}\,,
\end{equation}
where $x_c$ is the maximum value of $x$ for which the quasi-distribution is taken to be non-zero (i.e.\ its reconstruction proceeds for $x\in[0,x_c]$). The parameter $\rho$ regularizes the matrix $\textbf{M}_K$.
This regularization was proposed by Tikhonov \cite{Tikhonov:1963}, and is suggested as a possible way to make $\textbf{M}_K$ invertible \cite{Ulybyshev:2017szp,Ulybyshev:2017ped,Karpie:2018zaz}).
The value of $\rho$ determines the resolution of the method and should be taken as rather small, in order to avoid a bias.
We use $\rho=10^{-3}$, which leads to reasonable resolution and is large enough to avoid oscillations in the final distributions related to the presence of small eigenvalues of $\textbf{M}_K$. We have checked that the dependence on $\rho$ is negligible. Additionally, we define a $d$-dimensional vector $\textbf{u}_K$, with elements
\begin{equation}
u_{K;z/a}=\int_0^{x_c} dx'\,K(x')_{z/a}\,.
\end{equation}
Applying the aforementioned optimization conditions leads to
\begin{equation}
\textbf{a}_K(x)=\frac{\textbf{M}_K^{-1}(x)\,\textbf{u}_K}{\textbf{u}_K^T\,\textbf{M}_K^{-1}(x)\,\textbf{u}_K}\,.
\end{equation}
Finally, the BG-reconstructed quasi-distributions are given by
\begin{equation}
G_q(x,\xi,t,\mu_0,P_3)=\frac{1}{2} \sum_{z/a} \left( a_{\rm cos}(x)_{z/a} \, {\rm Re}\, F_G(z,\xi,t,P_3,\mu_0) +  a_{\rm sin}(x)_{z/a} \, {\rm Im}\, F_G(z,\xi,t,P_3,\mu_0) \right)\,.
\end{equation}

\subsection{Matching Procedure}
\label{sec:matching}

Following the reconstruction of the $x$-dependence of the quasi-GPDs, we proceed with obtaining the light-cone GPDs. Contact between the physical GPDs and the quasi-GPDs is established through a perturbative matching procedure. The general factorization formula reads
\begin{eqnarray}
\label{eq:matching}
G_q(x,\xi,t,\mu_0,(\mu_0)_3,P_3) &=& \int_{-1}^1 \frac{dy}{|y|}\, C_G \left(\frac{x}{y},\frac{\xi}{y},\frac{\mu}{y P_3},\frac{(\mu_0)_3}{y P_3},r\right) G(y,t,\xi,\mu)+\mathcal{O}\left(\frac{m^2}{P_3^2},\frac{t}{P_3^2},\frac{\Lambda_{\rm QCD}^2}{x^2P_3^2}\right)\,,
\end{eqnarray} 
where $C_G$ is the matching kernel and is known to one-loop level in perturbation theory. The involved renormalization scales are: $\mu_0$ -- RI renormalization scale, its $z$-component $(\mu_0)_3$ (with $r=\mu_0^2/(\mu_0)_3^2$), and $\mu$ -- final $\MSb$ scale; here we choose $\mu=2$ GeV. This formula establishes that quasi-distributions are equal to light-cone distributions up to power-suppressed corrections (nucleon mass ($m$) corrections and higher-twist corrections). The matching coefficient for the GPDs was first derived for flavor non-singlet unpolarized and helicity quasi-GPDs in Ref.\ \cite{Ji:2015qla} and for transversity quasi-GPDs in Ref.~\cite{Xiong:2015nua}, using the transverse momentum cutoff scheme. Recently, a matching formula was also derived for all Dirac structures \cite{Liu:2019urm} relating quasi-GPDs renormalized in a variant of the RI/MOM scheme to $\MSb$ light-cone PDFs (minimal projector of Eq.~\eqref{eq:PT}). In these calculations, it was shown that the matching for GPDs at zero skewness is the same as for PDFs. It was also demonstrated that, to one-loop level, the $H$-type and $E$-type GPDs have the same matching formula. The matching kernel for the transversity GPDs and parton momentum $p_3$ reads
\begin{align}
\label{e:bare_matching}
C_G\left(\sigma^{3j};x,\xi,\frac{p_3}{\mu},\frac{p_3}{(\mu_0)_3},r\right) &= \delta(x-1) + \frac{\alpha_s C_F}{2\pi}\left\{
\begin{array}{lc}
G_1(\sigma^{3j};x,\xi)_+				& x<-\xi\\[1.5ex]
G_2(\sigma^{3j};x,\xi,p_3/\mu)_+		& |x|<\xi\\[1.5ex]
G_3(\sigma^{3j};x,\xi,p_3/\mu)_+		& \xi<x<1\\[1.5ex]
-G_1(\sigma^{3j};x,\xi)_+				& x>1
\end{array}\right.\nonumber\\[2ex]
&-\frac{\alpha_s C_F}{2\pi}\left|\frac{p_3}{(\mu_0)_3}\right|f_{{\cal P}_T}\left(\sigma^{3j};\frac{p_3}{(\mu_0)_3}(x-1)+1,r\right)_+ +\frac{\alpha_s C_f}{4\pi}\delta(x-1)\ln \left(\frac{\mu^2}{(\mu_0)_3^2} \right)\,. 
\end{align}

\noindent
The functions $G_1, G_2,G_3$ for the matching of bare quasi-GPDs can be found in Ref.~\cite{Liu:2019urm}, while the one-loop RI counterterm $f_{{\cal P}_T}$ for the RI/MOM variant that we employ (minimal projector, ${\cal P}_T$) is given in Ref.~\cite{Liu:2018uuj}. The plus prescription is defined as
\begin{equation}
    f(x)_+	= f(x) - \delta(x-1)\int dy f(y)
\end{equation}
and it combines the so-called ``real'' (vertex) and ``virtual'' (self-energy) corrections. Below we give the expressions for the functions $G_i$ for completeness:
\begin{align}
& G_1(\sigma^{3j};x,\xi)=-\frac{x+\xi}{(x-1)(1+\xi)}\ln \frac{x-1}{x+\xi}+(\xi\rightarrow -\xi)\,,\\[2.5ex]&
G_2(\sigma^{3j};x,\xi)=\frac{x+\xi}{(1-x)(1+\xi)}\left[\ln \frac{4(1-x)^2(x+\xi)p_3^2}{(\xi-x)\mu^2}-1 \right]+\frac{2\xi}{1-\xi^2}\ln \frac{\xi-x}{1-x}\,,\\[2.5ex]&
G_3(\sigma^{3j};x,\xi)=\frac{2(x-\xi^2)}{(1-x)(1-\xi^2)}\left[\ln \frac{4\sqrt{x^2-\xi^2}(1-x)p_3^2}{\mu^2}-1 \right] + \frac{\xi}{1-\xi^2}\ln \frac{x+\xi}{x-\xi}\,.
\end{align}

\section{Numerical Results}
\label{sec:results}

\begin{figure}[h!]
    \centering
    \includegraphics[scale=0.55]{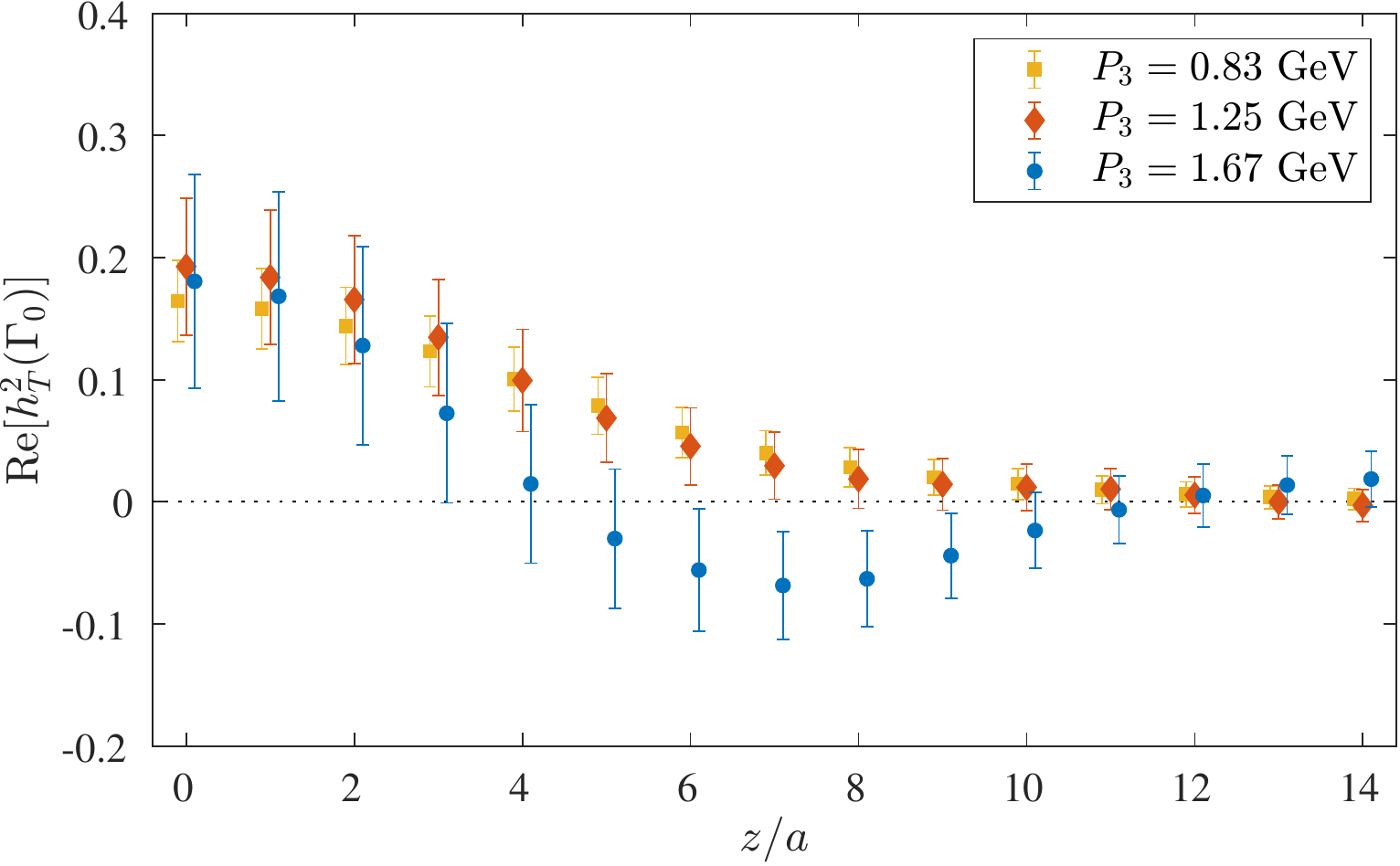}\hspace{0.05cm}
    \includegraphics[scale=0.55]{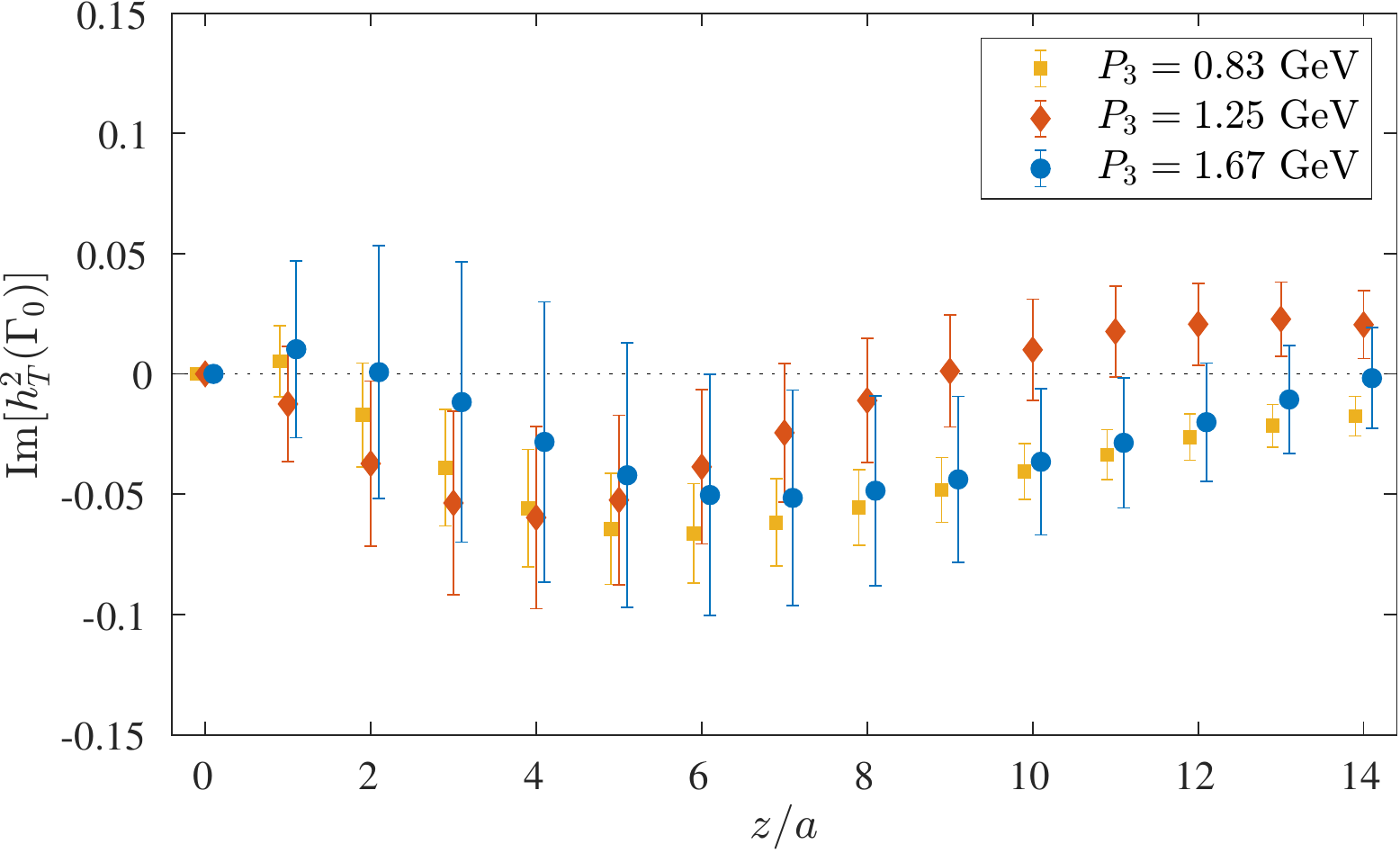}

    \includegraphics[scale=0.55]{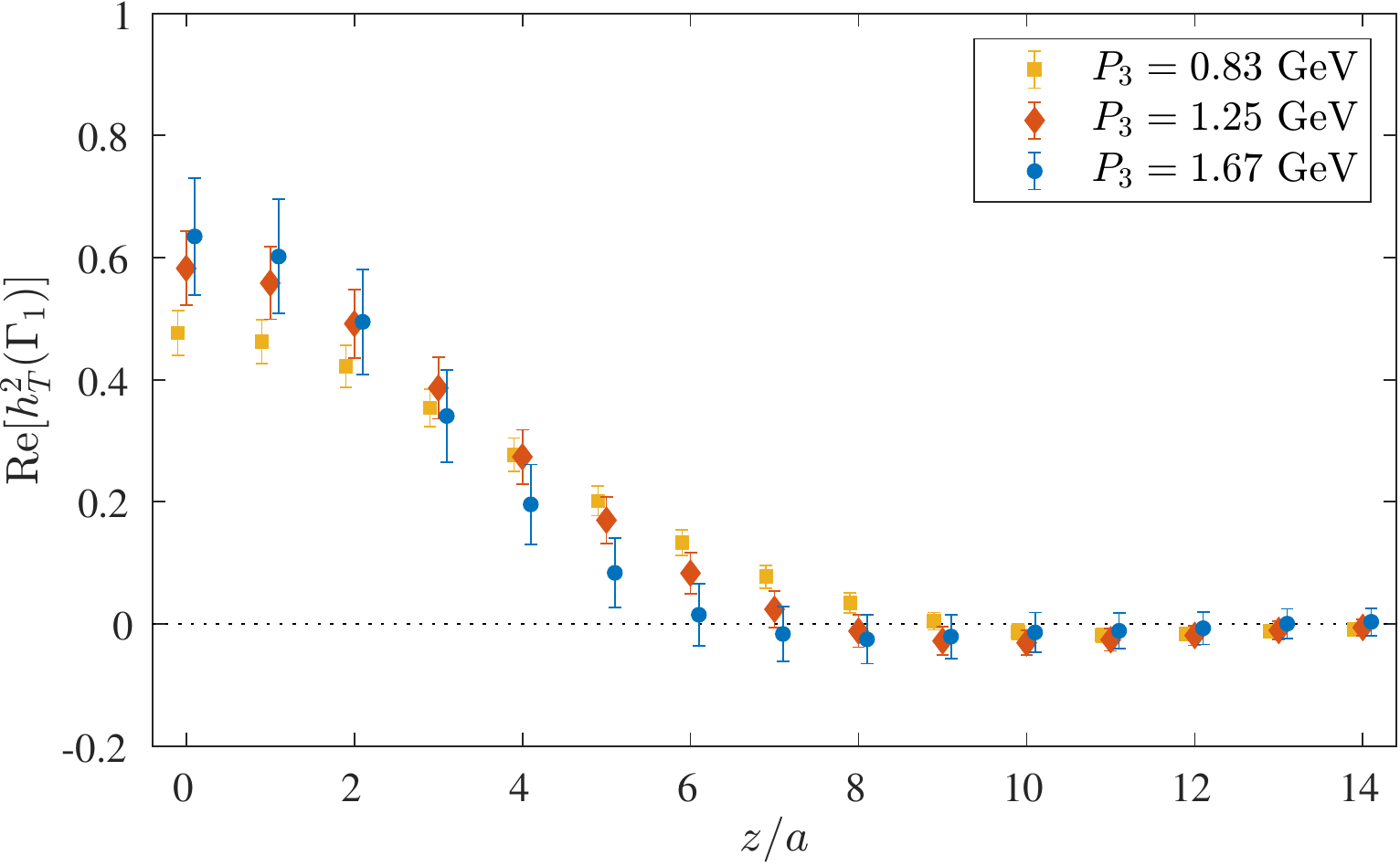}\hspace{0.05cm}
    \includegraphics[scale=0.55]{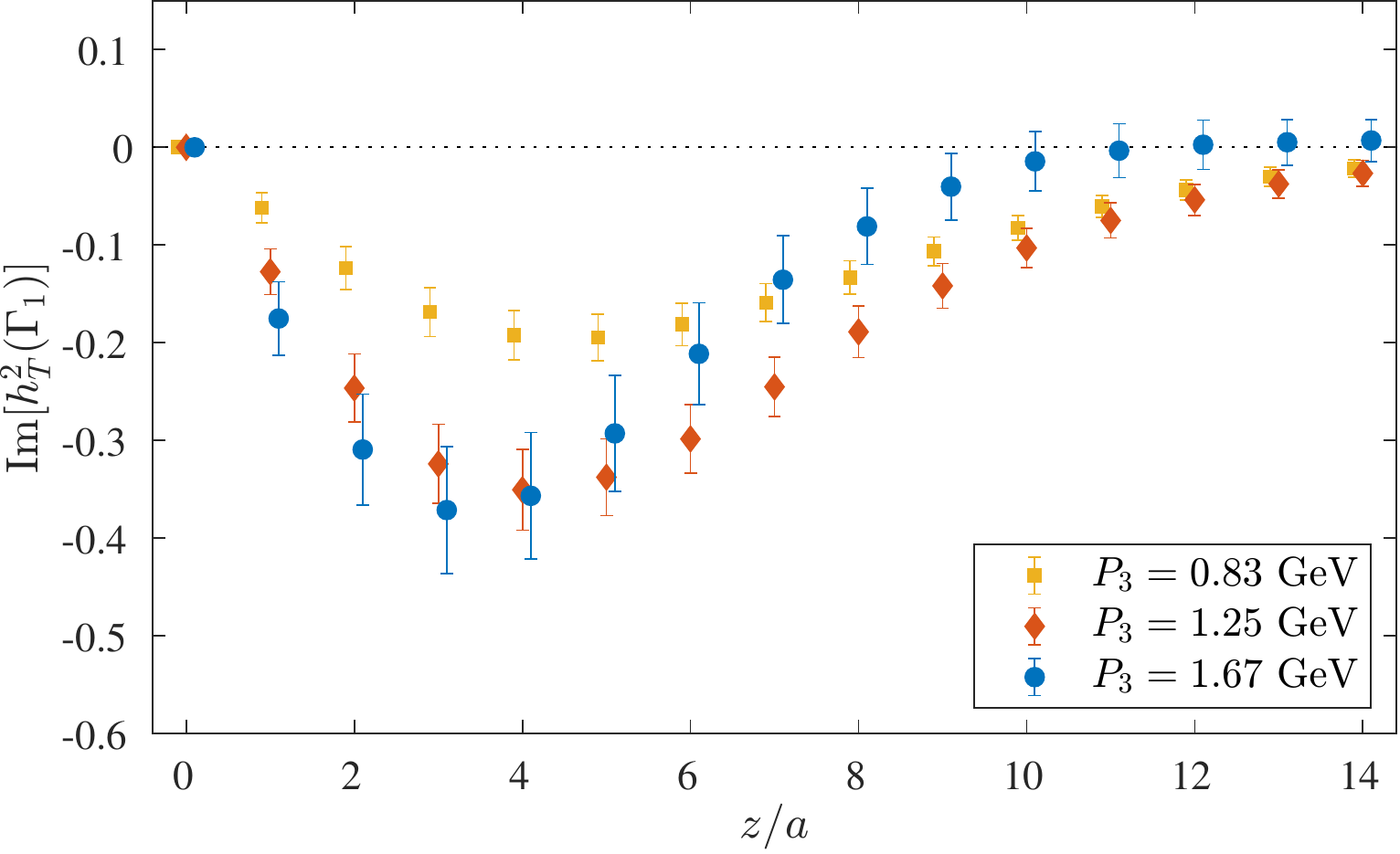}
    
    \includegraphics[scale=0.55]{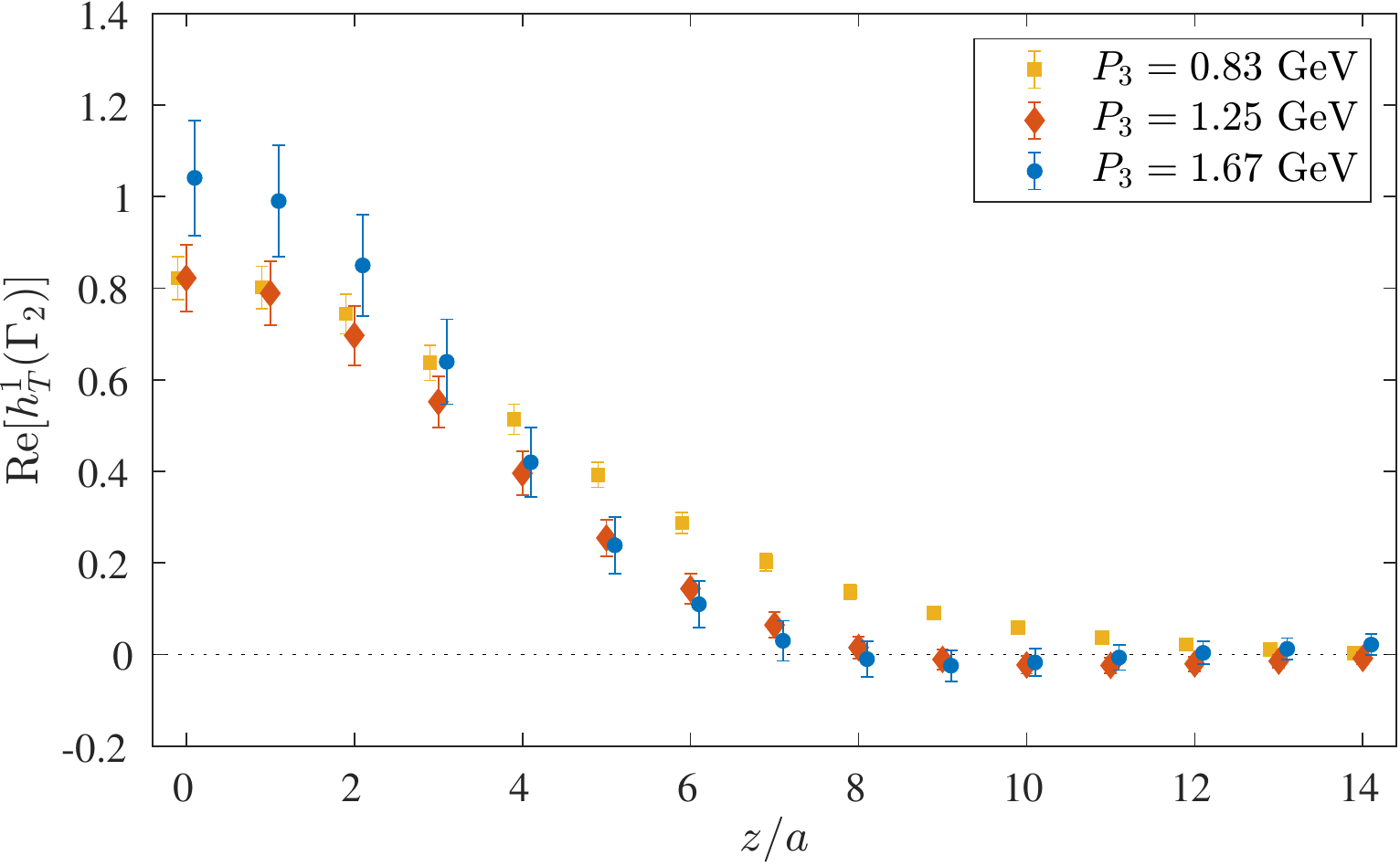}\hspace{0.05cm}
    \includegraphics[scale=0.55]{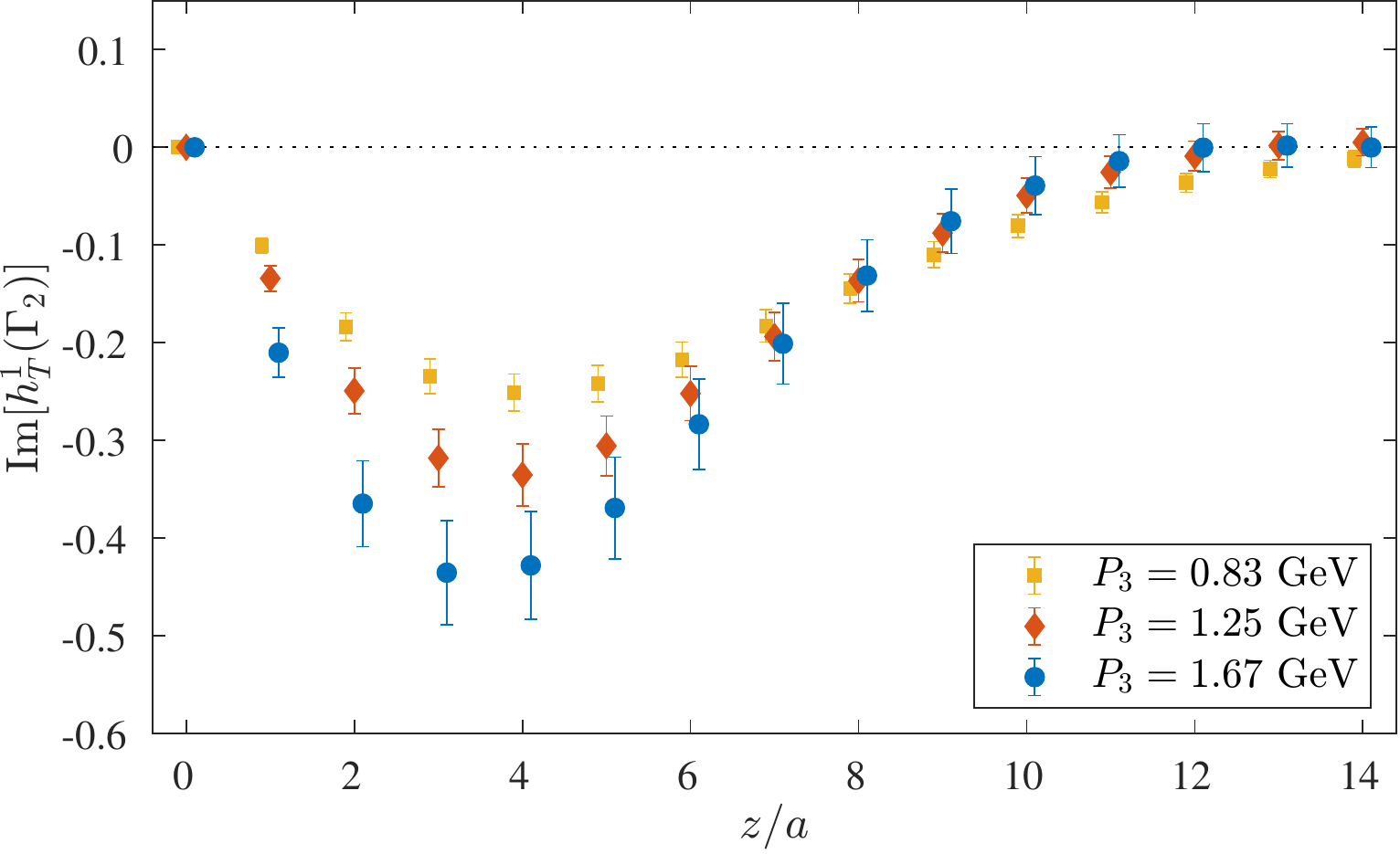}
    
    \includegraphics[scale=0.55]{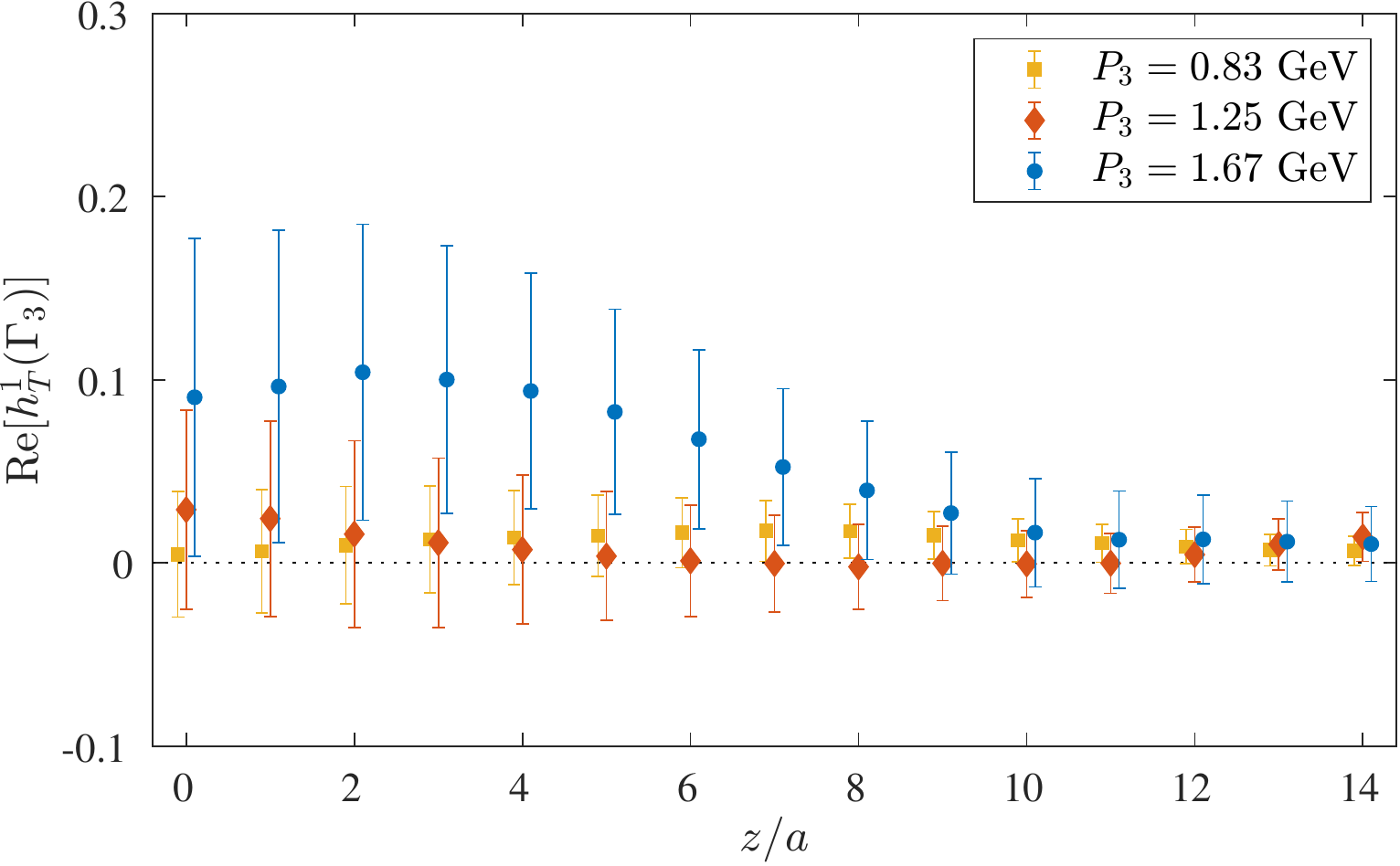}\hspace{0.05cm}
    \includegraphics[scale=0.55]{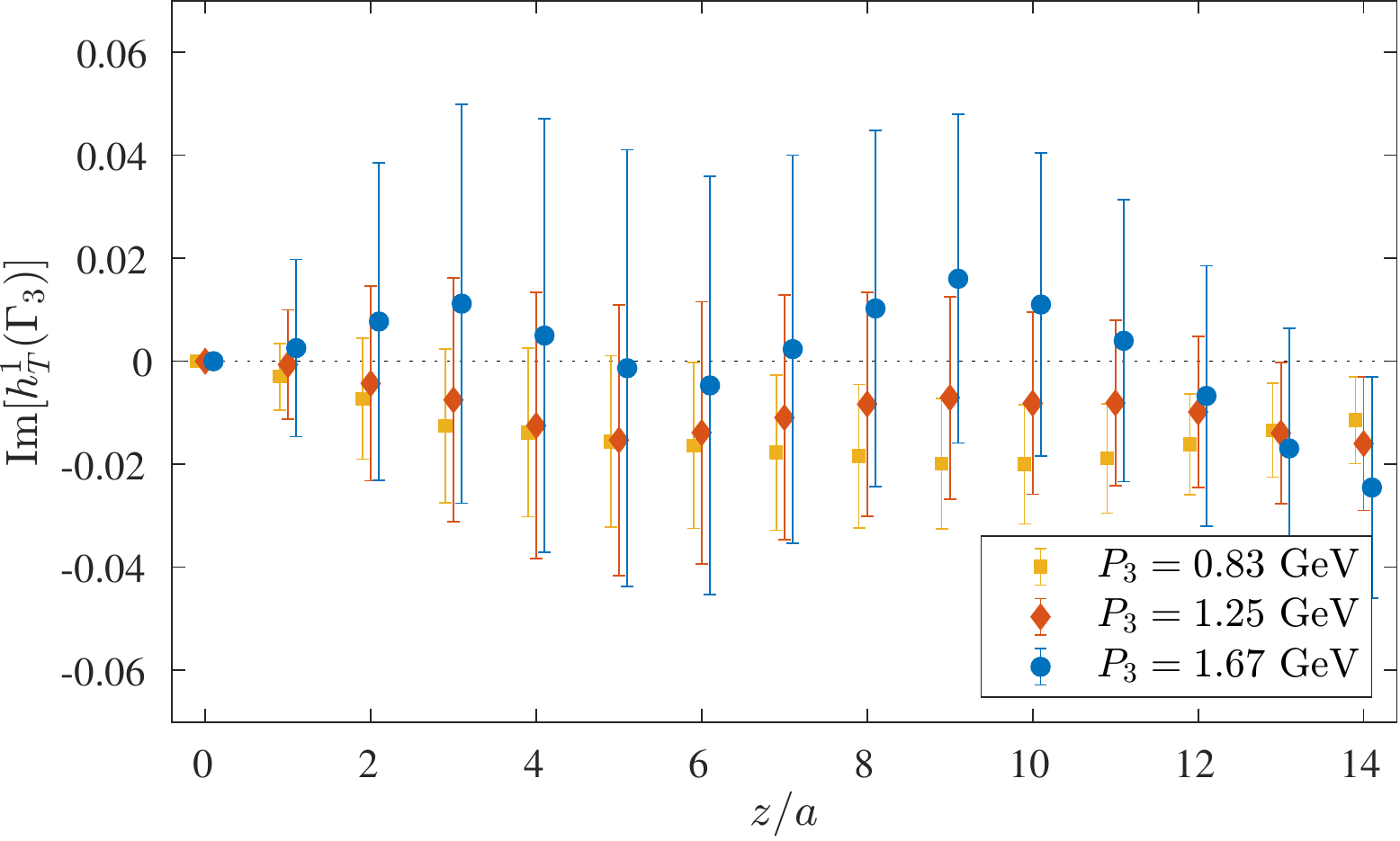}
    \caption{Bare matrix elements $h^j_T(\Gamma)$ for the four different projectors, $\Gamma_0$, $\Gamma_1$, $\Gamma_2$, $\Gamma_3$, at $t=-0.69$~GeV$^2$ and $\xi=0$. We compare results at three nucleon boosts: $P_3=0.83$~GeV (yellow squares), $P_3=1.25$~GeV (red diamonds) and $P_3=1.67$~GeV (blue circles).}
    \label{fig:ME_hT_xi0}
\end{figure}

We begin our presentation with the bare matrix elements for the ground state, as extracted from Eq.~\eqref{Eq:ratio2}. In Fig.~\ref{fig:ME_hT_xi0}, we plot the four matrix elements contributing to $\mathbf{\Delta}=\frac{2\pi}{32}(0,2,0)$ ($t=-0.69$ GeV$^2$), that is Eqs.~\eqref{eq:h12_zero} - \eqref{eq:h21_zero}. We compare the signal for the three values of $P_3$ employed. As expected, the statistical uncertainties increase with the momentum. We find that the matrix elements of $h^1_T(\Gamma_2)$ have the most dominant contributions in both the real and imaginary parts, followed closely by $h^2_T(\Gamma_1)$. We remind the reader that $h^1_T(\Gamma_2)$ is directly related to the leading $H_T$-GPD, see Eq.~\eqref{eq:h12_zero}. $h^2_T(\Gamma_0)$ has a smaller signal than the above matrix elements, but it is clearly non-negligible. On the contrary, the matrix element contributing to $\widetilde{E}_T$, $h^1(\Gamma_3)$, has negligible contribution for both the real and imaginary parts, with the exception of the real part for $P_3=1.67$ GeV, which slightly deviates from zero. We note that a convergence with respect to $P_3$ is not necessarily anticipated in the matrix elements, but rather at the level of the final matched GPDs. As can be seen from Eqs.~\eqref{eq:h12_zero} - \eqref{eq:h21_xi}, there is a dependence on the kinematic setup, which includes $P_3$ through the energies, and in some cases, directly. We remind the reader that the matching also contains the momentum boost $P_3$.

\begin{figure}[h!]
    \centering
    \includegraphics[scale=0.557]{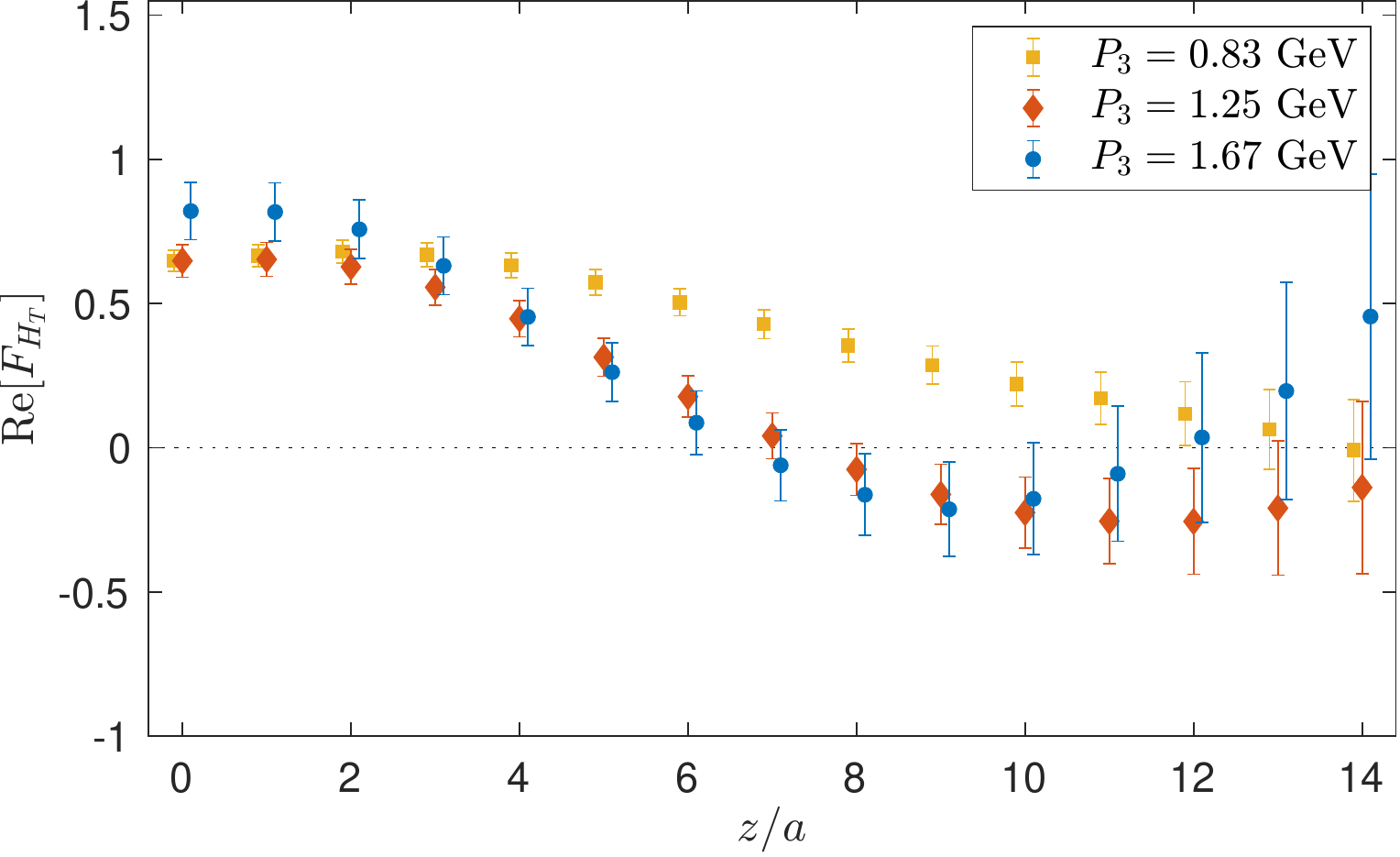}\hspace{0.05cm}
    \includegraphics[scale=0.557]{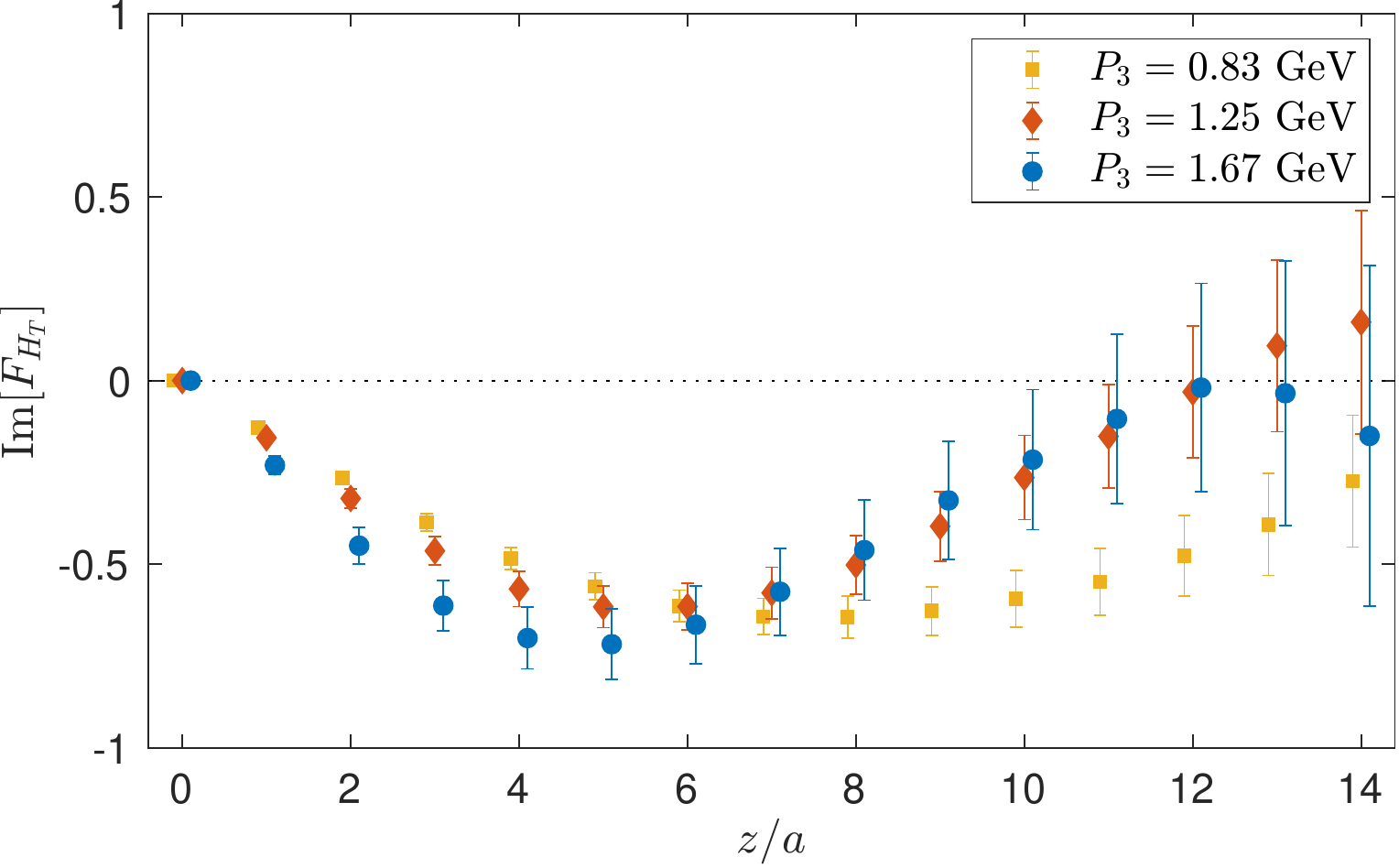}
    
     \includegraphics[scale=0.557]{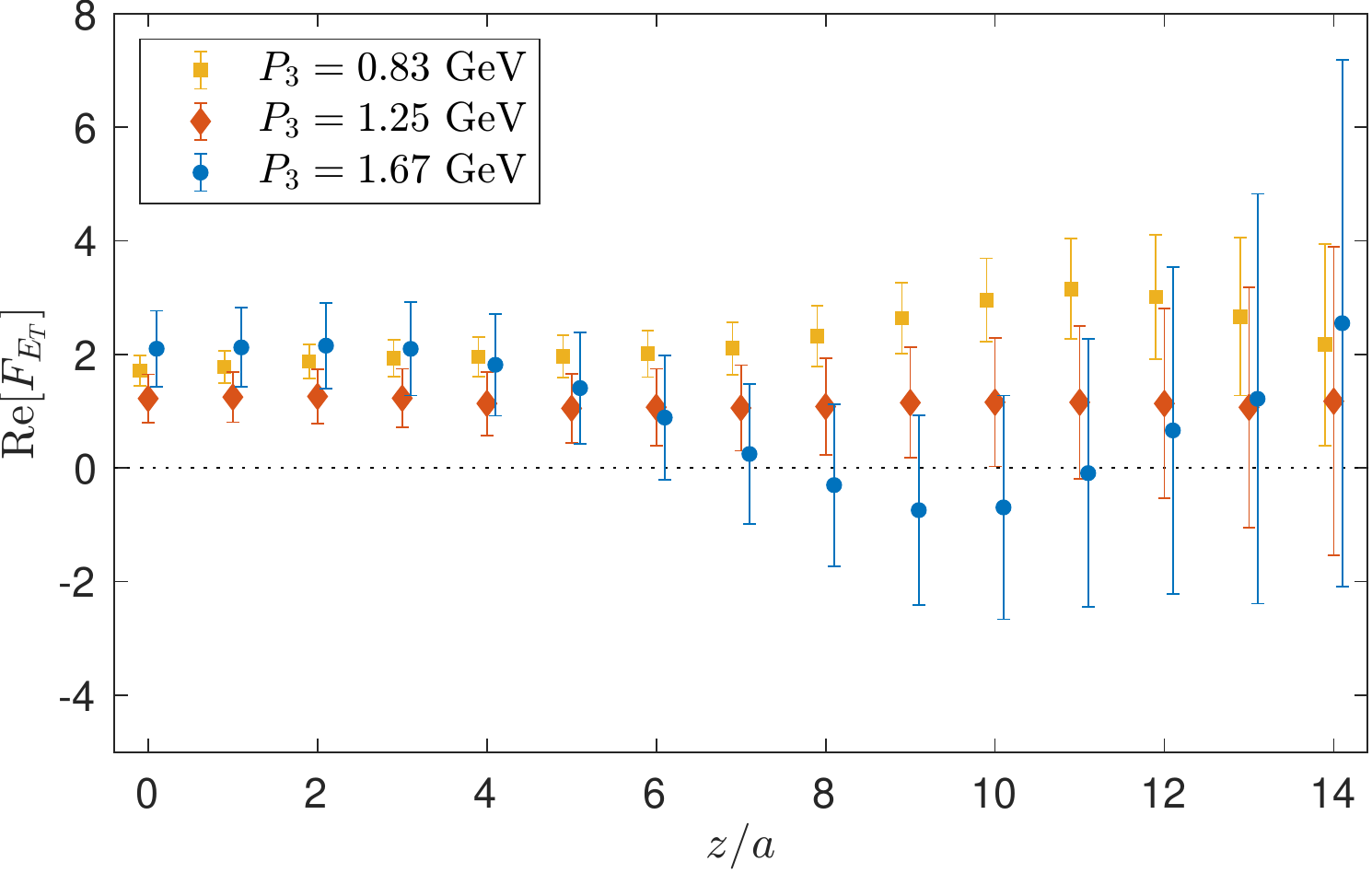}\hspace{0.05cm}
    \includegraphics[scale=0.557]{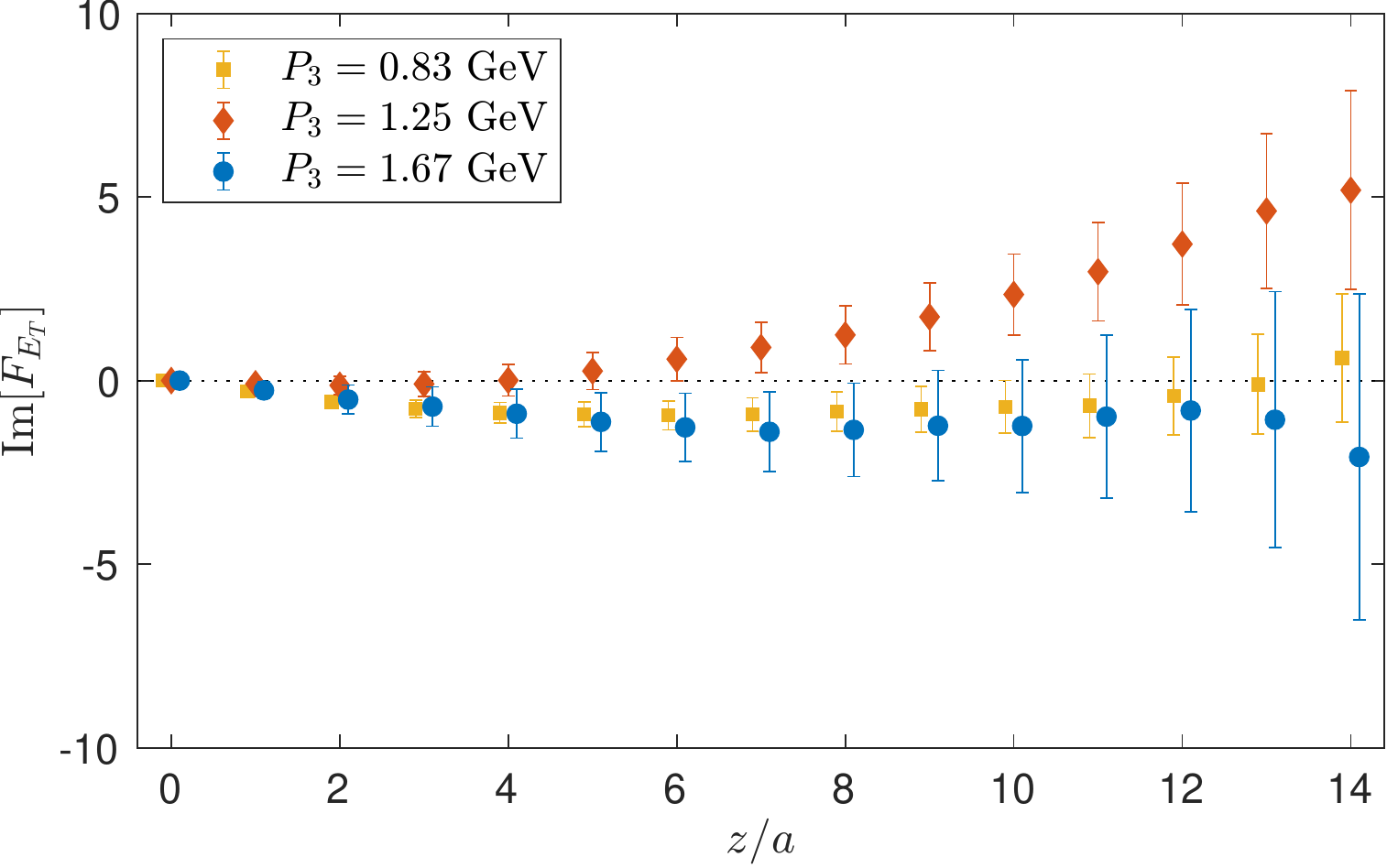}
    
    \includegraphics[scale=0.557]{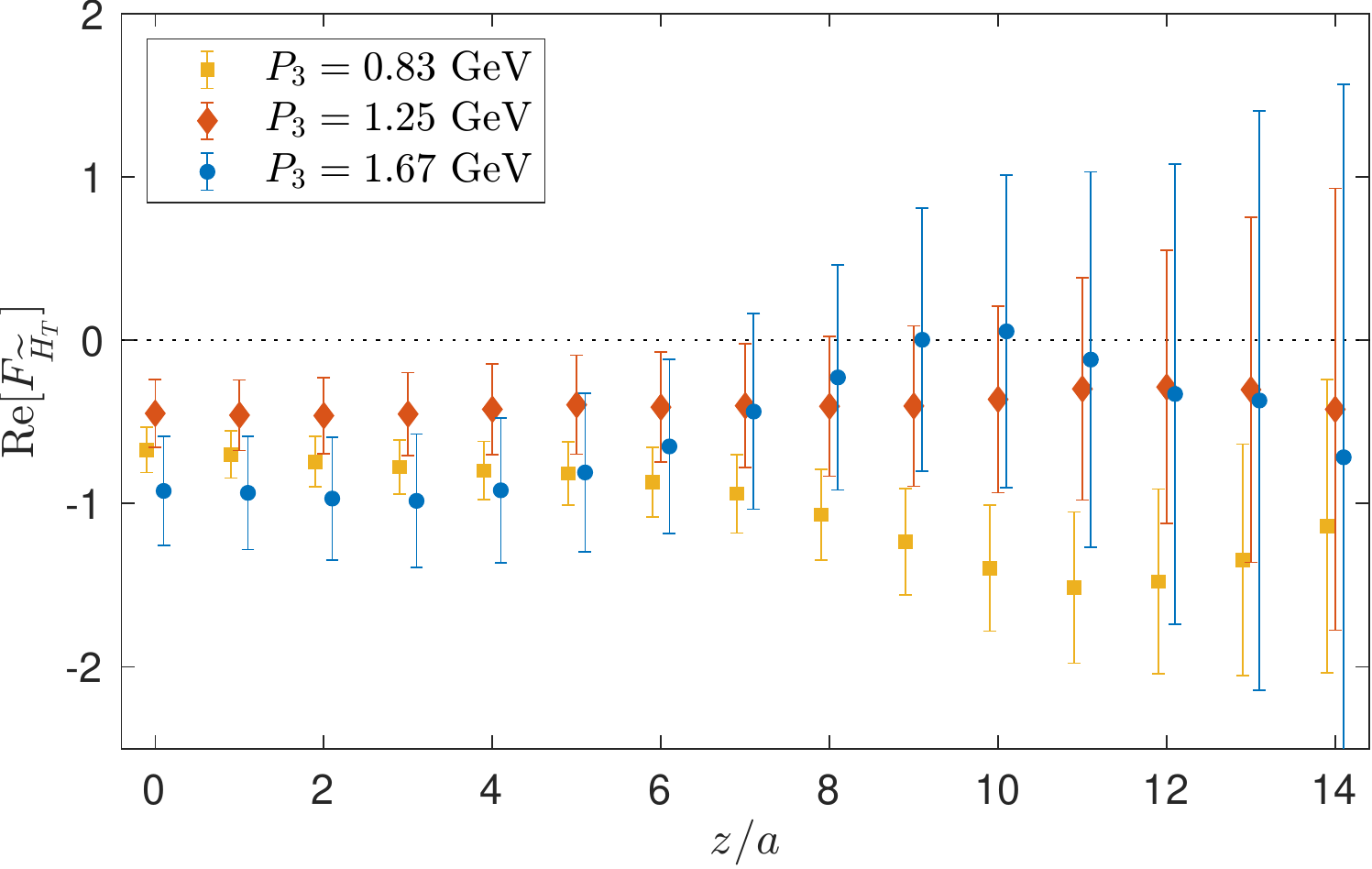}\hspace{0.05cm}
    \includegraphics[scale=0.557]{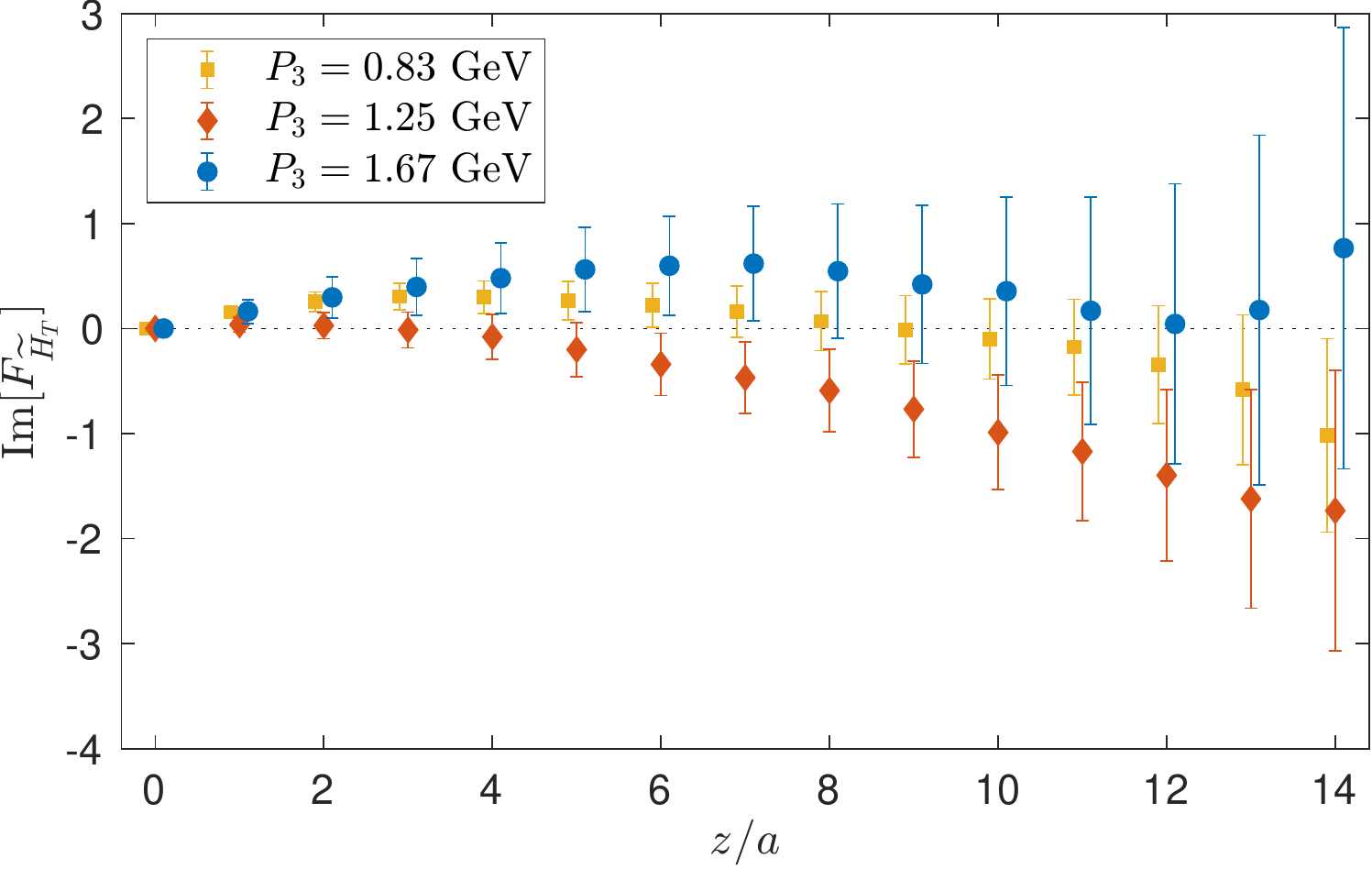}
    
    \includegraphics[scale=0.557]{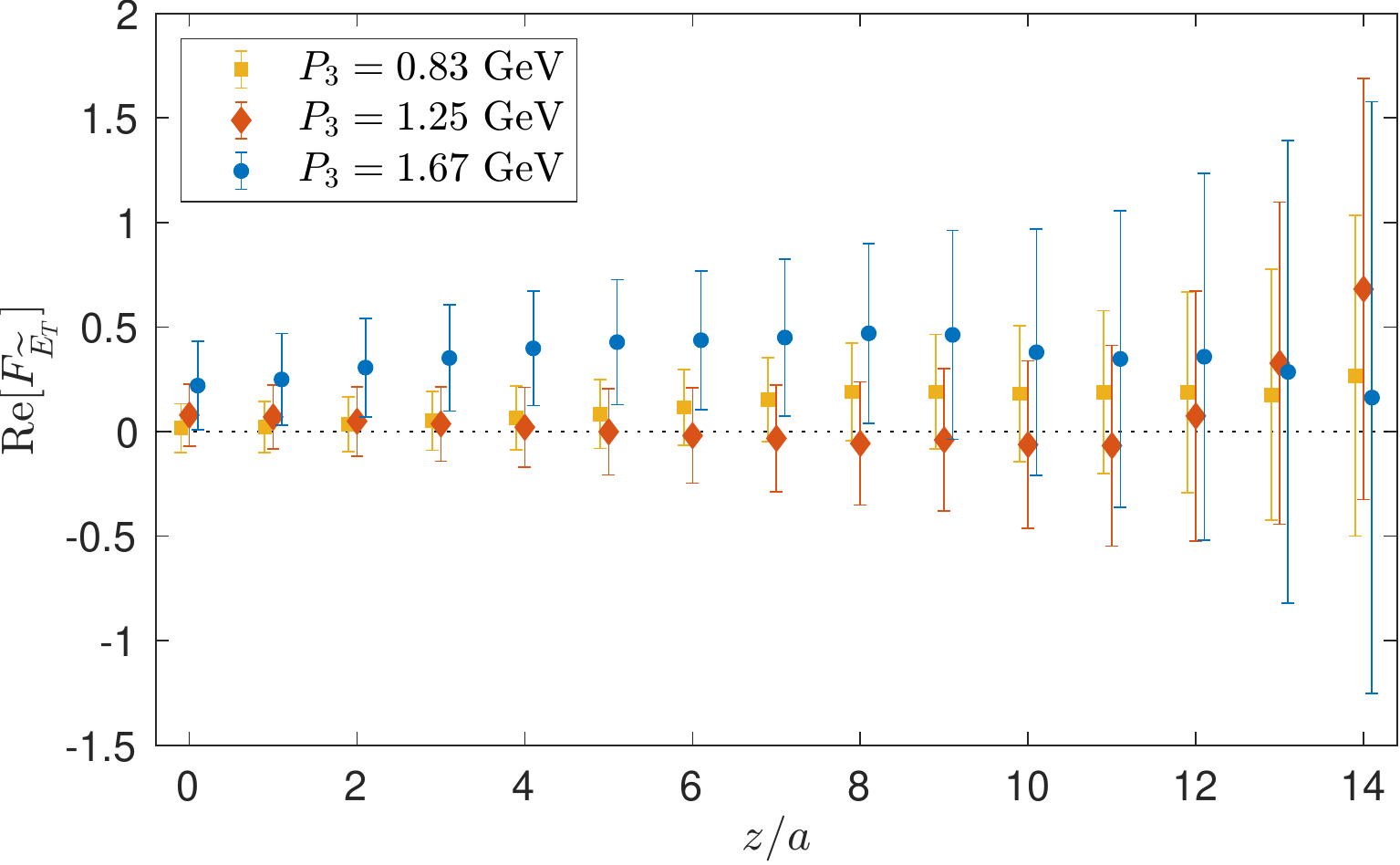}\hspace{0.05cm}
    \includegraphics[scale=0.557]{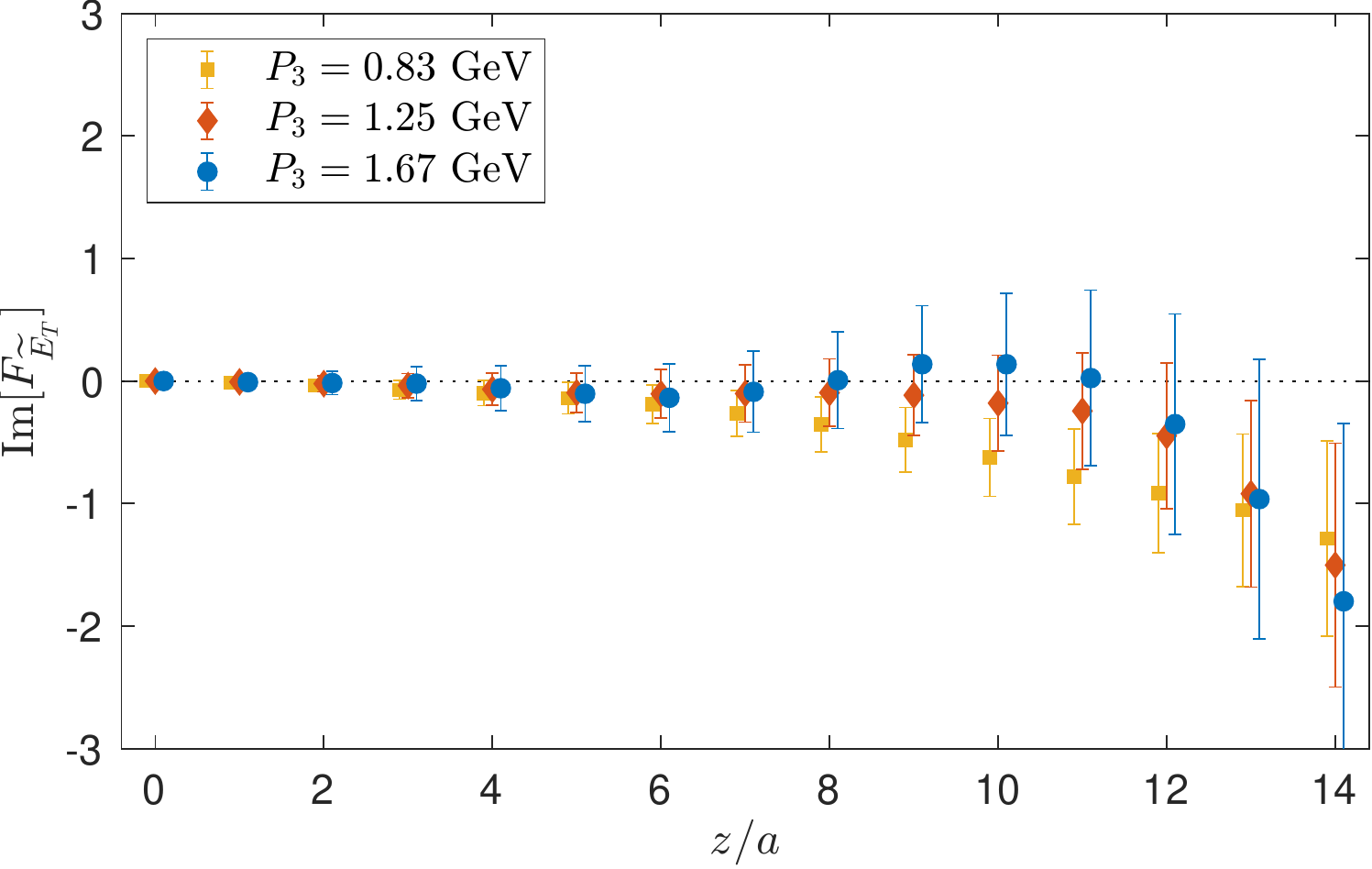}
    \caption{Renormalized $F_{H_T}$, $F_{E_T}$, $F_{\widetilde{H}_T}$ and $F_{\widetilde{E}_T}$ (from top to bottom), extracted using Eqs.~\eqref{eq:h12_zero} - \eqref{eq:h21_zero}. Results correspond to $t=-0.69$~GeV$^2$, $\xi=0$ and three nucleon momenta. The color notation is the same as in Fig.~\ref{fig:ME_hT_xi0}.}
    \label{fig:ME_GPDs_xi0}
\end{figure}

\newpage
Upon renormalization of the matrix elements of Fig.~\ref{fig:ME_hT_xi0}, we disentangle the four $F_G$ that will be eventually matched to each transversity GPD. We demonstrate the dependence of $F_G$ on $P_3$ in Fig.~\ref{fig:ME_GPDs_xi0}. For $z=0$, $F_G$ are independent of $P_3$; this does not hold for $z\ne 0$ due to the breaking of Lorentz invariance. In fact, the values at $z=0$ correspond to the tensor form factors, which are the lowest moments of the transversity GPDs. Further discussion can be found in Sec.~\ref{sec:moments}. Focusing on the highest momentum, we find signal for $F_{H_T}$, $F_{E_T}$ and $F_{\widetilde{H}_T}$. As expected from the behavior of  $h^1(\Gamma_3)$, $F_{\widetilde{E}_T}$ is suppressed compared to the other ones. The imaginary part of $F_{E_T}$ and $F_{\widetilde{H}_T}$ is also zero within uncertainties.

\begin{figure}[h!]
    \centering
    \includegraphics[scale=0.58]{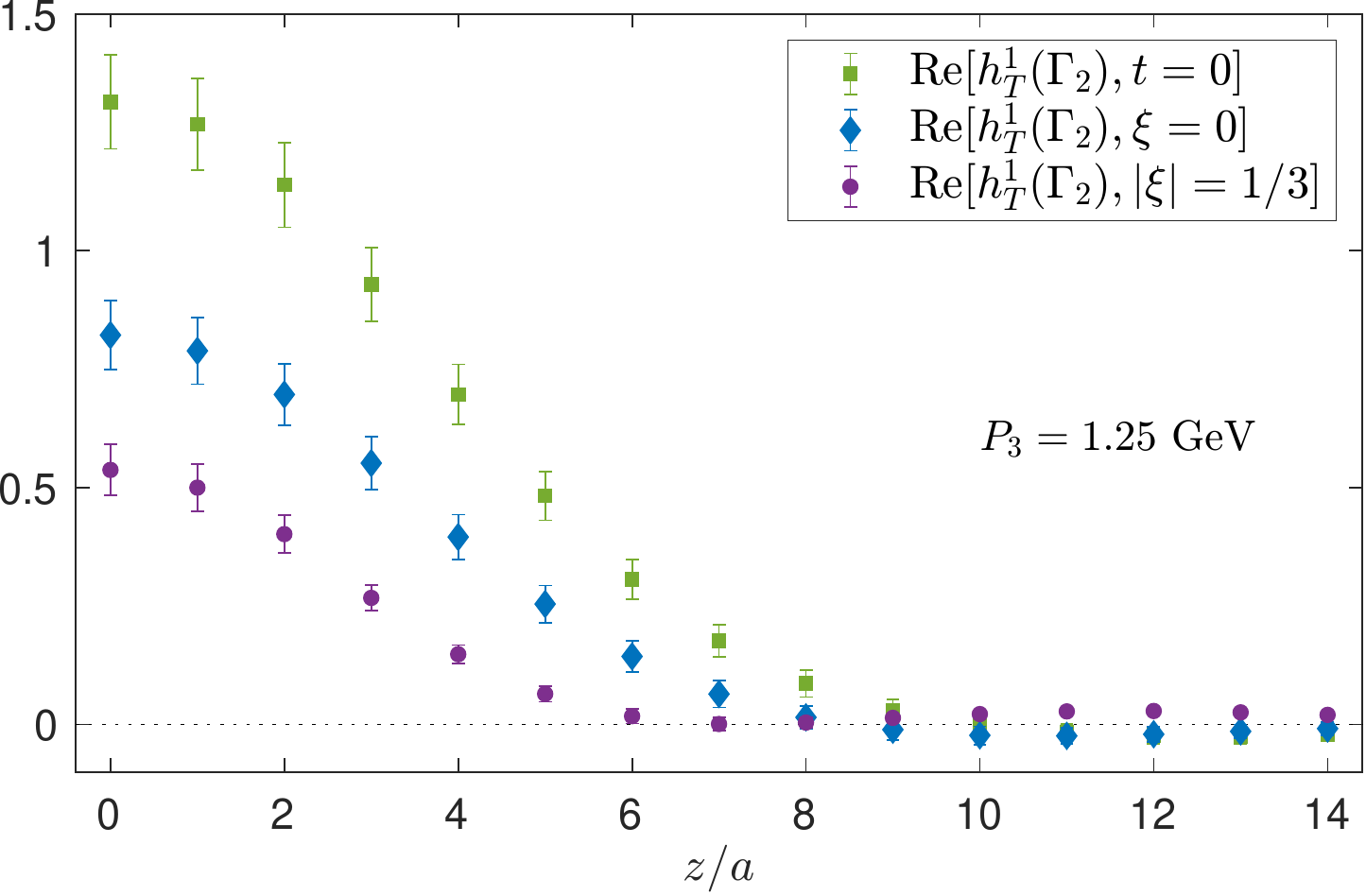}\hspace{0.15cm}
    \includegraphics[scale=0.58]{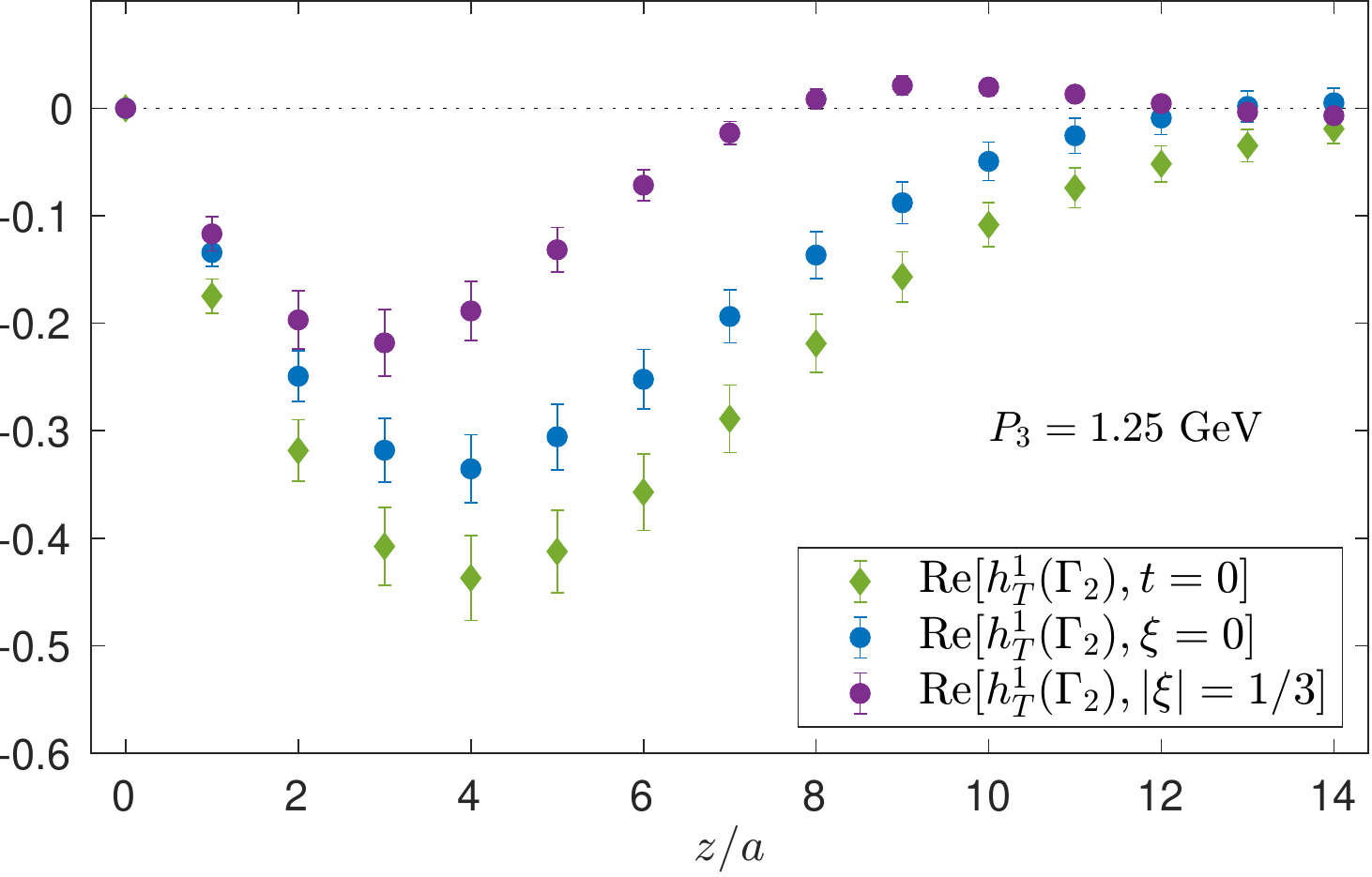}
    \caption{Bare matrix elements computed using the projector $\Gamma_2$ at zero momentum transfer (green squares), $\xi=0$ and $-t=0.69$ GeV$^2$ (blue diamonds), and $|\xi|=1/3$ at $-t=1.02$ GeV$^2$ (purple circles). The nucleon momentum is $P_3=1.25$ GeV.}
    \label{fig:ME_bare_PDF_GPD}
    \end{figure}

It is interesting to compare the matrix elements contributing to $H_T$ for different values of the momentum transfer. In Fig.~\ref{fig:ME_bare_PDF_GPD} we show $h_T^1(\Gamma_2)$ at $P_3=1.25$ GeV for $-t=0\,,0.69,\,1.02$ GeV$^2$. For the case of $\xi=0$ ($-t=0\,,0.69$ GeV$^2$), the matrix element is proportional to $F_{H_T}$, while for $\xi\ne0$ ($-t=1.02$ GeV$^2$) it receives contributions from $F_{E_T}$ and $F_{\widetilde{E}_T}$. The most notable feature of $h_T^1(\Gamma_2)$ is the lowering of its value with the increasing of $-t$ for both the real and imaginary part. The real part becomes compatible with zero at $z/a=9,\,8,\,6$ for $-t=0\,,0.69,\,1.02$ GeV$^2$, respectively. For the imaginary part, we find that compatibility with zero is at $z/a=14,\,12,\,8$ for $-t=0\,,0.69,\,1.02$ GeV$^2$, respectively.

\begin{figure}[h!]
    \centering
    \includegraphics[scale=0.57]{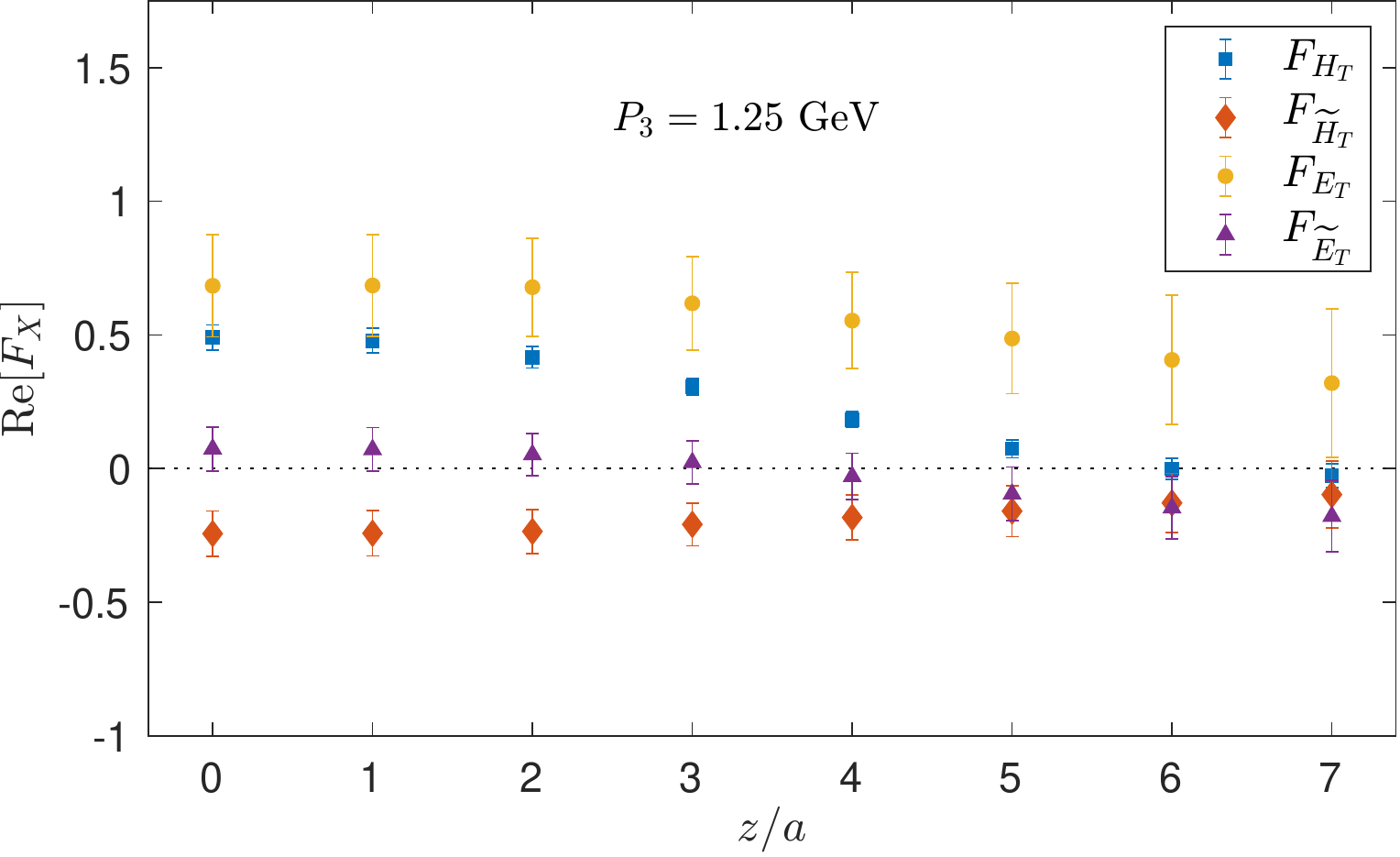}\hspace{0.05cm}
    \includegraphics[scale=0.57]{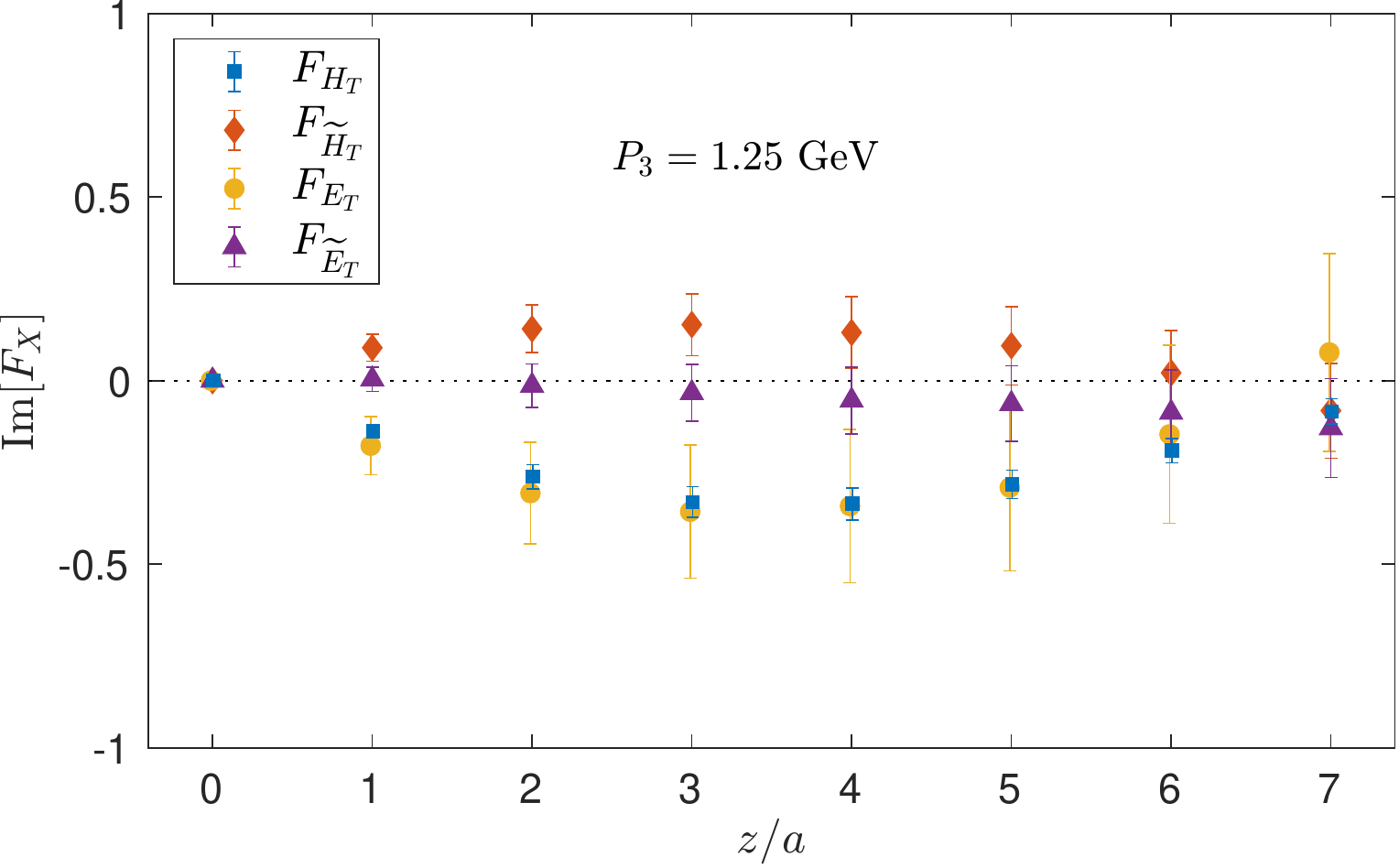}
    \caption{Renormalized $F_{H_T}$ (blue squares), $F_{{\widetilde{H}_T}}$ (red diamonds), $F_{E_T}$ (yellow circles), $F_{\widetilde{E}_T}$ (purple triangles) at $|\xi|=1/3$, $-t=1.02$ GeV$^2$ and nucleon momentum $P_3=1.25$ GeV.}
    \label{fig:ME_GPDs_nonzeroxi}
    \end{figure}

The decomposed renormalized $F_G$ are shown in Fig.~\ref{fig:ME_GPDs_nonzeroxi} for $|\xi|=1/3$ and $-t=1.02$ GeV$^2$. $F_{E_T}$ and $F_{H_T}$ are the most dominant contributions in the matrix elements, followed by  $F_{\widetilde{H}_T}$. The $\xi$-odd $F_{\widetilde{E}_T}$ is compatible with zero. We also find that $F_{H_T}$ ($F_{E_T}$) has the highest (lowest) signal-to-noise ratio. Based on these results, we expect that the final $\widetilde{E}_T$-GPDs will have a signal compatible with zero, and $E_T$ will have enhanced statistical uncertainties as compared to $H_T$.

\section{$x$-dependence of GPDs}
\label{sec:xGPDs}

As mentioned in Secs.~\ref{sec:method} - \ref{sec:latt_details}, the quasi-distribution approach relates the lattice data at a given value of the momentum boost to the light-cone GPDs. Therefore, the final light-cone GPDs should be momentum-independent. Practically, this argument is not exact, because the matching kernel is known only to one-loop level, and there are systematic effects, such as higher-twist contamination. In this work, we use three value of $P_3$ to check for convergence in the final GPDs with respect to the momentum boost. Choosing the right value for the cutoff $z_{\rm max}$ in the reconstruction of the $x$-dependence is also an important aspect of the analysis. The criterion is not unique, and one can use the $z$-behavior of each $F_G$ as a guidance. Based on our results, we choose $z_{\rm max}$ such that the functions $F_G(z_{\rm max})$ become zero. According to this criterion, we find that appropriate choices for $H_T$ at $P_3=0.83,\,1.25,\,1.67$ GeV and $\xi=0$ are $z_{\rm max}/a=13,\,9,\,7$, respectively. This holds for both the real and imaginary part. As expected, the increase of $P_3$ results in a faster decrease of the matrix elements. For the real part of $E_T$ and $\widetilde{H}_T$, we choose $z_{\rm max}/a=7$. Our results for $E_T$ and $\widetilde{H}_T$ indicate that the imaginary part is compatible with zero within errors, and is, hence, neglected. Some fluctuations at large $z_{\rm max}$ are due to the rapid increase of the renormalization functions. For all the GPDs at $|\xi|=1/3$, we use $z_{\rm max}/a=7$ for both the real and imaginary parts. We remind the reader that the distribusions at nonzero skewness, $G(x,1/3,t)$ as already been combined with $G(x,-1/3,t)$, which is symmetric for the three GPDs we show here.

\begin{figure}[h!]
    \centering
    \includegraphics[scale=0.575]{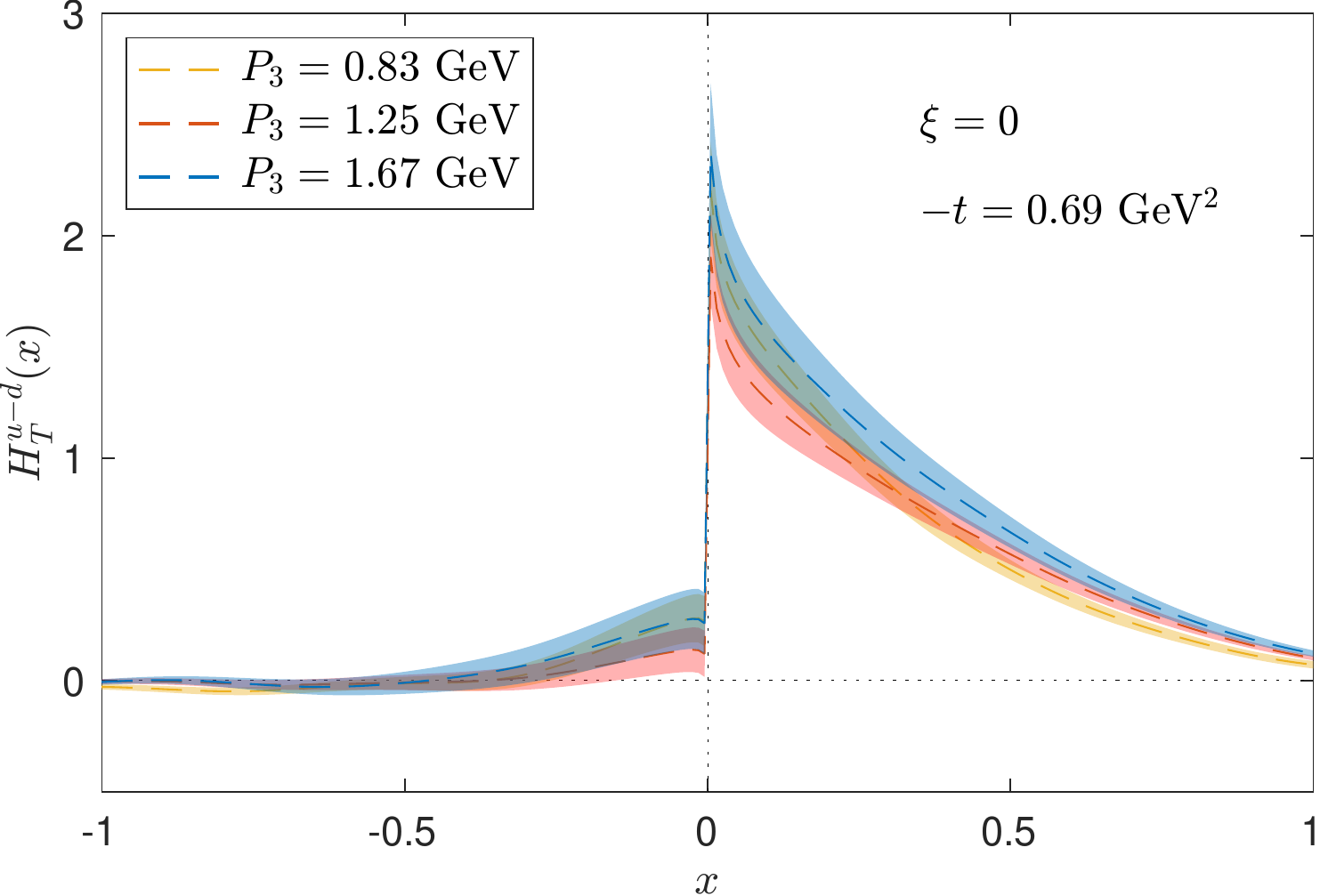}\hspace{0.2cm}
    \includegraphics[scale=0.575]{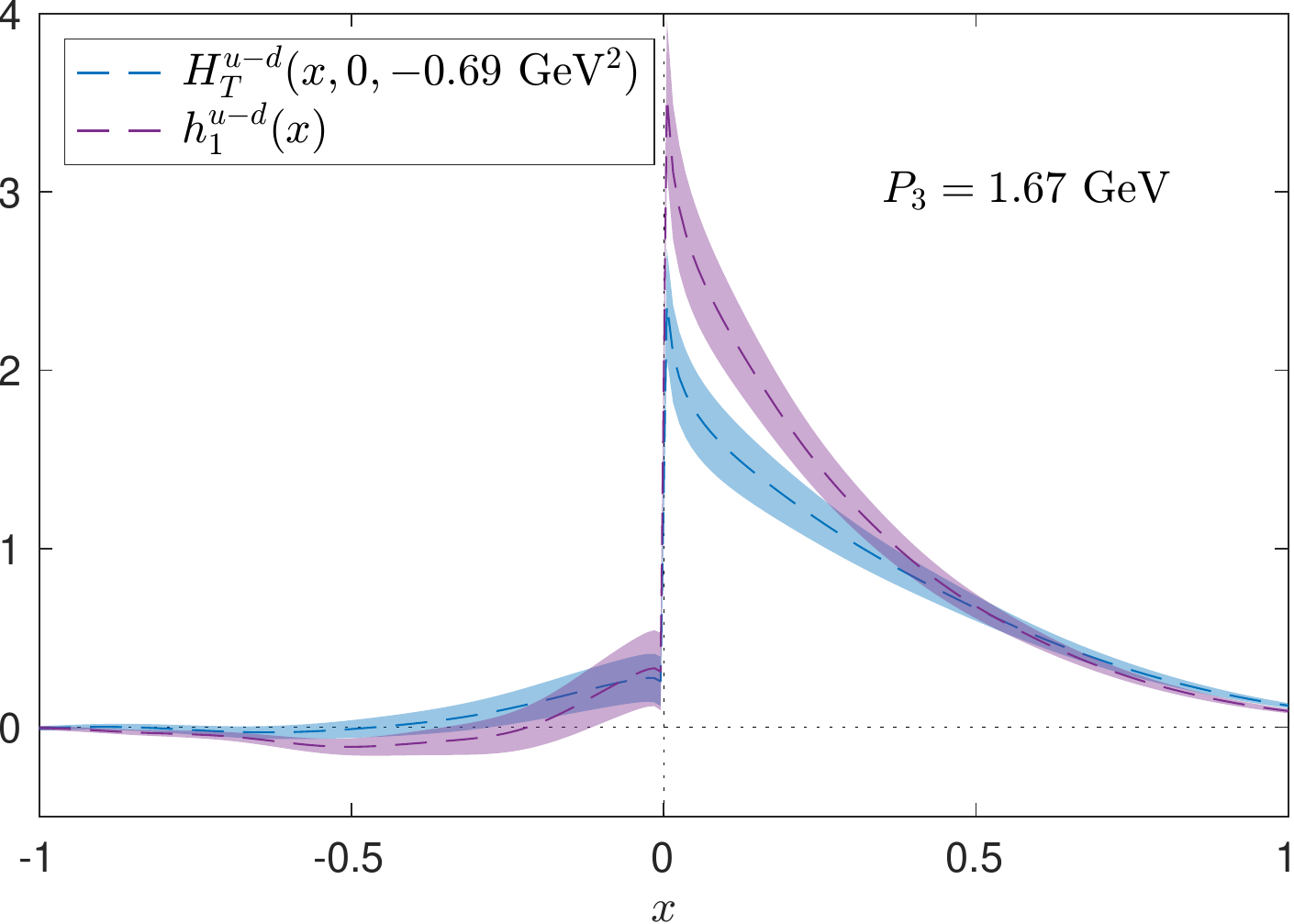}
    \caption{Left: $H_T$ for $\xi=0$ and $-t=0.69$ GeV$^2$, as a function of the proton momentum boost. $P_3=0.83,\,1.25,\,1.67$ GeV is shown with yellow, red and blue bands, respectively. Right: Comparison of $h_1(x)$ (violet band) and $H_T(x,0,-0.69\,{\rm GeV}^2)$ (blue band) for $P_3=1.67$ GeV.   Results are given in the $\overline{\rm MS}$ at a scale of 2 GeV.}
    \label{fig:HT_mom_dependence_xi0}
    \end{figure}

The convergence of $H_T$ is shown in the left panel of Fig.~\ref{fig:HT_mom_dependence_xi0} for $\xi=0$ and $t=-0.69$ GeV$^2$. The bands include only statistical uncertainties. 
We find that convergence is achieved for the two highest values of $P_3$, implying that the reconstructed $H_T$ is momentum-independent even when the matrix elements have a momentum boost of 1.25 GeV. This conclusion is based on the current statistical uncertainties and the one-loop truncation of the matching formalism. For $H_T$, a momentum of 0.83 GeV is also compatible with the higher momenta up to around $x=0.4$. Beyond that point, the distribution is lower than its value for the higher momenta. In the right panel of Fig.~\ref{fig:HT_mom_dependence_xi0}, we compare, at the highest momentum $P_3$, $H_T(x,0,-0.69\,{\rm GeV}^2)$ with its forward limit, $h_1(x)$. We find that for the small and intermediate $x$ region, $h_1(x)$ is higher than $H_T$, which is expected. After $x=0.4$ the two distributions are compatible. The same equality seems to hold numerically for the whole anti-quark region. The large-$x$ behavior of PDFs and GPDs for the unpolarized case has been studied using a power counting analysis~\cite{Yuan:2003fs}. While similar arguments do not exist for the transversity GPDs, our data indicate that there is no $t$ dependence for $x \to 1$, similar to the unpolarized $H$-GPD, but unlike the helicity $\widetilde{H}$-GPD \cite{Alexandrou:2020zbe}. 
    
   \begin{figure}[h!]
    \centering
    \includegraphics[scale=0.575]{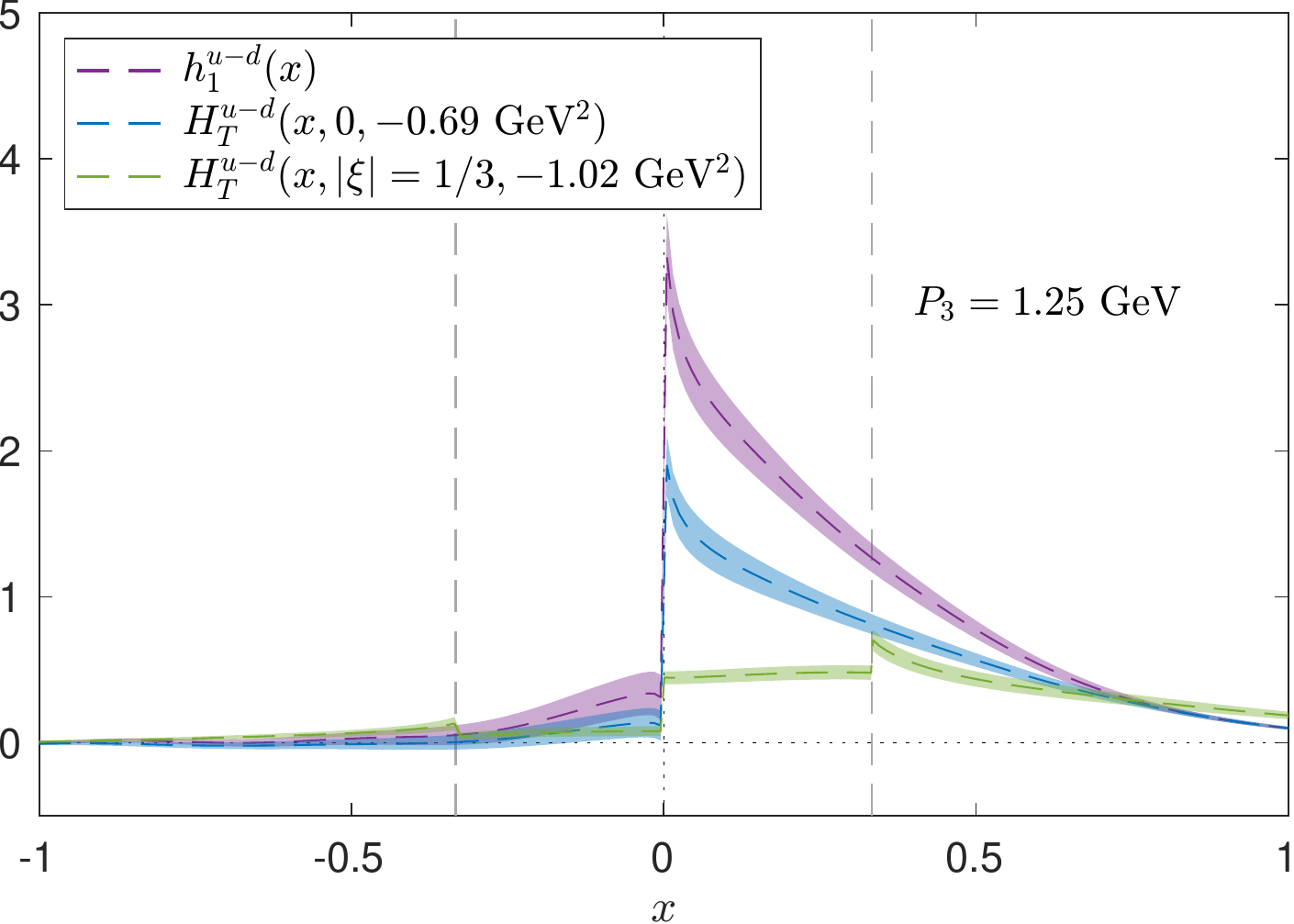}
    \caption{Comparison of $h_1(x)$ (violet band),  $H_T(x,0,-0.69\,{\rm GeV}^2)$ (blue band) and $H_T(x,1/3,-1.02{\rm\,GeV}^2)$ (green band) for $P_3=1.25$ GeV.   Results are given in the $\overline{\rm MS}$ at a scale of 2 GeV.}
    \label{fig:HT_mom_dependence_xinonzero}
    \end{figure}  
    
At $P_3=1.25$ GeV, we have results for both zero and nonzero skewness, which are compared in Fig.~\ref{fig:HT_mom_dependence_xinonzero}. In the ERBL region, there is a significant decrease of the distribution as $-t$ increases. However, the distribution in the DGLAP region shows less sensitivity in $t$. We note that the discontinuity at $x=\pm \xi$ is not physical, as twist-2 GPDs are continuous functions at the boundaries of the ERBL region~\cite{Bhattacharya:2018zxi,Bhattacharya:2019cme}. The observed effect is due to uncontrolled higher-twist contamination, which cannot be treated by the matching formalism as it contains only the leading twist.

\begin{figure}[h!]
    \centering
    \includegraphics[scale=0.56]{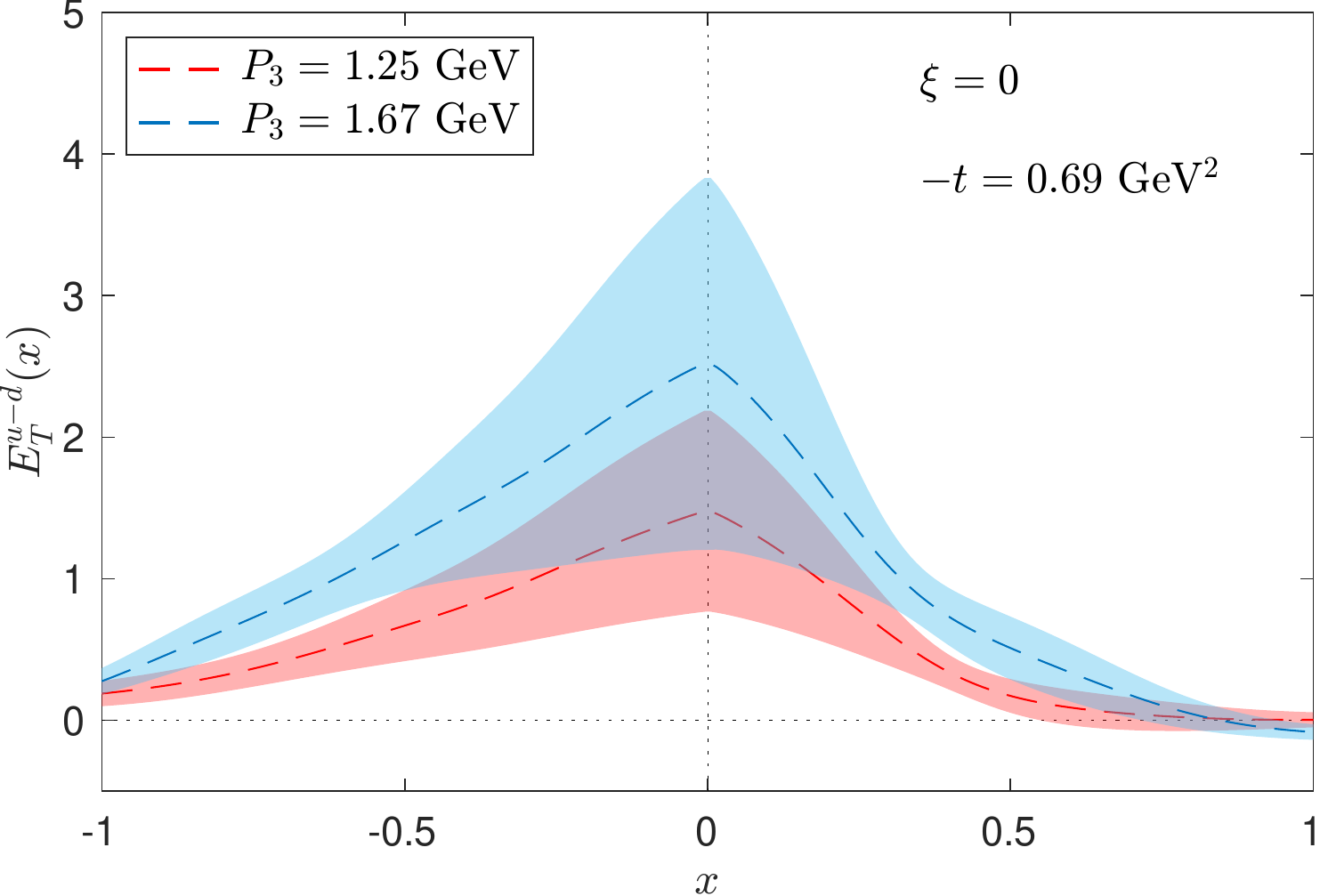}\hspace{0.05cm}
    \includegraphics[scale=0.56]{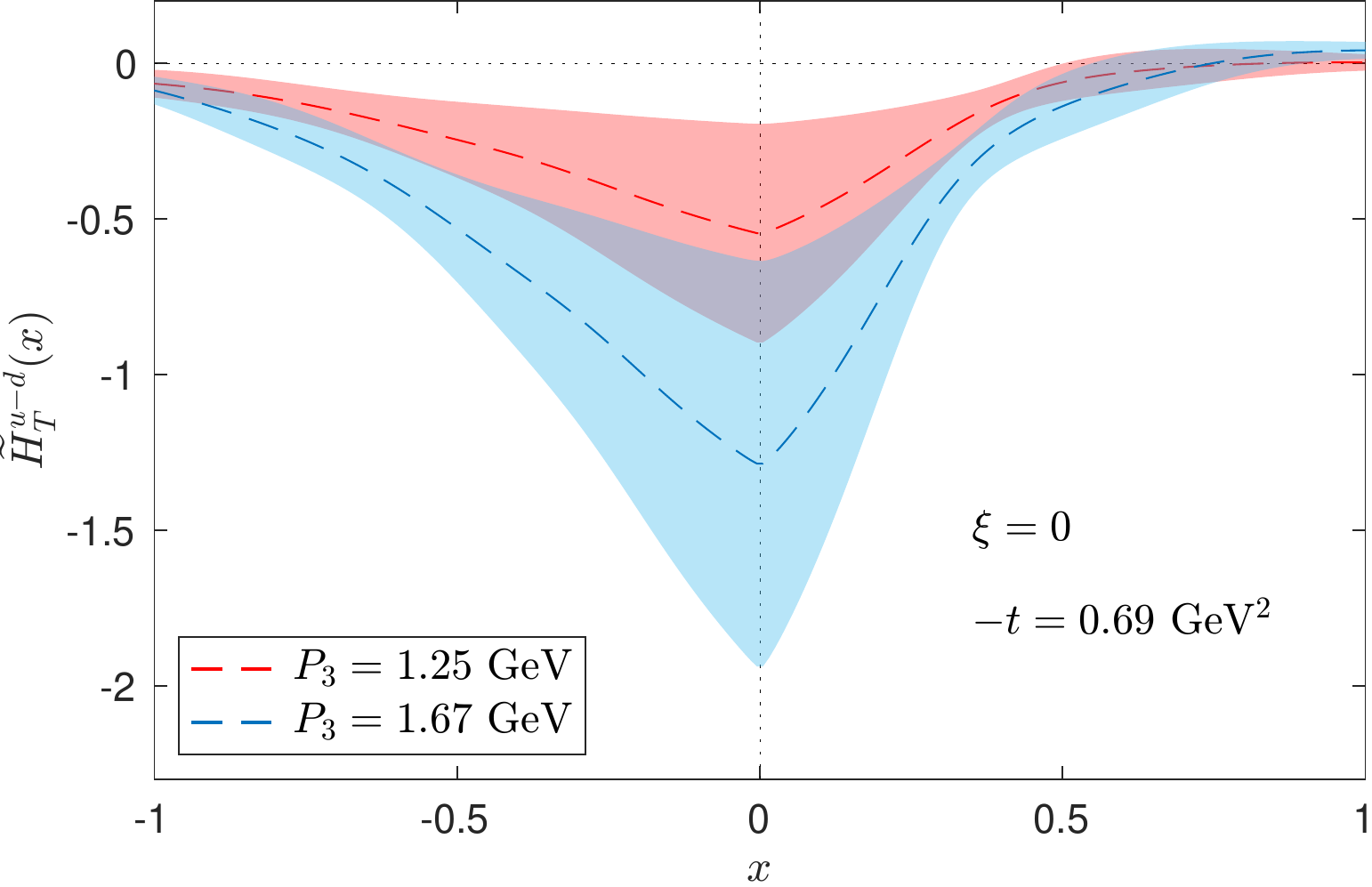}
    \caption{$E_T$ (left) and $\widetilde{H}_T$ (right) for momentum boost of 1.25 GeV (red band) and 1.67 GeV (blue band) at $\xi=0$ and $-t=0.69$ GeV$^2$.   Results are given in the $\overline{\rm MS}$ at a scale of 2 GeV.} 
    \label{fig:HTtilde_ET_mom_dependence_xi0}
    \end{figure}

The data for $E_T$ and $\widetilde{H}_T$ at $\xi=0$ are shown in Fig.~\ref{fig:HTtilde_ET_mom_dependence_xi0} for the two higher momenta. For these GPDs, we do not show results for $P_3=0.83$ GeV, as the matrix elements for $F_{E_T}$ and $F_{\widetilde{H}_T}$ do not decay to zero. This is an indication that a boost of 0.83 GeV is not large enough. As expected from the decomposition of the matrix elements in coordinate space, the uncertainties on these quantities are significantly enhanced compared to $H_T$. Thus, one will need considerably larger statistics to address them in the future. At the present stage, the qualitative conclusion that can be drawn is the approximate symmetry between the quark and antiquark regions, originating from the imaginary part of the respective matrix elements being compatible with zero (see Fig.~\ref{fig:ME_GPDs_xi0}). This also implies a much larger magnitude of the antiquark part for these two GPDs as compared to $H_T$. We also find that $\widetilde{H}_T$ is negative. Similar qualitative conclusions are observed in the scalar diquark model of Ref.~\cite{Bhattacharya:2019cme}. Comparing the distributions for the two momenta, we find compatibility within the large uncertainties.

\begin{figure}[h!]
    \centering
    \includegraphics[scale=0.56]{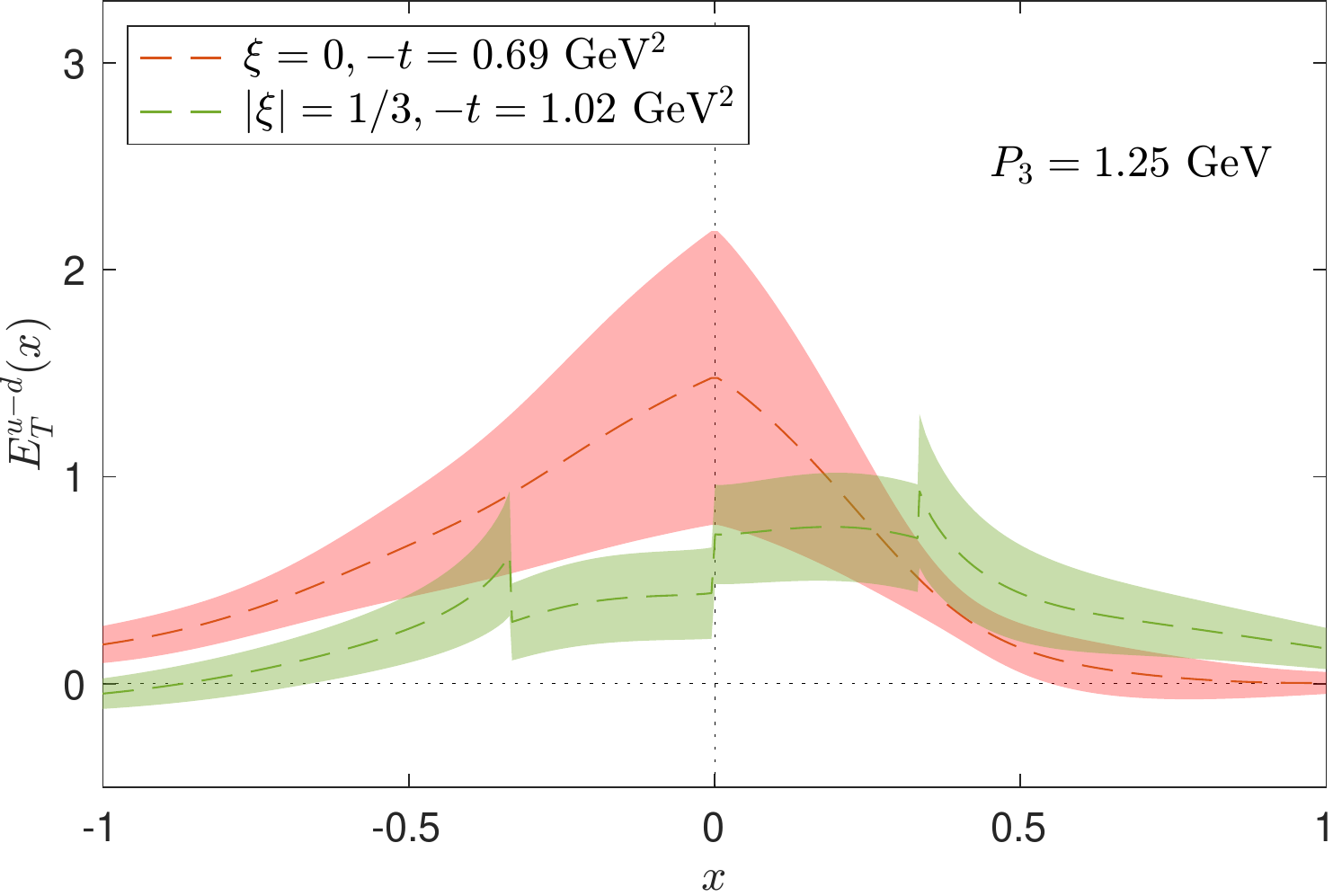}\hspace{0.05cm}
    \includegraphics[scale=0.56]{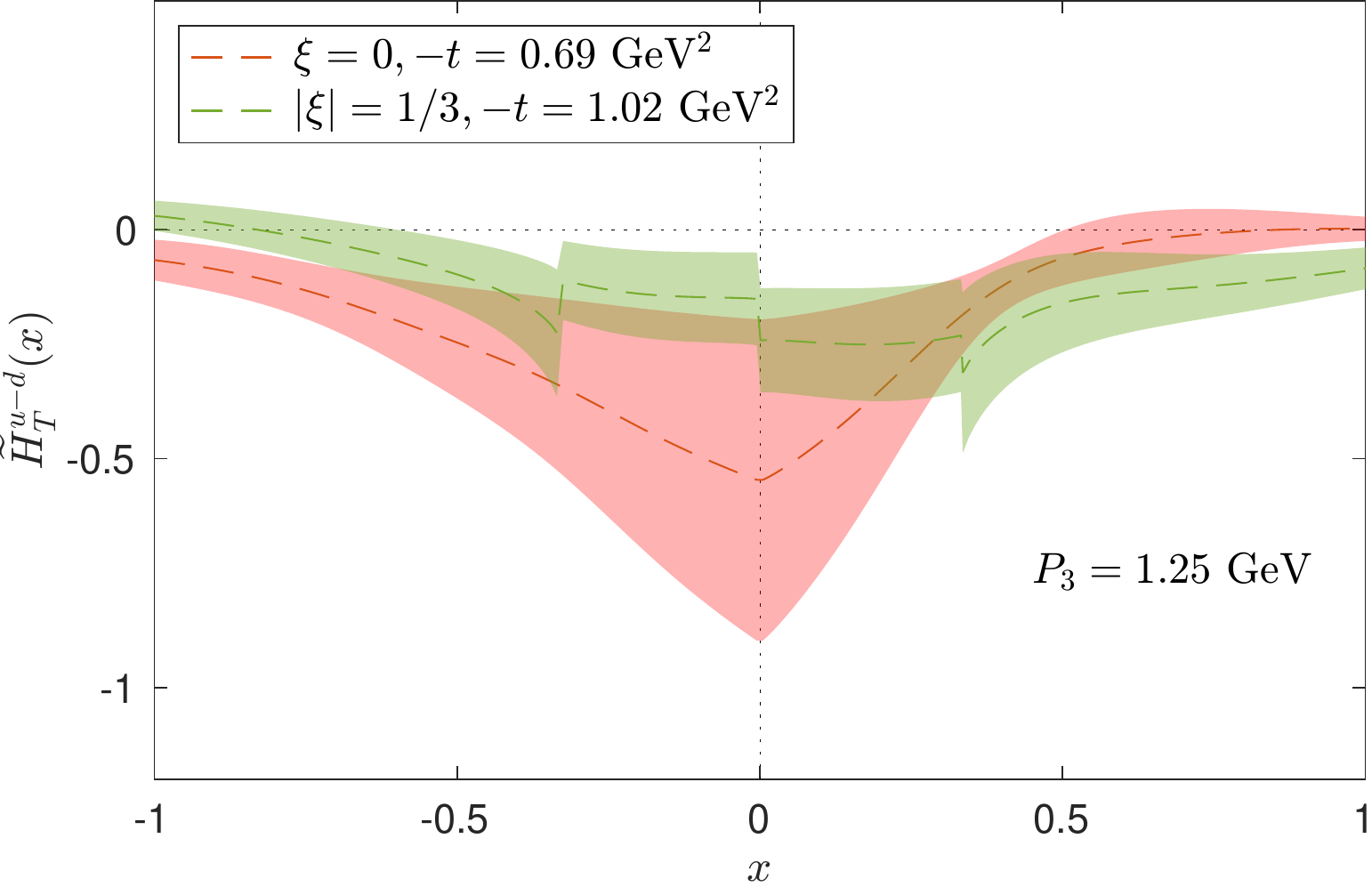}
    \caption{$E_T$ (left) and $\widetilde{H}_T$ (right) for momenta 1.25 GeV, with $\{\xi,t\}=\{0,-0.69\,{\rm GeV}^2\}$ (red band) and $\{|\xi|,t\}=\{1/3,-1.02\,{\rm GeV}^2\}$ (green band).   Results are given in the $\overline{\rm MS}$ at a scale of 2 GeV.}
    \label{fig:HTtilde_ET_nonzero_xi}
    \end{figure}
    
Fig.~\ref{fig:HTtilde_ET_nonzero_xi} compares $E_T$ (left) and $\widetilde{H}_T$ (right) at the same boost of 1.25 GeV for zero and nonzero skewness. The behavior is similar to $H_T$, that is, the increase of $-t$ reduces the magnitude of the distributions, and the introduction of skewness leads to nonphysical discontinuities at $x=\pm\xi$ due to higher-twist effects. However, due to the large uncertainties in $E_T$ and $\widetilde{H}_T$, the function left and right of the boundaries $x=\pm\xi$ appears to be continuous within uncertainties. Here we do not show $\widetilde{E}_T$, as the signal is weak and zero within uncertainties (see, e.g., Fig.~\ref{fig:ME_GPDs_nonzeroxi}). 

\begin{figure}[h!]
    \centering
    \includegraphics[scale=0.56]{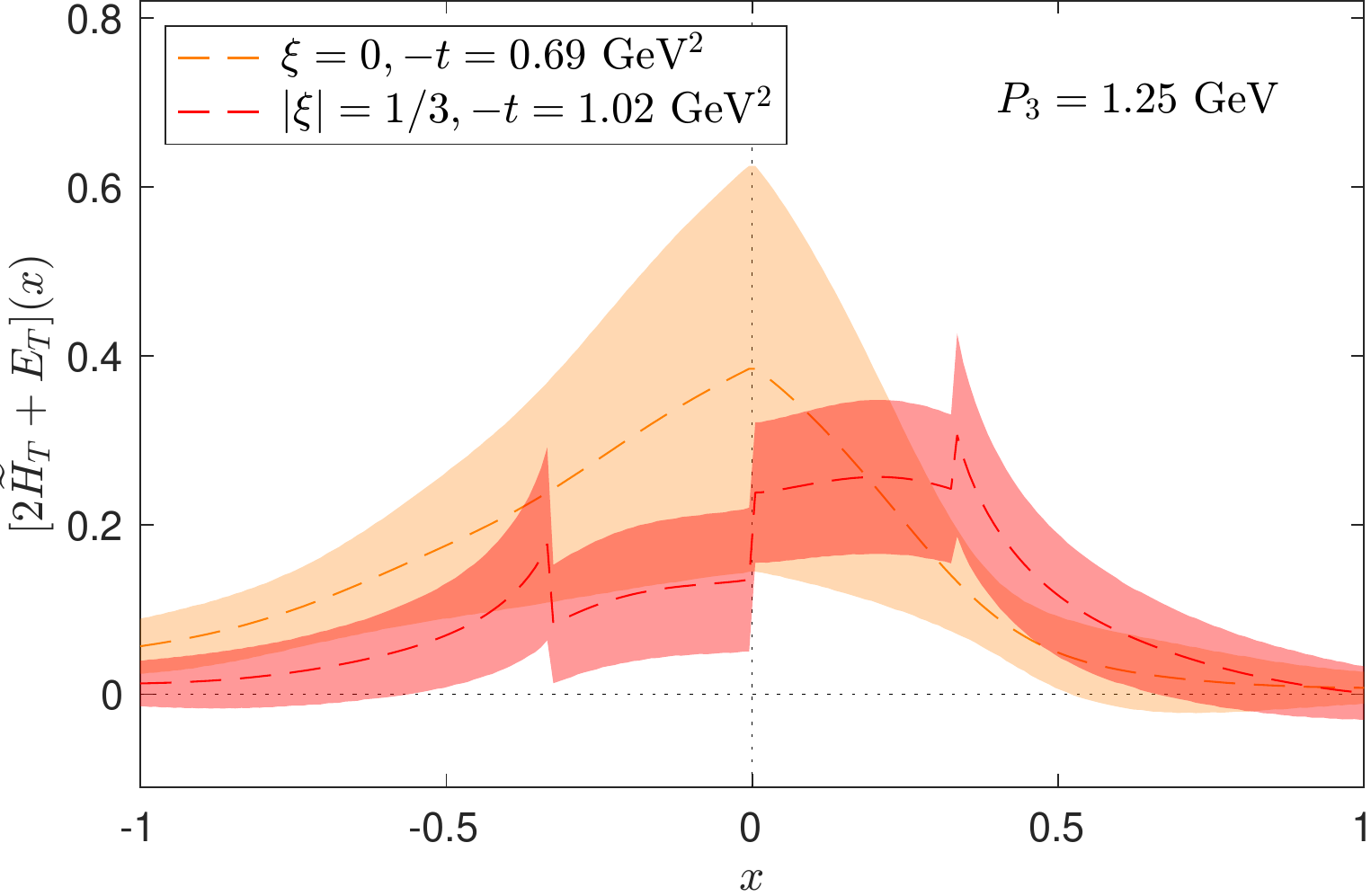}
    \caption{$E_T+ 2 \widetilde{H}_T$ for momentum boost of 1.25 GeV, with $\{\xi,t\}=\{0,-0.69\,{\rm GeV}^2\}$ (orange band) and $\{|\xi|,t\}=\{1/3,-1.02\,{\rm GeV}^2\}$ (red band).   Results are given in the $\overline{\rm MS}$ at a scale of 2 GeV.}
    \label{fig:ETplus2HtildeT}
    \end{figure}

We also explore the combination $E_T+ 2 \widetilde{H}_T$ which is related to the transverse spin structure of the proton, and is considered a more fundamental quantity than $E_T$~\cite{Diehl:2005jf}. The $E_T+ 2 \widetilde{H}_T$ combination has the physical interpretation of the lateral deformation in the distribution of transversely polarized quarks in an unpolarized proton. Also, according to Ref.~\cite{Burkardt:2005hp}, the lowest Mellin moment ($n=0$ in Eq.~\eqref{eq:Mellin}) of $E_T + 2 \widetilde{H}_T$ in the forward limit is the transverse spin-flavor dipole moment in an unpolarized target~\cite{Burkardt:2005hp}, $k_T$. The first non-trivial moment of  $E_T + 2 \widetilde{H}_T$ ($n=1$ in Eq.~\eqref{eq:Mellin}) is related to the transverse-spin quark angular momentum in an unpolarized proton. In Fig.~\ref{fig:ETplus2HtildeT}, we show the combination $E_T+ 2 \widetilde{H}_T$ for $P_3=1.25$ GeV at zero and nonzero skewness. Our results for the two values of $\xi$ are compatible within uncertainties, which are rather large. We do find that the distribution for $|\xi|=1/3$ tends to be systematically lower than the one at $\xi=0$, but further study is needed to control the uncertainties and reach more meaningful conclusions.

\begin{figure}[h!]
    \centering
    \includegraphics[scale=0.58]{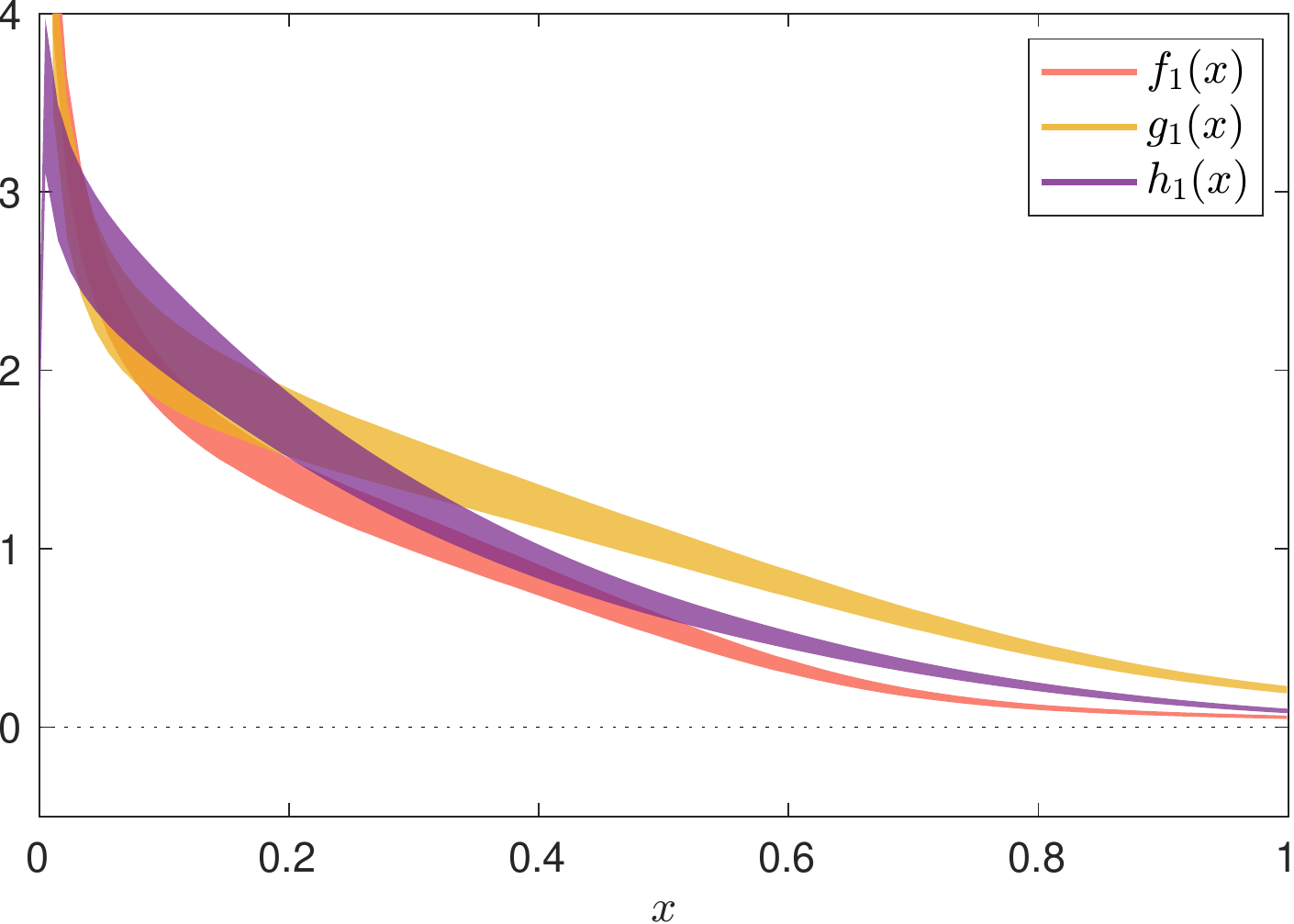}
    \vspace{0.08cm}
    
  \includegraphics[scale=0.58]{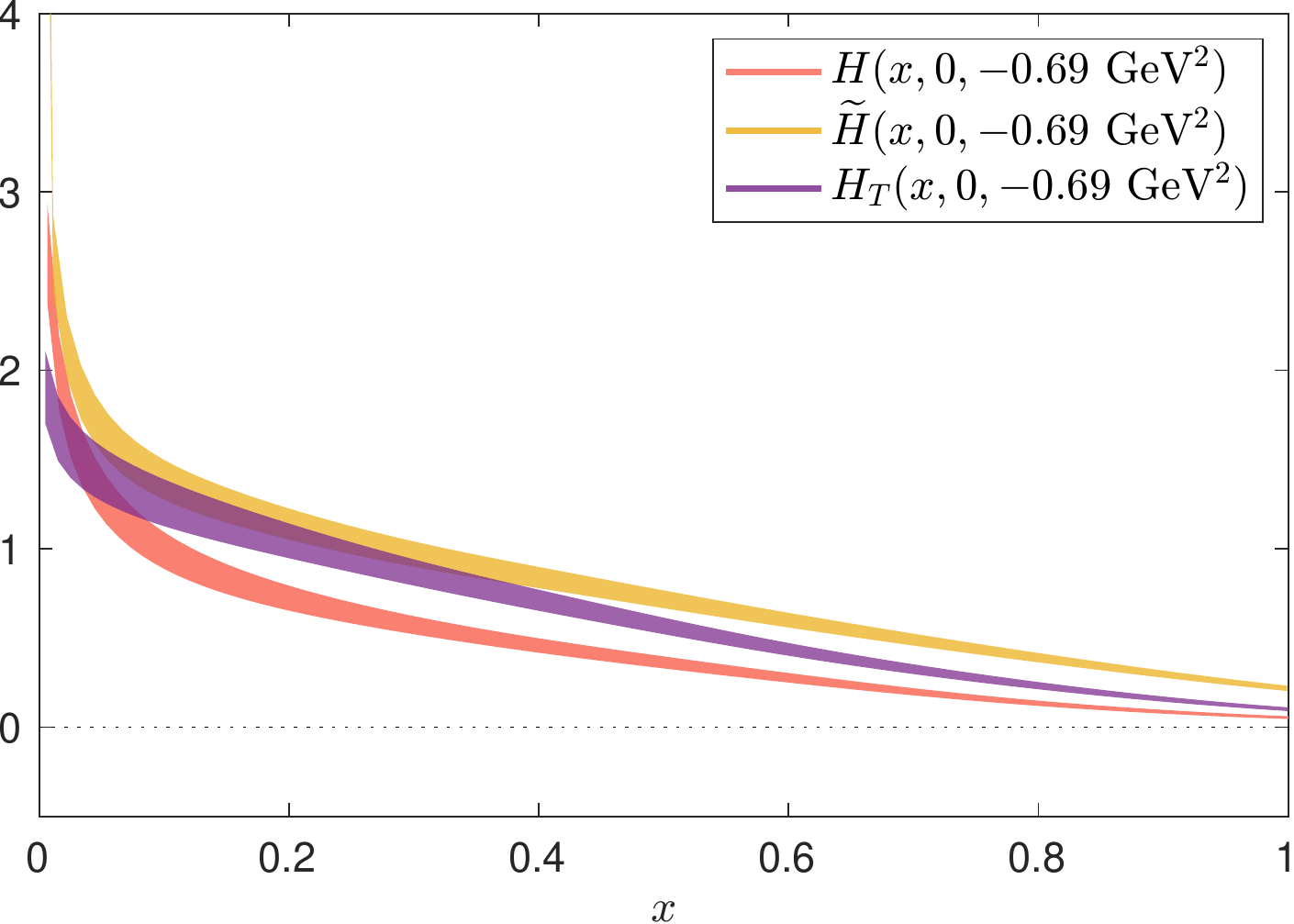}\hspace{0.15cm}
   \includegraphics[scale=0.58]{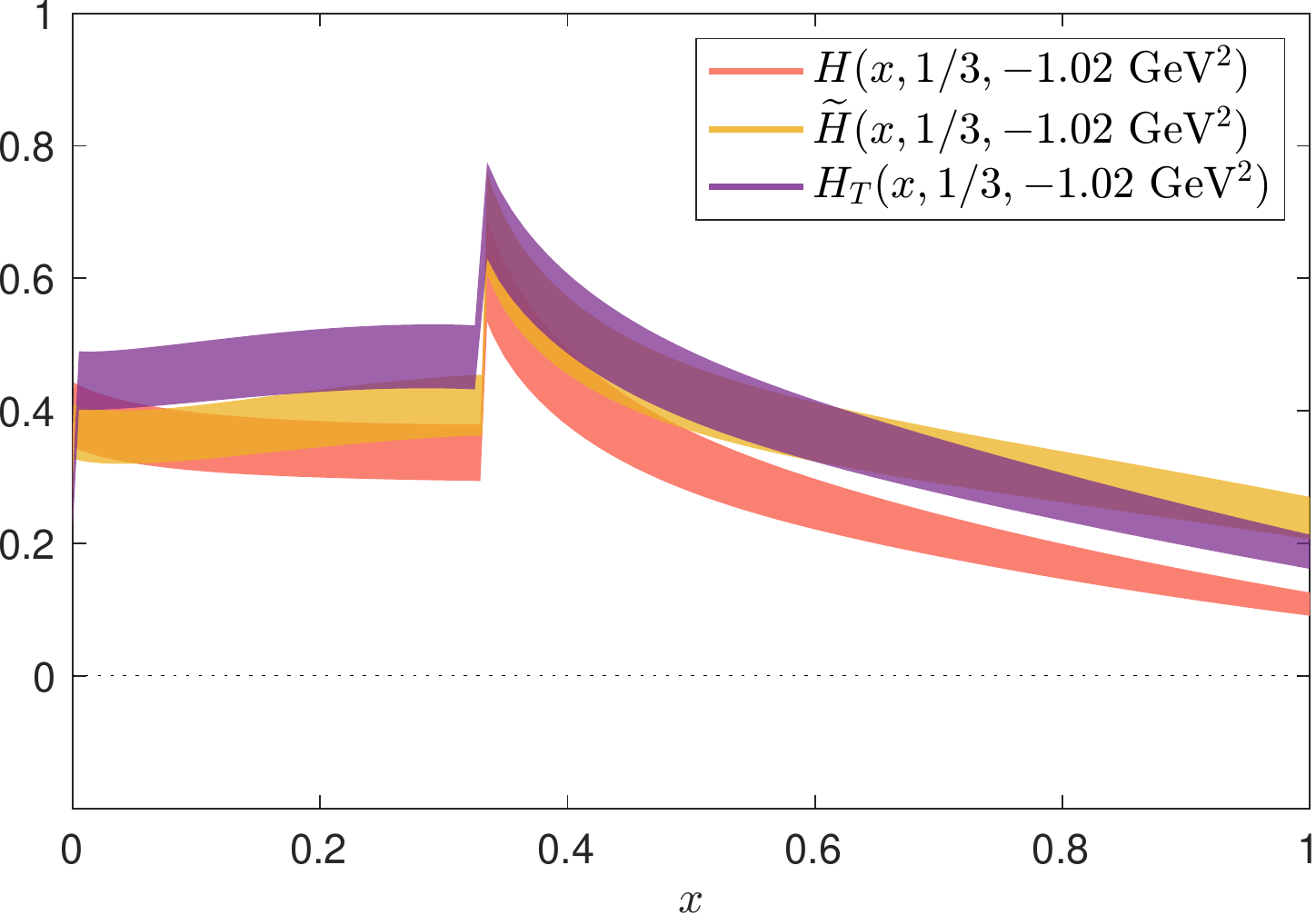}
    \caption{Top plot: the PDFs $f_1(x)$, $g_1(x)$ and $h_1(x)$ for $P_3=1.67$ GeV. Bottom left plot: the GPDs $H(x,0,-0.69\, {\rm GeV}^2)$, $\widetilde{H}(x,0,-0.69\, {\rm GeV}^2)$ and $H_T(x,0,-0.69\, {\rm GeV}^2)$ for  $P_3=1.67$ GeV. Bottom right plot: the GPDs $H(x,1/3,-1.02\, {\rm GeV}^2)$, $\widetilde{H}(x,1/3,-1.02\, {\rm GeV}^2)$ and $H_T(x,1/3,-1.02\, {\rm GeV}^2)$ for  $P_3=1.25$ GeV. The unpolarized, helicity and transversity distributions are shown with red, yellow and purple bands, respectively.   Results are given in the $\overline{\rm MS}$ at a scale of 2 GeV.}
    \label{fig:leading_GPDs}
    \end{figure}

Since we have results for the unpolarized and helicity GPDs on the same ensemble and kinematic setup~\cite{Alexandrou:2020zbe}, it is interesting to compare how the momentum transfer affects these distributions. To this end, we plot the unpolarized ($f_1(x)$), helicity ($g_1(x)$) and transversity ($h_1(x)$) PDFs for $P_3=1.67$ GeV in the top panel of Fig.~\ref{fig:leading_GPDs}. All distributions are of similar magnitude and shape. $f_1(x)$ and $h_1(x)$ decay faster to zero as $x$ increases, while $g_1(x)$ has a comparatively slower decay. The slope of $f_1(x)$ and $h_1(x)$ in the small and intermediate $x$ region is similar. For $\{\xi,t,P_3\}=\{0,-0.69\, {\rm GeV}^2, 1.67\, {\rm GeV}\}$ (lower left plot of Fig.~\ref{fig:leading_GPDs}), we observe that all distributions are suppressed compared to the PDFs. In particular, the decrease is more significant for the unpolarized case, that is $H(x,0,-0.69\, {\rm GeV}^2)$ is lower than $H_T(x,0,-0.69\, {\rm GeV}^2)$ in the small- and intermediate-$x$ regions. Their difference in the large-$x$ region remains the same. Furthermore, $\widetilde{H}(x,0,-0.69\, {\rm GeV}^2)$ and $H_T(x,0,-0.69\, {\rm GeV}^2)$ are compatible for $x<0.4$. Further increase of the momentum transfer, $\{|\xi|,t,P_3\}=\{1/3,-0.69\, {\rm GeV}^2, 1.25\, {\rm GeV}\}$, suppresses the GPDs even more (lower right plot of Fig.~\ref{fig:leading_GPDs}). We note that this plot corresponds to the maximum available momentum, $P_3=1.25$ GeV. However, we observed convergence with momentum for the leading GPDs and their PDFs, so comparison with the upper and lower left plots is acceptable. As in the previous two plots, the unpolarized $H(x,1/3,-1.02\, {\rm GeV}^2)$ is lower than the other two distributions. In the ERBL region, the transversity is slightly higher than $H$ and $\widetilde{H}$. In the DGLAP region, $\widetilde{H}$ and $H_T$ are compatible, while $H$ is a bit lower. We want to emphasize again that these observations are only qualitative.

\section{Moments of GPDs}
\label{sec:moments}

The Mellin moments of GPDs are defined via
\begin{equation}
\label{eq:Mellin}
\int_{-1}^{1} dx \, x^n\, G(x,\xi,t)\,,\qquad n=0,\,1,\,\cdots\,\,.
\end{equation}
These are interesting in their own right, as they are related to form factors and towers of generalized form factors. Recently, there has been exploration of the Mellin moments of quasi-GPDs, and their relation to the moments of GPDs. Of particular relevance is the work of Ref.~\cite{Bhattacharya:2019cme}, which derives relations for the Mellin moments of the transversity quasi-GPDs and GPDs using model-independent arguments. Also, a numerical analysis is presented using the diquark spectator model. The following model-independent relations are given for the $n=0$ Mellin moments for both the GPDs and the quasi-GPDs~\footnote{Here, the integral of the quasi-GPDs does not contain the kinematic factor shown in Ref.~\cite{Bhattacharya:2019cme}, because all factors are included in our definition of the matrix elements $h^j_T$.},

\begin{eqnarray}
\label{eq:mom0HT}
\int_{-1}^{1} dx \, H_T(x,\xi,t) &=&  \int_{-\infty}^{\infty} dx \,  H_{Tq}(x,\xi,t,P_3)  = A_{T10}(t) \,,\\[2.5ex]
\int_{-1}^{1} dx \, E_T(x,\xi,t) &=&  \int_{-\infty}^{\infty} dx \,  E_{Tq}(x,\xi,t,P_3) = B_{T10}(t)\,, \\[2.5ex]
\label{eq:mom0HtT}
\int_{-1}^{1} dx \, \widetilde{H}_T(x,\xi,t) &=& \int_{-\infty}^{\infty} dx \,  \widetilde{H}_{Tq}(x,\xi,t,P_3)  = \widetilde{A}_{T10}(t) \,,  \\[2.5ex]
\int_{-1}^{1} dx \, \widetilde{E}_T(x,\xi,t) &=& \int_{-\infty}^{\infty} dx \, \widetilde{E}_{Tq}(x,\xi,t,P_3) = 0\,.
\label{eq:mom0EtT}
\end{eqnarray}
As can be seen, the lowest moments of GPDs are independent of $\xi$, and the lowest moments of quasi-GPDs are, in addition, $P_3$-independent. The form factors $A_{T10},\,B_{T10},\,\widetilde{A}_{T10}$ are extracted from the matrix element of the local tensor operator as defined in Ref.~\cite{Diehl:2005jf}. Note that the lowest moment of $\widetilde{E}_T$ is zero due to time reversal symmetry~\cite{Diehl:2001pm}. 

The corresponding relation for the $n=1$ Mellin moments of the transversity GPDs are related to the generalized form factor of the one-derivative tensor operator, that is~\cite{Diehl:2005jf}
\begin{eqnarray}
\label{eq:momHT}
\int_{-1}^{1} dx \, x\, H_T(x,\xi,t) &=& A_{T20}(t) \,, \\[2.5ex]
\int_{-1}^{1} dx \, x\, E_T(x,\xi,t) &=& B_{T20}(t) \,,\\[2.5ex]
\int_{-1}^{1} dx \, x\, \widetilde{H}_T(x,\xi,t) &=& \widetilde{A}_{T20}(t) \,, \\[2.5ex]
\label{eq:mom1EtildeT}
\int_{-1}^{1} dx \, x\, \widetilde{E}_T(x,\xi,t) &=& 2 \xi \widetilde{B}_{T21}(t)\,.
\end{eqnarray}
For the two lowest moments, that is, $n=0,1$ (Eqs.~\eqref{eq:mom0HT} - \eqref{eq:mom1EtildeT}), a $\xi$-dependence appears only in the $n=1$ moment of $\widetilde{E}_T$, as it is the only $\xi$-odd GPD. $H_T$, $E_T$ and $\widetilde{H}_T$ are even functions of $\xi$.

Here, we calculate the moments of the transversity GPDs as a consistency check of our results. Our goal is not to provide numerical results for the form factors and generalized form factors, as the calculation of these quantities are still at an exploratory stage, but to perform a number of checks for the Mellin moments using our results:
\begin{itemize}
\item[\textbf{1.\,}] $P_3$-independence of the $n=0$ Mellin moments of quasi-GPDs;
\item[\textbf{2.\,}] Relation between the $n=0$ Mellin moments of GPDs and quasi-GPDs;
\item[\textbf{3.\,}] $t$-dependence of the form factors and generalized form factors;
\item[\textbf{4.\,}] Relation between the $n=0$ and $n=1$ Mellin moments of a given GPD;
\item[\textbf{5.\,}] Comparison between the $n=0$ Mellin moments with the corresponding value of the matrix element at $z=0$.
\end{itemize}

\noindent
For the $H_T$-GPD, we find the following values using our lattice data

\begin{align}
\label{eq:qHTn0}
\hspace*{-0.25cm}
 \int_{-2}^{2} dx  H_{Tq}(x,0,{-}0.69\,{\rm GeV}^2,P_3) {=} \{0.65(4),\, 0.64(6),\,0.81(10)\} \,, 
\,\,\, & \,\,\,  \int_{-2}^{2}  dx  H_{Tq}(x,\frac{1}{3},-1.02\,{\rm GeV}^2,1.25\,{\rm GeV}) {=} 0.49(5) \,, \\[1.5ex]
 \int_{-1}^{1} dx \, H_T(x,0,-0.69\,{\rm GeV}^2) = \{0.69(4),\,0.67(6),\,0.84(10)\} \,, 
\,\,\, & \,\,\,   \int_{-1}^{1} dx \, H_{T}(x,\frac{1}{3},-1.02\,{\rm GeV}^2) = 0.45(4) \,, \\[1.5ex]
 \int_{-1}^{1} dx \, x\, H_T(x,0,-0.69\,{\rm GeV}^2) = \{0.20(2),\,0.21(2),\,0.24(3)\} \,, 
\,\,\, & \,\,\,   \int_{-1}^{1} dx \, x \, H_{T}(x,\frac{1}{3},-1.02\,{\rm GeV}^2) = 0.15(2) \,.
\end{align}

\noindent
For the $E_T$-GPD, we have

\begin{align}
\label{eq:qETn0}
 \int_{-2}^{2} dx \, E_{Tq}(x,0,-0.69\,{\rm GeV}^2,P_3) = \{   1.20(42) ,\, 2.05(65) \} \,,
\quad & \quad  \int_{-2}^{2}  dx \, E_{Tq}(x,\frac{1}{3},-1.02\,{\rm GeV}^2,1.25\,{\rm GeV}) = 0.67(19)\,, \\[1.5ex]
 \int_{-1}^{1} dx \, E_T(x,0,-0.69\,{\rm GeV}^2) = \{   1.15(43) ,\, 2.10(67)\} \,,
\quad & \quad  \int_{-1}^{1} dx \, E_{T}(x,\frac{1}{3},-1.02\,{\rm GeV}^2) = 0.73(19)  \,,  \hspace*{1cm}  \\[1.5ex]
 \int_{-1}^{1} dx \, x\, E_T(x,0,-0.69\,{\rm GeV}^2) =  \{  0.06(4),\, 0.13(5) \} \,,
\quad & \quad  \int_{-1}^{1}  dx \, x \, E_{T}(x,\frac{1}{3},-1.02\,{\rm GeV}^2) = 0.11(11) \,,   \hspace*{0.5cm}
\end{align}

\noindent
and for the $\widetilde{H}_T$-GPD, we find

\begin{align}
\label{eq:qHtTn0}
 \int_{-2}^{2} dx \, \widetilde{H}_{Tq}(x,0,-0.69\,{\rm GeV}^2,P_3) =  \{  -0.44(20),\, -0.90(32) \} \,, \quad & \quad  \int_{-2}^{2}  dx \, \widetilde{H}_{Tq}(x,\frac{1}{3},-1.02\,{\rm GeV}^2,1.25\,{\rm GeV}) = -0.26(9) \,,  \\[1.5ex]
\int_{-1}^{1} dx \, \widetilde{H}_T(x,0,-0.69\,{\rm GeV}^2) =  \{  -0.42(21),\, -0.92(33)  \} \,, \quad & \quad \int_{-1}^{1} dx \, \widetilde{H}_{T}(x,\frac{1}{3},-1.02\,{\rm GeV}^2) = -0.27(9) \,, \\[1.5ex]
\int_{-1}^{1} dx \, x\, \widetilde{H}_T(x,0,-0.69\,{\rm GeV}^2) = \{   -0.17(8) ,\, -0.30(10) \}\,, \quad & \quad   \int_{-1}^{1} dx \, x \, \widetilde{H}_{T}(x,\frac{1}{3},-1.02\,{\rm GeV}^2) = -0.05(5) \,.
\end{align}

\noindent
The numbers in the curly brackets correspond to $P_3=\lbrace{ 0.83,\,1.25,\,1.67\rbrace}$ GeV for $H_T$, respectively. For $E_T$ and $\widetilde{H}_T$, we only show results for $P_3=\lbrace{1.25,\,1.67\rbrace}$ GeV as explained in the previous section. For the quasi-GPDs, we integrate in the region $x \in [-2,+2]$, but we checked that extending the interval gives compatible results. The $n=1$ moment of $\widetilde{E}_T$ is zero within uncertainties for $\xi=0$, which is consistent with Eq.~\eqref{eq:mom1EtildeT}. 
Before commenting further on the above results, let us also provide the values of the form factors, as extracted from the matrix elements at $z=0$,
\begin{eqnarray}
\label{eq:AT10}
A_{T10}(-0.69\,{\rm GeV}^2) = \{0.65(4)\,, 0.65(6)\,, 0.82(10)\} \,, &\qquad\qquad& A_{T10}(-1.02\,{\rm GeV}^2) = 0.49(5) \,, \\[2ex]
B_{T10}(-0.69\,{\rm GeV}^2) = \{1.71(28)\,, 1.22(43)\,, 2.10(67)\} \,, &\qquad\qquad& B_{T10}(-1.02\,{\rm GeV}^2) = 0.68(19) \,, \\[2ex]
\widetilde{A}_{T10}(-0.69\,{\rm GeV}^2) = \{-0.67(14)\,, -0.45(21)\,, -0.92(33)\} \,, &\qquad\qquad& \widetilde{A}_{T10}(-1.02\,{\rm GeV}^2) =  -0.24(8) \,,
\label{eq:AtT10}
\end{eqnarray}
Similar to the Mellin moments of GPDs, the form factors do not depend on the momentum boost of the proton. Since Eqs.~\eqref{eq:AT10} - \eqref{eq:AtT10} are extracted directly from the matrix elements, we can also provide estimates for $B_{T10}$ and 
$\widetilde{A}_{T10}$ at $P_3=0.83$ GeV.

\newpage
Based on the results shown in Eqs.~\eqref{eq:momHT} - \eqref{eq:AtT10}, we conclude the following
\begin{itemize}
\item[\textbf{1.\,}] For the quasi-GPDs at $t=-0.69$ GeV$^2$ and $\xi=0$, we have three momenta for $H_{Tq}$ (Eq.~\eqref{eq:qHTn0}) and two momenta for $E_{Tq}$ and $\widetilde{H}_{Tq}$ (Eq.~\eqref{eq:qETn0} and Eq.~\eqref{eq:qHtTn0}). The two lowest $P_3$ for $H_T$ are in agreement, and in slight tension with $P_3=1.67$ GeV. The values for  $E_{Tq}$ between $P_3=1.25$ GeV and $P_3=1.67$ GeV are consistent within the uncertainties. It should be mentioned, however, that the uncertainties are much larger than for $H_{Tq}$. Similar conclusions to the ones for $E_{Tq}$ are also valid for $\widetilde{H}_{Tq}$. 

We observe that the agreement of both the $n=0,1$ moments of the GPDs for different $P_3$ values is better than for the quasi-GPDs. This is an indication that the matching procedure removes the bulk of the $P_3$-dependence.
\item[\textbf{2.\,}] The $n=0$ moments of quasi-GPDs for a given value of $P_3$ are fully compatible with the results of the $n=0$ moment of the corresponding GPDs for the same value of $P_3$.
\item[\textbf{3.\,}] For all the $n=0$ moments that we present here, we find that the values at $t=-1.02$ GeV$^2$ are lower than those at $t=-0.69$ GeV$^2$, as expected. For $n=1$, we observe a flatter behavior with increase of $-t$ in $H_T$. This is similar to the $t$-dependence of past calculations, for example, of $A_{T20}$ using one-derivative operators~\cite{Gockeler:2005cj,Alexandrou:2013wka}. For $E_T$ and $\widetilde{H}_T$, the signal decays to zero at $t=-1.02$ GeV$^2$.
\item[\textbf{4.\,}] Another outcome of the numerical analysis is the fact that the $n=1$  moment of a given GPD is suppressed compared to $n=0$. This is expected, as the higher moments have support at higher values of $x$, where the GPDs decay.
\item[\textbf{5.\,}] Finally, we compare the $n=0$ moments (Eqs.~\eqref{eq:mom0HT} - \eqref{eq:mom0HtT}) with the value of the matrix elements at $z=0$ (Eqs.~\eqref{eq:AT10} - \eqref{eq:AtT10}). We find that these are in excellent agreement with the values obtained from the integrals, for both $t=-0.69$ GeV$^2$ and $t=-1.02$ GeV$^2$. Regarding the case of $\widetilde{E}_T$, the results are consistent with zero.
\end{itemize}
 The above conclusions are highly nontrivial~\footnote{We note that the equality of the zeroth Mellin moments of quasi-GPDs and GPDs should be trivially satisfied due to the use of the full plus function of Eq.~\eqref{e:bare_matching}.}, as the extraction of the Mellin moments from the final GPDs includes the reconstruction of the $x$-dependence and the matching. Therefore, these results serve as very important cross-checks of the validity of our results.

\section{Summary}
\label{sec:summary}

In this paper, we present the first lattice QCD calculation of transversity GPDs for the proton, employing the quasi-distribution approach. GPDs are defined in the Breit frame, which we employ in this work. We use kinematic setups for both zero and nonzero skewness. In particular, we present results for $\{\xi,t\}=\{0,-0.69\,{\rm GeV}^2\}$ using momentum boosts $P_3=0.83,1.25,1.67$ GeV. For nonzero skewness, we have $\{|\xi|,t\}=\{1/3,-1.02\,{\rm GeV}^2\}$ for $P_3=1.25$ GeV. The matrix elements are renormalized in position space using a variation of the RI-MOM scheme, the so-called minimal projector. The choice of the projector is such that it isolates the tree-level contributions from the vertex functions of the operator. This is necessary, as the available matching formulas have been developed for that scheme~\cite{Liu:2019urm}. To compute the $x$-dependence of the GPDs, we apply the Backus-Gilbert method to obtain the quasi-GPDs. Finally, we apply the  perturbative matching equations to extract the light-cone GPDs. In particular, the analytic equations of the matching relate the quasi-GPDs defined in the RI scheme at a scale $\mu_0$, to the physical GPDs in the $\overline{\rm MS}$ scheme at 2 GeV.

We use a combination of operators, momentum source and sink, as well as parity projectors, so that we can disentangle the four transversity GPDs, $H_T$, $E_T$, $\widetilde{H}_T$ and $\widetilde{E}_T$. For the latter, we find zero signal within uncertainties, as it is suppressed compared to the other GPDs. The $P_3$-dependence, at fixed $-t=0.69$~GeV$^2$ and $\xi=0$, is investigated boosting the proton at $P_3=0.83, 1.25$ and  $1.67$~GeV. Our results in Fig.~\ref{fig:HT_mom_dependence_xi0} and Fig.~\ref{fig:HTtilde_ET_mom_dependence_xi0} show that momentum convergence in $H_T$ is observed at the two highest boosts. A much larger statistics is needed to fully establish such a conclusion for $E_T$ and $\widetilde{H}_T$, that suffer from large statistical errors. Nevertheless, there is qualitative agreement between our lattice results and the analysis of GPDs in the scalar diquark model of  Ref.~\cite{Bhattacharya:2019cme}, where, for example, $\widetilde{H}_T$ is negative, in agreement with our findings. At $P_3=1.25$~GeV, we also extract the GPDs at $|\xi|=1/3$ and $-t=1.02$~GeV$^2$. At non-zero $\xi$, there is a non-trivial distinction between the ERBL ($-\xi<x<+\xi$) and DGLAP ($-1<x<-\xi$, $\xi < x <1$) regions, and we find that the $t$-dependence of the GPDs is more prominent in the ERBL region. 

In addition to the individual GPDs, we extract the combination $E_T+2\widetilde{H}_T$ (see Fig.~\ref{fig:ETplus2HtildeT}), both at zero and non-zero skewness and for the two $t$-values considered in this work. This quantity provides the transverse spin-flavor dipole moment in an unpolarized target, $k_T$, through its lowest moment and in the forward limit ($t=0$). At the present stage, we are in no position to estimate $k_T$, because that would require the knowledge of $E_T$ and $\widetilde{H}_T$ for multiple $t$-values to extract their values at $t=0$ through fits. This is certainly a very interesting direction, that we will pursue in the future.

Our results for the transversity GPDs are combined with the unpolarized and helicity GPDs from Ref.~\cite{Alexandrou:2020zbe}, that were calculated on the same ensemble and for the same kinematic setup. We compare the three types of PDFs, and the effect of introducing momentum transfer and nonzero skewness (see Fig.~\ref{fig:leading_GPDs}). As expected, the GPDs, $H$, $\widetilde{H}$ and $H_T$,  are suppressed compared to their PDF counterparts. 

Another aspect of our analysis is the calculation of the two lowest Mellin moments for the GPDs. We also extract the lowest moment of quasi-GPDs using the relations of Ref.~\cite{Bhattacharya:2019cme}. This is an important part of this work, leading to a number of conclusions that are consistent with the expected relations. In a nutshell, we find that the $n=0$ Mellin moments of quasi-GPDs do not depend on $P_3$, even though the quasi-GPDs have an explicit dependence on $P_3$. In addition, the moments of GPDs obtained at different momenta are consistent. The expectation that the $n=0$ Mellin moments of GPDs and quasi-GPDs are the same, is confirmed by our results numerically. Also, the Mellin moments have the expected $t$-dependence, that is, they decrease as $-t$ increases. Another conclusion is that going from $n=0$ to $n=1$ results in decreasing the values for the moment. Last, but not least, the $n=0$ moments are fully consistent with their extraction from the matrix elements at $z=0$. These conclusions hold for all transversity GPDs, except $\widetilde{E}_T$, which is consistent with zero within our precision.

The calculation presented here is the first of a series of studies aiming at the calculation of GPDs on several ensembles, in order to quantify systematic uncertainties such as pion mass dependence and discretization effects. Having results from larger-volume ensembles will allow us to obtain the GPDs for several values of $t$ and fit the $t$-dependence. As previously mentioned, this is important for obtaining the forward limit for the GPDs that drop out of the matrix element at $t=0$.
In this way, lattice QCD can provide a robust way of probing the three-dimensional structure of the nucleon and complement the rich experimental programs aiming at unraveling this structure.

\vspace*{1cm}
\begin{acknowledgements}
We would like to thank all members of ETMC for their constant and pleasant collaboration. M.C. thanks S. Bhattacharya and Y. Zhao for useful discussions. K.C.\ is supported by the National Science Centre (Poland) grant SONATA BIS no.\ 2016/22/E/ST2/00013. M.C. acknowledges financial support by the U.S. Department of Energy Early Career Award under Grant No.\ DE-SC0020405. K.H. is supported by the Cyprus Research and Innovation Foundation under grant POST-DOC/0718/0100. 
F.S.\ was funded by the NSFC and the Deutsche Forschungsgemeinschaft (DFG, German Research
Foundation) through the funds provided to the Sino-German Collaborative Research Center TRR110 “Symmetries and the Emergence of Structure in QCD” (NSFC Grant No. 12070131001, DFG Project-ID 196253076 - TRR 110).
Partial support is provided by the European Joint Doctorate program STIMULATE of the European Union’s Horizon 2020 research and innovation programme under grant agreement No. 765048. Computations for this work were carried out in part on facilities of the USQCD Collaboration, which are funded by the Office of Science of the U.S. Department of Energy. This research was supported in part by PLGrid Infrastructure (Prometheus supercomputer at AGH Cyfronet in Cracow). 
Computations were also partially performed at the Poznan Supercomputing and Networking Center (Eagle supercomputer), the Interdisciplinary Centre for Mathematical and Computational Modelling of the Warsaw University (Okeanos supercomputer) and at the Academic Computer Centre in Gda\'nsk (Tryton supercomputer). The gauge configurations have been generated by the Extended Twisted Mass Collaboration on the KNL (A2) Partition of Marconi at CINECA, through the Prace project Pra13\_3304 "SIMPHYS". 
\end{acknowledgements}

\newpage
\bibliography{references}

\end{document}